\documentclass[jair,twoside,11pt,theapa]{article}
\usepackage{jair, theapa, rawfonts}

\usepackage{latexsym}
\usepackage{graphicx}
\usepackage{subfig}
\usepackage{times}

\ShortHeadings{Learning Asynchronous-Time Information Diffusion Models}
{Saito, Kimura, Ohara, \& Motoda}
\firstpageno{1}

\newcommand{\bbr}{\mbox{\boldmath $r$}}
\newcommand{\bbq}{\mbox{\boldmath $q$}}
\newcommand{\bbp}{\mbox{\boldmath $p$}}

\newcommand{\bbtheta}{\mbox{\boldmath $\Theta$}}

\begin{document}

\title{Learning Asynchronous-Time Information Diffusion Models\\ and its
Application to Behavioral Data Analysis\\ over Social Networks}

\author{\name Kazumi Saito \email k-saito@u-shizuoka-ken.ac.jp\\
\addr School of Administration and Informatics\\
University of Shizuoka\\
Shizuoka 422-8526, Japan
\AND
\name Masahiro Kimura \email kimura@rins.ryukoku.ac.jp \\
\addr Department of Electronics and Informatics\\
Ryukoku University\\
Shiga 520-2194, Japan
\AND
\name Kouzou Ohara \email ohara@it.aoyama.ac.jp \\
\addr Department of Integrated Information Technology\\
Aoyama Gakuin University\\
Kanagawa 229-8558, Japan
\AND
\name Hiroshi Motoda \email motoda@ar.sanken.osaka-u.ac.jp \\
\addr
Institute of Scientific and Industrial Research\\
Osaka University\\
Osaka 567-0047, Japan}

\maketitle

\begin{abstract}
One of the interesting and important problems of information diffusion
over a large social network is to identify an appropriate model from a
limited amount of diffusion information. There are two contrasting
approaches to model information diffusion. One is a push type model,
known as Independent Cascade (IC) model and the other is a pull type
model, known as Linear Threshold (LT) model. We extend these two models
(called AsIC and AsLT in this paper) to incorporate asynchronous time
delay and investigate 1) how they differ from or similar to each other
in terms of information diffusion, 2) whether the model itself is
learnable or not from the observed information diffusion data, and 3)
which model is more appropriate to explain for a particular topic
(information) to diffuse/propagate.  We first show that there can be
variations with respect to how the time delay is modeled, and derive the
likelihood of the observed data being generated for each model. Using
one particular time delay model, we show that the model parameters are
learnable from a limited amount of observation. We then propose a method
based on predictive accuracy by which to select a model which better
explains the observed data. Extensive evaluations were performed using
both synthetic data and real data. We first show using synthetic data
with the network structures taken from four real networks that there are
considerable behavioral differences between the AsIC and the AsLT models,
the proposed methods accurately and stably learn the model parameters,
and identify the correct diffusion model from a limited amount of
observation data. We next apply these methods to behavioral analysis of
topic propagation using the real blog propagation data, and show that
there is a clear indication as to which topic better follows which model
although the results are rather insensitive to the model selected at the
level of discussing how far and fast each topic propagates from the
learned parameter values. The correspondence between the topic and the
model selected is well interpretable considering such factors as
urgency, popularity and people's habit.
\end{abstract}


\section{Introduction}

The growth of Internet has enabled to form various kinds of large-scale
social networks, through which a variety of information including
innovation, hot topics and even malicious rumors can be propagated in
the form of so-called ``word-of-mouth'' communications. Social networks
are now recognized as an important medium for the spread of information,
and a considerable number of studies have been made
~\cite{newman:physrev02,newman:siam,gruhl:sigkdd,domingos:ieee,leskovec:ec,romero:www11,bakshy:wsdm11,mathioudakis:kdd11}.

Widely used information diffusion models in these studies are the {\em
  independent cascade (IC)}~\cite{goldenberg,kempe:kdd,kimura:tkdd} and
the {\em linear threshold (LT)}~\cite{watts,watts:ccr}. They have been
used to solve such problems as the {\em influence maximization
  problem}~\cite{kempe:kdd,yang:kdd,kimura:dmkd} and the {\em
  contamination minimization problem}~\cite{kimura:tkdd}. These two
models assume different mechanisms for information diffusion which are
based on two opposite views.  In the IC model each active node {\em
  independently} influences its inactive neighbors with given diffusion
probabilities ({\em information push style model}). In the LT model a
node is influenced by its active neighbors if their total weight exceeds
the threshold for the node ({\em information pull style model}). Which
model is more appropriate depends on the situation and selecting the
appropriate one for a particular problem is an interesting and important
problem.  To answer this question, first of all, we have to understand
the behavioral difference between there two models. 

Both models have parameters that need be specified in advance: diffusion
probabilities for the IC model, and weights for the LT model. However,
their true values are not known in practice, which poses a challenging
problem of estimating them from a limited amount of information
diffusion data that are observed as time-sequences of influenced
(activated) nodes. Fortunately this falls in a well defined parameter
estimation problem in machine learning setting.  Given a generative
model with its parameters and the independent observed data, we can
calculate the likelihood that the data are generated and can estimate
the parameters by maximizing the likelihood. This approach has a
thorough theoretical background. The way the parameters are estimated
depends on how the generative model is given. To the best of our
knowledge, we were the first to follow this line of research. We
addressed this problem first for the basic IC model~\cite{saito:kes08,kimura:sbp09}
and then its variant that incorporates asynchronous time delay (referred
to as the AsIC model)~\cite{saito:acml09}. We further applied this to a
variant of the LT model that also incorporates asynchronous time delay
(referred to as the AsLT model)~\cite{saito:sbp10,saito:ecml10}.

\citeA{gruhl:sigkdd} also challenged the same problem of estimating the
parameters and proposed an EM-like algorithm, but they did not formalize
the likelihood and it is not clear what is being optimized in deriving
the parameter update formulas. \citeA{goyal:wsdm} attacked this problem
from a different angle. They employed a variant of the LT model and
estimated the parameter values by four different methods, all of which
are directly computed from the frequency of the events in the observed
data. Their approach is efficient, but it is more likely ad hoc and
lacks in theoretical evidence.  \citeA{bakshy:ec} addressed the problem
of diffusion of user-created content (asset) and used the maximum
likelihood method to estimate the rate of asset adoption. However, they
only modeled the rate of adoption and did not consider the diffusion
model itself. Their focus was data analysis. \citeA{rodriguez:kdd}
proposed an efficient method of inferring a network from the observed
diffusion sequences based on the continuous time version of the IC
model, assuming the probability that a node affects its child node is a
function of the difference of the activation times between the two
nodes. Their focus is inferring the structure of the network rather than
inferring the best predictive model for a known network. They fixed a
model and approximated the likelihood function in such a way that the
simplified likelihood function can be maximized by adding a link in each
iteration. Recent work of \citeA{myers:nips10} is close to ours. They
used a model similar to but different in details from the AsIC model and
showed that the liklihood maximization problem can effectively be
transformed to a convex programming for which a global solution is
guaranteed\footnote{We discuss the difference between their model and
  our model in Section \ref{discussion}.}. Their focus was also
inferring the structure of the network.

In this paper, we first detail the Asynchronous Independent Cascade
Model and the Asynchronous Linear Threshold Model as two contrasting
information diffusion models. Both are extensions of the basic
Independent Cascade Model and Linear Threshold Model that incorporate
time delay in an asynchronous way. Especially we focus on the liklihood
derivation of these models. We show that there are a few variations of
time delay and different time delay models result in different liklihood
formulations.  We then show for a particular time delay model how to
obtain the parameter values that maximize the respective liklihood by
deriving an EM-like iterative approach using the observed sequence
data.  Indeed, being able to cope with asynchronous time delay is
indispensable to do realistic analysis of information diffusion because,
in the real world, information propagates along the continuous time
axis, and time-delays can occur during the propagation
asynchronously. In fact, the time stamps of the observed data are not
equally spaced. This means that the proposed learning method has to
estimate not only the diffusion parameters (diffusion probabilities for
the AsIC model and weights for the AsLT model) but also the time-delay
parameters from the observed data.  We identified that there are
basically two types of delay: {\em link delay} and {\em node delay}. The
former corresponds to the delay associated with information propagation,
and the latter corresponds to the delay associated with human action
which is further divided into two types: {\em non-override} and {\em
  override}. We choose {\em link delay} to explain the learning
algorithms and perform the experiments on this model. For the other time
delay models we only derive the likelihood functions that are required
for the learning algorithms.  Incorporating time-delay makes the
time-sequence observation data structural, which makes the analysis of
diffusion process difficult because there is no way of knowing which
node has activated which other node from the observation data sequence.

Knowing the optimal parameter values does not mean that the observation
follows the model well. We have to decide which model better explains
the observation and select the right (or more appropriate) model.  We
solve this problem by comparing the predictive accuracy of each
model. We use a variant of hold-out method applied to a set of
sequential data, which is similar to the leave-one-out method applied to
a multiple time sequence data, i.e., we use a part of the data, train
the model, predict the activation probability at one step later and
compare it with the observation. We repeat this by changing the size of
the training data.  

In summary, we want to 1) clarify how the AsIC model and the AsLT model
differ from or similar to each other in terms of information diffusion,
2) propose a method to learn the model parameters from a limited number
of observed data and show that the method is effective, and 3) show that
how the information diffuses depend on the topic and the proposed method
can identify which model is more appropriate to explain for a particular
topic (information) to diffuse/propagate.

We have performed extensive experiments to verify the proposed
approaches using both synthetic data and real data. Experiments using
synthetic data generated by the models (AsIC and AsLT) with network
structures taken from four real networks revealed that there are
considerable behavioral difference between the AsIC and the AsLT models,
and the difference can be explained by the diffusion mechanism
qualitatively. It is also shown that the proposed liklihood maximization
methods accurately and stably learn the model parameters, and identify
the correct diffusion model from a limited amount of observation
data. Experiments of behavioral analysis of topic propagation using the
real blog data show that the results are rather insensitive to the model
selected at an abstract level of discussing how relatively far and fast
each topic propagates from the learned parameter values but still there
is a clear indication as to which topic better follows which model.  The
correspondence between the topic and the model selected is well
interpretable considering such factors as urgency, popularity and
people's habit.

The paper is organized as follows. In Section 2, we introduce the two
contrasting information diffusion models (AsIC and AsLT) we used in
this paper, and in Section 3, we detail how the likelihood functions can
be formulated for various variations of time delay model and in Appendix
how the parameters can be obtained using one particular model of time
delay (link delay). In Section 4, we show the detailed analysis results
of behavioral difference between AsIC and AsLT obtained by using four
real network structures. In Section 5 we detail the learning performance
(accuracy of parameter learning and influential node ranking) using the
synthetic data obtained by the same four real network structure. In
Section 6 we focus on model selection using both synthetic data and a
real blog network data. In Section 7 we discuss some of the important
issues regarding the related work and those for future work. We end the
paper by summarizing what has been achieved in Section 8.

\section{Information Diffusion Models}
\label{models}

\subsection{Two Contrasting Diffusion Models}

It is quite natural to bring in the notion of information sender and
receiver. The IC model is sender-centered. It is motivated by epidemic
spread in which the disease carrier is the information sender. If a
person gets infected, his or her neighbors also get infected, {\em
  i.e.}, the information sender tries to push information to its
neighbors. The LT model is receiver-centered. It is based on the view
that the receiver has a control over the information flow. This models
the way innovation propagates. For example, a person is attempted to buy
a new tablet PC if many of his or her neighbors have purchased it and said
that it is good, {\em i.e.}, the information receiver tries to pull
information.

Both models have respective reasons for their working mechanisms, but
they are quite contrasting to each other. We are interested in 1) how
they differ from or similar to each other in terms of information
diffusion, 2) whether the model itself is learnable or not from the
observed information diffusion data, and 3) which model is more
appropriate to explain for a particular topic (information) to
diffuse/propagate. Both models have parameters, {\em i.e.}, diffusion
probability attached to each directional link in the IC model and weight
attached to each directional link in the LT model. As shown later in
Section \ref{learning AsLT}, the weight is equivalent to a
probability. Thus, intuitively both models appear to be comparative in
terms of the average influence degree if the parameter values are
comparable. The simulation results, however, show that these two models
behave quite differently. We will explain why they are different in
Section \ref{behavior-difference-results}.

In the following two subsections we will describe the two diffusion
models that we use in this paper: the {\em asynchronous independent
  cascade (AsIC) model}, first introduced by \citeA{saito:acml09}, and
the {\em asynchronous linear threshold (AsLT) model}, first introduced
by \citeA{saito:sbp10}. They differ from the basic IC and LT models in
that they explicitly handle the time delay. The diffusion process
evolves with time. The basic models deal with time by allowing nodes to
change their states in a synchronous way at each discrete time step,
{\em i.e.}, no time delay is considered, or one can say that every state
change is uniformly delayed exactly by one discrete time step. Their
asynchronous time delay versions explicitly treat the time delay of each
node independently. We discuss the notion of time delay in more depth in
Section \ref{notion-of-time-delay}. 

The models we explain in the following two sub sections and the learning
algorithms we describe in Section \ref{learning} are based on a
particular time-delay model, which we call {\em link delay}. This is the
model that the time delay is caused by the communication channel, e.g.,
network traffic and/or some malfunction, and as soon as the information
arrives at the destination, the node responds without delay.

Before we explain the models, we give the definition of a graph and
children and parents of a node. A graph we use is a directed graph $G =
(V, E)$ without self-links, where $V$ and $E$ $(\subset V \times V)$
stand for the sets of all the nodes and links, respectively. For each
node $v$ in the network $G$, we denote $F(v)$ as a set of child nodes of
$v$, i.e.,
$$
F(v) = \{w \in V; (v, w) \in E\}.
$$
Similarly, we denote $B(v)$ as a set of parent nodes of $v$, i.e.,
$$
B(v) = \{u \in V; (u, v) \in E\}.
$$ 
We call nodes {\em active} if they have been influenced with the
information. In the following models, we assume that nodes can switch
their states only from inactive to active, but not the other way around,
and that, given an initial active node set $S$, only the nodes in $S$
are active at an initial time.

\subsection{Asynchronous Independent Cascade Model}
We first recall the definition of the IC model according to the work of
\citeA{kempe:kdd}, and then introduce the AsIC model. In the IC model,
we specify a real value $p_{u,v}$ with $0 < p_{u,v} < 1$ for each link
$(u, v)$ in advance. Here $p_{u, v}$ is referred to as the {\em
  diffusion probability} through link $(u,v)$. The diffusion process
unfolds in discrete time-steps $t \geq 0$, and proceeds from a given
initial active set $S$ in the following way. When a node $u$ becomes
active at time-step $t$, it is given a single chance to activate each
currently inactive child node $v$, and succeeds with probability
$p_{u,v}$. If $u$ succeeds, then $v$ will become active at time-step
$t+1$. If multiple parent nodes of $v$ become active at time-step $t$,
then their activation attempts are sequenced in an arbitrary order, but
all performed at time-step $t$. Whether or not $u$ succeeds, it cannot
make any further attempts to activate $v$ in subsequent rounds. The
process terminates if no more activations are possible.

In the AsIC model, we specify real values $r_{u,v}$ with $r_{u, v} > 0$
in advance for each link $(u, v) \in E$ in addition to $p_{u,v}$, where
$r_{u,v}$ is referred to as the {\em time-delay parameter} through link
$(u,v)$.  The diffusion process unfolds in continuous-time $t$, and
proceeds from a given initial active set $S$ in the following way.
Suppose that a node $u$ becomes active at time $t$. Then, $u$ is given a
single chance to activate each currently inactive child node $v$. We
choose a delay-time $\delta$ from the exponential distribution\footnote{
  Similar formulation can be derived for other distributions such as
  power-law and Weibull.}  with parameter $r_{u,v}$. If $v$ has not been
activated before time $t + \delta$, then $u$ attempts to activate $v$,
and succeeds with probability $p_{u,v}$. If $u$ succeeds, then $v$ will
become active at time $t + \delta$. Said differently, whichever parent
$u$ that succeeds in satisfying the activation condition and for which
the activation time is the earliest considering the time delay
associated with each link can actually activate the node.  Under the
continuous time framework, it is unlikely that $v$ is activated
simultaneously by its multiple parent nodes exactly at time $t +
\delta$. So we do not consider this
possibility.
Whether or not $u$ succeeds, it cannot make any further attempts to
activate $v$ in subsequent rounds.
The process terminates if no more activations are possible.

\subsection{Asynchronous Linear Threshold Model}

Same as the above, we first recall the LT model. In this model, for
every node $v$ $\in$ $V$, we specify a {\em weight} $(q_{u, v} >
0)$ from its parent node $u$ in advance such that 
$$
\sum_{u \in B(v)} q_{u, v} \leq 1.
$$ The diffusion process from a given initial active set $S$ proceeds
according to the following randomized rule.  First, for any node $v$
$\in$ $V$, a {\em threshold} $\theta_v$ is chosen uniformly at random
from the interval $[0, 1]$. At time-step $t$, an inactive node $v$ is
influenced by each of its active parent nodes, $u$, according to weight
$q_{u, v}$. If the total weight from active parent nodes of $v$ is no
less than $\theta_v$, that is,
$$
\sum_{u \in B_t(v)} q_{u, v} \geq \theta_v,
$$
then $v$ will become active at time-step
$t+1$. Here, $B_t(v)$ stands for the set of all the parent nodes of $v$
that are active at time-step $t$. The process terminates if no more
activations are possible.

The AsLT model is defined in a similar way to the AsIC. In the AsLT
model, in addition to the weight set $\{ q_{u, v} \}$, we specify real
values $r_{u,v}$ with $r_{u,v} > 0$ in advance for each link $(u,v)$.
Same as for AsIC, we refer to $r_{u,v}$ as the {\em time-delay
  parameter} through link $(u,v)$.  The diffusion process unfolds in
continuous-time $t$, and proceeds from a given initial active set $S$ in
the following way. Each active parent $u$ of the node $v$ exerts its
effect on $v$ with the time delay $\delta$ drawn from the exponential
distribution with the delay parameter $r_{u,v}$.  Suppose that the
accumulated weight from the active parents of node $v$ has become no
less than $\theta_v$ at time $t$ for the first time. Then, the node $v$
becomes active at $t$ without any delay and exerts its effect on its
child with a delay associated with its link. This process is repeated
until no more activations are possible.

\section{Learning Algorithms}
\label{learning}

We define the diffusion parameter vector $\bbp$ and the time-delay
parameter vector $\bbr$ by
$$
\bbp = {(p_{u, v})}^{}_{(u, v) \in E} \ \ \ \ \
\bbr = {(r_{u, v})}^{}_{(u, v) \in E}
$$
for the AsIC model, and the weight parameter vector $\bbq$ and the time-delay parameter vectors
$\bbr$ by 
$$
\bbq = {(q_{u, v})}^{}_{(u, v) \in E}  \ \ \ \ \
\bbr = {(r_{u, v})}^{}_{(u, v) \in E}
$$ 
for the AsLT model.  We next consider an observed data set of $M$
independent information diffusion results,
$$
\{D_m; \ m = 1, \cdots, M \}.
$$
Here, each $D_m$ is a set of pairs of 
active node and its activation time 
in the $m$-th diffusion result, 
$$
D_m = \{ (u, t_{m,u}), \ (v, t_{m,v}), \ \cdots \}.
$$ 
We denote by $t_{m, v}$ the activation time of node $v$
for the $m$-th diffusion result.
For each $D_m$, we denote the observed initial time by 
$$
t_m = \min \{t_{m,v}; \ (v, t_{m,v}) \in D_m \},
$$ 
and the observed final time by 
$$
T_m \geq \max \{t_{m,v}; \ (v, t_{m,v}) \in D_m \}.
$$ 
Note that $T_m$ is not necessarily equal to the final activation time. 
Hereafter, we express our observation data by 
$$
{\cal D}_M = \{(D_m, T_m); \ m = 1, \cdots, M \}.
$$
For any $t \in [t_m, T_m]$, we set
$$
C_m(t) = \{v \in V; \ (v, t_{m,v}) \in D_m, \ t_{m,v} < t \}.
$$ 
Namely, $C_m(t)$ is the set of active nodes before time $t$ in the
$m$-th diffusion result. For convenience sake, 
we use $C_m$ as referring to the set of all the active nodes 
in the $m$-th diffusion result, i.e.,
$$
C_m = \bigcup_{t \geq t_m} C_m(t).
$$ 
Moreover, we define a set of non-active nodes with at least 
one active parent node for each by 
$$
\partial C_m = \{v \in V; \ (u,v) \in E, \ u \in C_m, \ v \notin C_m \}.
$$ 
For each node $v \in C_m \cup \partial C_m$, 
we define the following subset of parent nodes, 
each of which had a chance to activate $v$. 
\begin{eqnarray*}
{\cal B}_{m,v} & = & \left \{ \begin{array}{ll} 
B(v) \cap C_m(t_{m,v}) & \mbox{if  $v \in C_m$}, \\ 
B(v) \cap C_m & \mbox{if  $v \in \partial C_m$}. \end{array} \right.
\label{active-parent}
\end{eqnarray*}

Note that the underlying model behind the observed data is not available
in reality. Thus, we investigate how the model affects the information
diffusion results, and consider selecting a model which better explains the
given observed data from the candidates, i.e., AsIC and AsLT models.  To
this end, we first have to estimate the values of $\bbr$ and $\bbp$
for the AsIC model, and the values of $\bbq$ and $\bbr$ for the AsLT
model for the given ${\cal D}_M$.

\subsection{Learning Parameters of AsIC Model}
\label{learning AsIC}

First, we propose a method of learning the model parameters from the
observed data for the AsIC model.
To estimate the values of $\bbr$ and $\bbp$ from ${\cal D}_M$ for the
AsIC model, we derive the likelihood function ${\cal L} (\bbr, \bbp;
{\cal D}_M)$ to use as the objective function.

First, for the $m$-th information diffusion result, 
we consider any node $v \in C_m$ with $t_{m,v} > t_m$, and
derive the probability density $h_{m,v}$ that the node $v$ is activated at time $t_{m,v}$.
Note that $h_{m, v} = 1$ if $t_{m,v} = t_m$. 
Let ${\cal X}_{m,u,v}$ 
denote the probability density that
a node $u$ $\in$ ${\cal B}_{m,v}$
activates the node $v$ at time $t_{m,v}$, that is,
\begin{equation}
{\cal X}_{m,u,v}
= p_{u,v} r_{u, v} \exp(- r_{u, v}(t_{m,v}-t_{m,u})).
\label{calA}
\end{equation}
Let ${\cal Y}_{m,u,v}$
denote
the probability that
the node $v$ is not
activated by a node $u$ $\in$ ${\cal B}_{m,v}$
within the time-period $[t_{m,u}, t_{m,v}]$,
that is,
\begin{eqnarray}
{\cal Y}_{m,u,v}
& = & 1 - p_{u,v} \int_{t_{m,u }}^{t_{m,v}}  r_{u, v} \exp(- r_{u, v}(t-t_{m,u})) dt
\nonumber\\
& = & p_{u,v} \exp(- r_{u, v}(t_{m,v}-t_{m,u})) + (1-p_{u,v}).
\label{calB}
\end{eqnarray}
If there exist multiple active
parents for the node $v$, i.e.,
$|{\cal B}_{m,v}| > 1$,
we need to consider possibilities that each parent node
succeeds in activating $v$ at time $t_{m,v}$.
However, in case of the continuous time delay model,
we don't have to consider simultaneous activations by multiple active parents
due to the continuous property.
Here, for any $u \in {\cal B}_{m,v}$, let $h_{m,v} (u)$ be the
probability density that the node $u$ activates $v$ at time $t_{m,v}$
but all the other nodes $z$ in ${\cal B}_{m,v}$ have failed in activating $v$
within the time-period $[t_m, t_{m,v}]$ for the $m$-th information
diffusion result.  Then, we have
$$
h_{m,v} (u) = {\cal X}_{m,u,v}
\prod_{z \in {\cal B}_{m,v} \setminus \{u\}} {\cal Y}_{m,z,v}.
$$
Since the probability density $h_{m, v}$ is given by $h_{m,v} = \sum_{u \in {\cal B}_{m,v}} h_{m,v} (u)$,
we have
\begin{eqnarray}
h_{m, v} & = & \sum_{u \in {\cal B}_{m,v}} {\cal X}_{m,u,v} \left (
\prod_{z \in {\cal B}_{m,v} \setminus \{u\}} {\cal Y}_{m,z,v} \right ).
\nonumber \\
& = & \prod_{z \in {\cal B}_{m,v}} {\cal Y}_{m,z,v}
\sum_{u \in {\cal B}_{m,v}} {\cal X}_{m,u,v} ({\cal Y}_{m,u,v})^{-1}.
\label{probability1}
\end{eqnarray}
Note that we are not able to know which node $u$ actually activated the
node $v$. This can be regarded as a hidden structure.

Next, for the $m$-th information diffusion result,
we consider any link $(v, w) \in E$ such that
$v \in C_m$ and $w$ $\notin$ $C_m$, and
derive the probability $g_{m,v}$ that the node $v$ fails to activate its child nodes.
Note that $g_{m,v} = 1$ if $F(v) \setminus C_m = \emptyset$.
Let $g_{m,v,w}$ denote the probability that
the node $w$
is not activated by the node $v$ within the observed time period $[t_m, T_m]$.
We can easily derive the following equation:
\begin{equation}
g_{m, v, w} = p_{v,w} \exp(- r_{v, w}(T_m-t_{m,v})) + (1-p_{v,w}).
\label{probability4}
\end{equation}
Here we can naturally assume that each information diffusion process
finished sufficiently earlier than the observed final time, i.e.,
$T_m \gg  \max\{t_{m,v}; (v, t_{m,v}) \in D_m \}$.
Thus,
as $T_m \to \infty$ in Equation~(\ref{probability4}),
we can assume
\begin{equation}
g_{m, v, w} \ = \ 1 - p_{v,w}.
\label{probability5}
\end{equation}
Therefore, the probability $g_{m, v}$ is given by
\begin{equation}
g_{m, v} \ = \ \prod_{w \in F(v)\setminus C_m} g_{m,v,w}.
\label{probability6}
\end{equation}

By using Equations~(\ref{probability1}) and (\ref{probability6}), and
the independence properties,
we can define the likelihood function ${\cal L} (\bbr, \bbp; {\cal D}_M)$
with respect to $\bbr$ and $\bbp$ by
\begin{eqnarray}
{\cal L}(\bbr, \bbp;{\cal D}_M) = 
\prod_{m = 1}^{M} \prod_{v \in C_m} \left( h_{m, v} \ 
g_{m, v} \right).
\label{objective}
\end{eqnarray}

In this paper, we focus on Equation~(\ref{probability5}) for simplicity,
but we can easily modify our method to cope with the general one (i.e.,
Equation~(\ref{probability4})). Thus, our problem is to obtain the
values of $\bbr$ and $\bbp$, which maximize Equation~(\ref{objective}).
For this estimation problem, we derive a method based on an iterative
algorithm in order to stably obtain its solution. The details of the
parameter update algorithm are given in Appendix A.

\subsection{Learning Parameters of AsLT Model}
\label{learning AsLT}

Next, we propose a method of learning the model parameters from the
observed data for the AsLT model.
Similarly to the AsIC model, we first derive the likelihood function
${\cal L} (\bbr, \bbq; {\cal D}_M)$ with respect to $\bbr$ and
$\bbq$.  For the sake of technical convenience, we introduce a slack
weight $q_{v, v}$ for each node $v \in V$ such that 
$$
q_{v, v} + \sum_{u \in B(v)} q_{u, v} = 1.
$$ 
Here note
that we can regard each weight $q_{*, v}$ as a multinomial
probability since a threshold $\theta_v$ is chosen uniformly at random
from the interval $[0, 1]$ for each node $v$.

First, for the $m$-th information diffusion result, 
we fix any node $v \in C_m$ with $t_{m,v} > t_m$, and
derive the probability density $h_{m,v}$ that the node $v$ is activated at time $t_{m,v}$.
Note that $h_{m, v} = 1$ if $t_{m,v} = t_m$. 
Suppose any parent node $z \in {\cal B}_{m,v}$ exerts its
effect on $v$ with a delay $\delta_{z,v}$. 
Further suppose that the threshold $\theta_v$ is first exceeded when the effect of 
$u \in {\cal B}_{m,v}$ reaches $v$ after the delay $\delta_{u,v}$. 
We define the subset ${\cal B}_{m,v}(u)$ of ${\cal B}_{m,v}$ by
$$
{\cal B}_{m,v} (u) = \{
z \in {\cal B}_{m,v}; \
t_{m,z} + \delta_{z,v} < t_{m,u} + \delta_{u,v} 
\}.
$$
Then, we have
$$
\sum_{z \in {\cal B}_{m,v}(u)} q_{z,v} \ < \ \theta_v \ \leq \
q_{u,v} + \sum_{z \in {\cal B}_{m,v}(u)} q_{z,v}.
$$
This implies that the probability that $\theta_v$ is chosen from this range is $q_{u,v}$.
Let ${\cal X}_{m,u,v}$ denote the probability density that node $u$ activates node $v$
at time $t_{m,v}$.
Then, we have
\begin{eqnarray}
{\cal X}_{m,u,v} = q_{u,v} r_{u,v} \exp(- r_{u, v}(t_{m,v}-t_{m,u})).
\label{calX-lt}
\end{eqnarray}
Since the probability density $h_{m,v}$ is given by
$h_{m,v} = \sum_{u \in {\cal B}_{m,v}} {\cal X}_{m,u,v}$, we have
\begin{eqnarray}
h_{m, v} & = & \sum_{u \in {\cal B}_{m,v}}
q_{u,v} r_{u,v} \exp(- r_{u, v}(t_{m,v}-t_{m,u})).
\label{AsLT:probability1}
\end{eqnarray}

Next, for the $m$-th information diffusion result,
we consider any node $v \in \partial C_m$, and
derive the probability $g_{m,v}$ that node $v$
is not activated within the observed time period $[t_m, T_m]$.
We can calculate $g_{m,v}$ as
\begin{eqnarray}
g_{m, v} & = & 1 - \hspace{-3pt}\sum_{u \in {\cal B}_{m,v}} q_{u,v}
\int_{t_{m,u}}^{T_m} r_{u,v} \exp(- r_{u,v}(t-t_{m,u})) dt \nonumber\\
& = &
 1 - \hspace{-3pt}\sum_{u \in {\cal B}_{m,v}} q_{u,v}
(1 - \hspace{-3pt}\exp(- r_{u,v}(T_m-t_{m,u}))) \nonumber \\ 
& = & q_{v,v} + \sum_{u \in B(v) \setminus {\cal B}_{m,v}} q_{u,v}
+ \sum_{u \in {\cal B}_{m,v}} q_{u,v} \exp(-r_{u,v}(T_m-t_{m,u})). 
\label{AsLT:probability5}
\end{eqnarray}
Therefore, by using Equations~(\ref{AsLT:probability1}) and 
(\ref{AsLT:probability5}), and the independence properties, we can 
define the likelihood function ${\cal L} (\bbr, \bbq; {\cal D}_M)$
with respect to $\bbr$ and $\bbq$ by
\begin{eqnarray}
{\cal L}(\bbr, \bbq;{\cal D}_M) = 
\prod_{m = 1}^{M} \left (
\prod_{v \in C_m} h_{m, v} \right ) \left (
\prod_{v \in \partial C_m} g_{m,v} \right ). 
\label{AsLT:objective}
\end{eqnarray}
Thus, our problem is to obtain the time-delay parameter vector $\bbr$
and the weight parameter vector $\bbq$, which together maximize
Equation~(\ref{AsLT:objective}). The details of the parameter update
algorithm are given in Appendix B.

\subsection{Alternative Time-delay models}\label{alternative-time-delay}
In Section \ref{models} we introduced one instance of time delay, i.e.,
{\rm link delay}. In this subsection we discuss time delay phenomena in
more depth for both the AsIC and the AsLT models.

\subsubsection{Notion of Time-delay}\label{notion-of-time-delay}

Each parent $u$ of a node $v$ can be activated independently of the
other parents and because the associated time delay from a parent to its
child is different for every single pair, which parent $u$ actually
affects the node $v$ in which order is more or less opportunistic.

To explicate the information diffusion process in a more realistic
setting, we consider two examples, one associated with blog posting and
the other associated with electronic mailing. In case of blog posting,
assume that some blogger $u$ posts an article. Then it is natural to
think that it takes some time before another blogger $v$ comes to notice
the posting. It is also natural to think that if the blogger $v$ reads
the article, he or she takes an action to respond (activated) because
the act of reading the article is an active behavior. In this case, we
can think that there is a delay in information diffusion from $u$ to $v$
(from $u$'s posting and $v$'s reading) but there is no delay in $v$
taking an action (from $v$'s reading to $v$'s posting). In case of
electronic mailing, assume that someone $u$ sends a mail to someone else
$v$. It is natural to think that the mail is delivered to the receiver
$v$ instantaneously. However, this does not necessarily mean that $v$
reads the mail as soon as it has been received because the act of
receiving a mail is a passive behavior. In this case, we can think that
there is no delay in information diffusion from $u$ to $v$ ($u$'s
sending and $v$'s receiving) but there is a delay in $v$ taking an
action (from $v$'s receiving to $v$'s sending). Further, when $v$
notices the mail, $v$ may think to respond to it later. But before $v$
responds, a new mail may arrive which needs a prompt response and $v$
sends a mail immediately. We can think of this as an update of acting
time.\footnote{Note that there are two actions here, reading and sending,
but the activation time in the observed sequence data corresponds to the
time $v$ sends a mail.} These are just two examples, but it appears
worth distinguishing the difference of these two kinds of time delay and
update scheme (override of decision) in a more general setting.

In view of the discussion above, we define two types of delay: link
delay and node delay. It is easiest to think that link delay corresponds
to propagation delay and node delay corresponds to action delay. We
further assume that they are mutually exclusive. This is a strong
restriction as well as a strong simplification by necessity because the
activation time of a node we can observe is a sum of the activation time
of its parent node and the two delays and we cannot distinguish between
these two delays. Thus we have to choose either one of the two as
occurring exclusively for the likelihood maximization to be feasible. In
addition, in case of node delay there are two types of activation:
non-override and override. The former sticks to the initial decision
when to activate and the latter can decide to update (override) the time
of activation multiple times to the earliest possible each time one of
the parents gets newly activated.  In summary, {\rm node delay} can go
with either override or non-override, and {\em link delay} can only go
with non-override.

Since we have already derived the likelihood function for link delay,
here we consider the likelihood function for node delay.  In this case,
the time delay parameter vector $\bbr$ is expressed as $\bbr = (r_v)_{v
  \in V}$.  The likelihood function ${\cal L}(\bbr, \bbp; {\cal D}_M)$
for the AsIC in the case of node delay is given by
Equation~(\ref{objective}), where $h_{m,v}$ is the probability density
that node $v$ is activated at time $t_{m,v}$ for the $m$-th information
result, and $g_{m,v}$ is the probability that node $v$ does not activate
its child nodes within the observed time period $[t_m, T_m]$ for the
$m$-th information result.  Note that $g_{m,v}$ remains the same as in
the case of link delay (see Equations~(\ref{probability5}) and
(\ref{probability6})).  The likelihood function ${\cal L}(\bbr, \bbq;
{\cal D}_M)$ for the AsLT in the case of node delay is given by
Equation~(\ref{AsLT:objective}), where the definition of $h_{m,v}$ is
the same as above, and $g_{m,v}$ is the probability that the node $v$ is
not activated within the observed time period $[t_m, T_m]$ for the $m$th
information result.  Note also that $g_{m,v}$ remains the same as in the
case of link delay (see Equation~(\ref{AsLT:probability5})).  Therefore,
our task now is: We fix any node $v \in C_m$ with $t_{m,v} > t_m$, and
present the probability density $h_{m,v}$ that node $v$ is activated at
time $t_{m,v}$ for the $m$-th information result in the case of node
delay, Here for simplicity, we order the active parent node $u \in {\cal
  B}_{m,v}$ of node $v$ according to the time $t_u$ it was activated,
and set
$$
{\cal B}_{m,v} = \{u_1, u_2, ..., u_J \}, \ \ \ 
t_{m, u_1} < t_{m, u_2} < \cdots < t_{m, u_J}.
$$

\subsubsection{Alternative Asynchronous Independent Cascade Model}\label{alt-asic}

First,
we derive $h_{m,v}$ for node delay with non-override and 
$h_{m,v}$ for node delay with override
in the case of the AsIC model.

\paragraph{Node delay with non-override}

There is no delay in propagating the information to the node $v$ from
the node $u$, but there is a delay $\delta$ before the node $v$ gets
actually activated. Assume that it is the node $u_i$ that first
succeeded in activating the node $v$ (more precisely satisfying the
activation condition). Since there is no link delay and no override, it
must be the case that all the other parents that had become active
before $t_{u_i}$ must have failed in activating $v$ (more precisely
satisfying the activation condition).  Since the node $v$ decides when
to actually activate itself at the time the node $u_i$ succeeded in
satisfying the activation condition and would not change its mind, other
nodes which may have been activated after the node $u_i$ got activated 
could do nothing on the node $v$.  Thus, the
probability density $h_{m,v}$ is given by

\begin{eqnarray*}
h_{m,v} & = & \sum_{j=1}^J {\cal X}_{m,u_j,v} \prod_{i=1}^{j-1} (1-p_{u_i,v}),
\label{nodedelay-nooverride-ic}
\end{eqnarray*}
where ${\cal X}_{m,u_j,v}$ is the probability density that node $u_j$ activates node $v$
at time $t_{m,v}$, and is obtained by 
\begin{equation}
{\cal X}_{m,u_j,v}
= p_{u_j, v} r_v \exp(- r_v (t_{m,v}-t_{m,u_j})),
\label{calX}
\end{equation}
(see Equation~(\ref{calA})).
Note that in comparison to Equation~(\ref{probability1}), 
the probability ${\cal Y}_{m,u_i,v}$ 
is replaced by $(1-p_{u_i,v})$. 

\paragraph{Node delay with override}

In this case the actual activation time is allowed to be updated. For example,
suppose that the node $u_i$ first succeeded in satisfying the activation
condition of the node $v$ and the node $v$ decided to activate itself at
time $t_{u_i}+\delta_i$. At some time later but before
$t_{u_i}+\delta_i$, other parent $u_j$ also succeeded in satisfying the
activation condition of the node $v$. Then the node $v$ is allowed to
change its actual activation time to time $t_{u_j}+\delta_j$ if it is
before $t_{u_i}+\delta_i$. Thus, the probability density $h_{m,v}$ is
given by
\begin{eqnarray*}
h_{m,v} & = & \sum_{j=1}^J {\cal X}_{m,u_j,v} \prod_{i=1,i\ne j}^{J} {\cal Y}_{m,u_i,v}.
\label{nodedelay-override-ic}
\end{eqnarray*}
Here, ${\cal X}_{m,u_j,v}$ is the probability density that node $u_j$ activates node $v$
at time $t_{m,v}$, and is obtained by Equation~(\ref{calX}).
Also,
${\cal Y}_{m,u_i,v}$ is the probability that node $v$ is not activated by node $u_i$
within the time-period $[t_{m, u_i}, t_{m,v}]$, and is obtained by
$$
{\cal Y}_{m,u_i,v}
= p_{u_i,v} \exp(- r_v (t_{m,v}-t_{m,u_i})) + (1-p_{u_i,v})
$$
(see Equation~(\ref{calB})).
Note that this formula $h_{m,v}$ is equivalent to Equation~(\ref{probability1}) 
except that the parameter $r_{u,v}$ is replaced by $r_{v}$.

\subsubsection{Alternative Asynchronous Linear Threshold Model}\label{alt-aslt}

Next, we derive $h_{m,v}$ for node delay with non-override and $h_{m,v}$
for node delay with override in the case of the AsLT model.

\paragraph{Node delay with non-override}

As soon as the parent node $u_i$ is activated, its effect is immediately
exerted to its child $v$. The delay depends on the node $v$'s choice.
Suppose the node $v$ first became activated for the $i$-th parent
according to the time $t_{u_i}$ ordering. Then by the same reasoning as
in Section \ref{learning AsLT}, the threshold $\theta_v$ is between
$\sum_{j=1}^{i-1}q_{u_j,v}$ and $\sum_{j=1}^{i-1}q_{u_j,v} +
q_{u_i,v}$, and the probability density $h_{m,v}$ can be expressed
as
\begin{eqnarray*}
h_{m, v} = \sum_{j=1}^J {\cal X}_{m,u_j,v},
\end{eqnarray*}
where ${\cal X}_{m,u_j,v}$ is the probability density that node $u_j$ activates node $v$
at time $t_{m,v}$, and is obtained by Equation~(\ref{calX-lt}).
Note that this formula is equivalent to Equation~(\ref{AsLT:probability1}) 
except that the parameter $r_{u,v}$ is replaced by $r_{v}$. 

\paragraph{Node delay with override}

Here, multiple updates of the activation time of the node $v$ is
allowed. Suppose that the node $v$'s threshold is first exceeded by
receiving the effect of the parent $u_j$. All the parents that have
become activated after that can still influence the updates. Among these
parents, let $u_i$ be the one which succeeded in activating the node $v$
and let $\{u_\zeta\}$ be the other parents that failed.  Then, the
probability density ${\cal X}_{m,u_j,v}$ that the node $v$ is activated
at time $t_{m,v}$ by the node $u_i$, which get activated later than
$u_j$ for which the threshold is first exceeded is given by
\begin{eqnarray*}
{\cal X}_{m,u_j,v} &=& q_{u_j,v}\sum_{i=j}^J r_v\exp(- r_{v}(t_{m,v}-t_{m,u_i}))
\prod_{\zeta=j,\zeta \ne i}^{J}\int_{t_{m,v}}^{\infty} r_{v}
\exp(-r_{v}(t-t_{m,u_\zeta}))dt \nonumber \\
&=& q_{u_j,v}(J-j+1)r_v \prod_{i=j}^J \exp(-r_{v}(t_{m,v}-t_{m,u_i})).
\label{calX-ls2}
\end{eqnarray*}
Thus, we obtain
\begin{eqnarray*}
 h_{m,v} = \sum_{j=1}^J {\cal X}_{m,u_j,v}.
\end{eqnarray*}
Note that this formula is substantially different from Equation~(\ref{AsLT:probability1}). 

\subsubsection{Summary of Different Time Delay Models}

We note that $h_{m,v}$ for {\em link delay} and {\em node delay with
  override} is identical for the AsIC model and that for {\em link
  delay} and {\em node delay with non-override} is identical for the
AsLT model, except for a minor notational difference in the time delay
parameter $r$ in both. Thus, there are basically two cases for each
model. We omit to show how different time delay models affect diffusion
phenomena. There are indeed some differences in transient time period
(for the first 10 to 30 time span in unit of average time
delay).\footnote{Note that difference in the time delay models vanishes
  when an equilibrium is reached.} The difference becomes larger as the
values for diffusion parameters become larger as expected. For more
details, see the work of \citeA{saito:acml10}. 

We only showed the parameter learning algorithms for the case of link
delay for both AsIC and AsLT models in Appendix. It is straightforward
to derive the similar algorithm for the other time delay models.

\subsection{Assumptions Introduced in Parameter Setting}\label{parameters}

The formulations so far assumed that the parameters ($p_{u,v}, q_{u,v}$
and $r_{u,v}$\footnote{To be more precise we assumed that $r_{u,v}=r_v$ in case of
  node-delay. Simplification in this case can also be made
  accordingly.})  that appear both in the AsIC and the AsLT models
depend on individual link $\{u,v\} \in E$. The number of parameters,
thus, is equal to the number of links, which is huge for any realistic
social network. This means that we need a prohibitively huge amount of
observation data that passes each link at least several times to obtain
accurate estimates for these parameters that do not overfit the
data. This is not realistic and we can introduce a few alternative
simplifying assumptions to avoid this overfitting problem.

The simplest one would be to assume that each of the parameters
$p_{u,v}, q_{u,v}$ and $r_{u,v}$ be represented by a single variable for
the whole network. For a diffusion probability, we assume a uniform
value $p_{u,v}=p$ for all links. For a weight we assume a uniform
coefficient $q$ such that $q_{u,v}= \frac{q}{|B(v)|}$, {\em i.e.}, the
weight $q_{u,v}$ is proportional to the reciprocal of the number of
$v$'s parents. This is the simplest realization to satisfy the
constraint $\sum_{u \in B(v)} q_{u,v} \le 1$. As can be shown later in
Section \ref{parameter-estimation}, this is a reasonable approximation
to discuss information diffusion for a specific topic. Next
simplification would be to divide $E$ (or $V$) into subsets $E_1, E_2,
..., E_{L_E}$ (or $V_1, V_2, ..., V_{L_V}$) and assign the same value
for each parameter within each subset. For example, we may divide the
nodes into two groups: those that strongly influence others and those
not, or we may divide the nodes into another two groups: those that are
easily influenced by others and those not. Links connecting these nodes
can accordingly be divided into subsets. If there is some background
knowledge about the node grouping, our method can make the best use of
it. Obtaining such background knowledge is also an important research
topic in the knowledge discovery from social networks. Yet another
simplification which looks more realistic would be to focus on the
attribute of each node and assume that there is a generic dependency
between the parameter values of a link and the attribute values of the
connected nodes and learn this dependency rather than learn the
parameter values directly from the data. In \citeA{saito:ismis11} we
adopted this approach assuming a particular class of attribute
dependency, and confirmed that the dependency can be correctly learned
even if the number of parameters is several tens of thousands. Learning
a function is much more realistic and does not require such a huge
amount of data. This way it is possible that the parameter values take
different values for each link (or node).

\section{Behavioral Difference between the AsIC and the AsLT Models}
\label{behavioral-difference}

\subsection{Data Sets and Parameter Setting}
\label{dataset1}

We employed four datasets of large real networks (all bidirectionally
connected).  The first one is a trackback network of Japanese blogs used
by \citeA{kimura:tkdd} and has $12,047$ nodes and $79,920$ directed
links (the blog network).  The second one is a network of people derived
from the ``list of people'' within Japanese Wikipedia, also used by
\citeA{kimura:tkdd}, and has $9,481$ nodes and $245,044$ directed links
(the Wikipedia network).  The third one is a network derived from the
Enron Email Dataset \cite{klimt} by extracting the senders and the
recipients and linking those that had bidirectional communications. It
has $4,254$ nodes and $44,314$ directed links (the Enron network).  The
fourth one is a coauthorship network used by \citeA{palla} and has
$12,357$ nodes and $38,896$ directed links (the coauthorship network).
These networks are confirmed to satisfy the typical characteristics of
social networks, {\em e.g.}, power law for degree distribution, higher
clustering coefficient, etc.

In this experiments, we set the value of diffusion probability (AsIC)
and the value of the link weight (AsLT) such that they are consistent in
the following sense under the simplest assumption to make a fair
comparison: $\sum_{(u,v) \in E} p_{u,v} = \sum_{(u,v) \in E} q_{u,v} =
|V|$. Thus, $p_{u,v} = 1 / \bar{d}$ and $q_{u,v} = 1 / |B(v)|$ for any
$(u, v) \in E$, where $\bar{d}$ is the average out-degree of the
network.  Thus, the value of $p_{u,v}$ ($(u, v) \in E$) is given as
0.15, 0.04, 0.1, and 0.32 for the Blog, the Wikipedia, the Enron, and
the Coauthorship networks, respectively.

We compare influence degree obtained by the AsIC and the AsLT models
from various angles. Here, the influence degree $\sigma(v)$ of a node
$v$ is defined to be the expected number of active nodes at the end of
information diffusion process that starts from a single initial activate
node $v$. Since the time-delay parameter vector $\bbr$ does not affect
the influence degree (because it is defined at the end of diffusion
process), that is, $\sigma (v)$ is invariant with respect to the value
of $\bbr$, we can evaluate the value of $\sigma (v)$ by the influence
degree of the corresponding basic IC or LT model.  We estimated the
influence degree by the bond percolation based
method~\cite{kimura:dmkd}, in which we used $300,000$ bond percolation
processes according to \citeA{kempe:kdd}, meaning that the expectation
is approximated by the empirical mean of $300,000$ independent
simulations.

\subsection{Experimental Results}
\label{behavior-difference-results}

\begin{figure*}[!t]
\centering
\subfloat[Blog network\label{blog_infl}]
{\includegraphics[width=7cm]{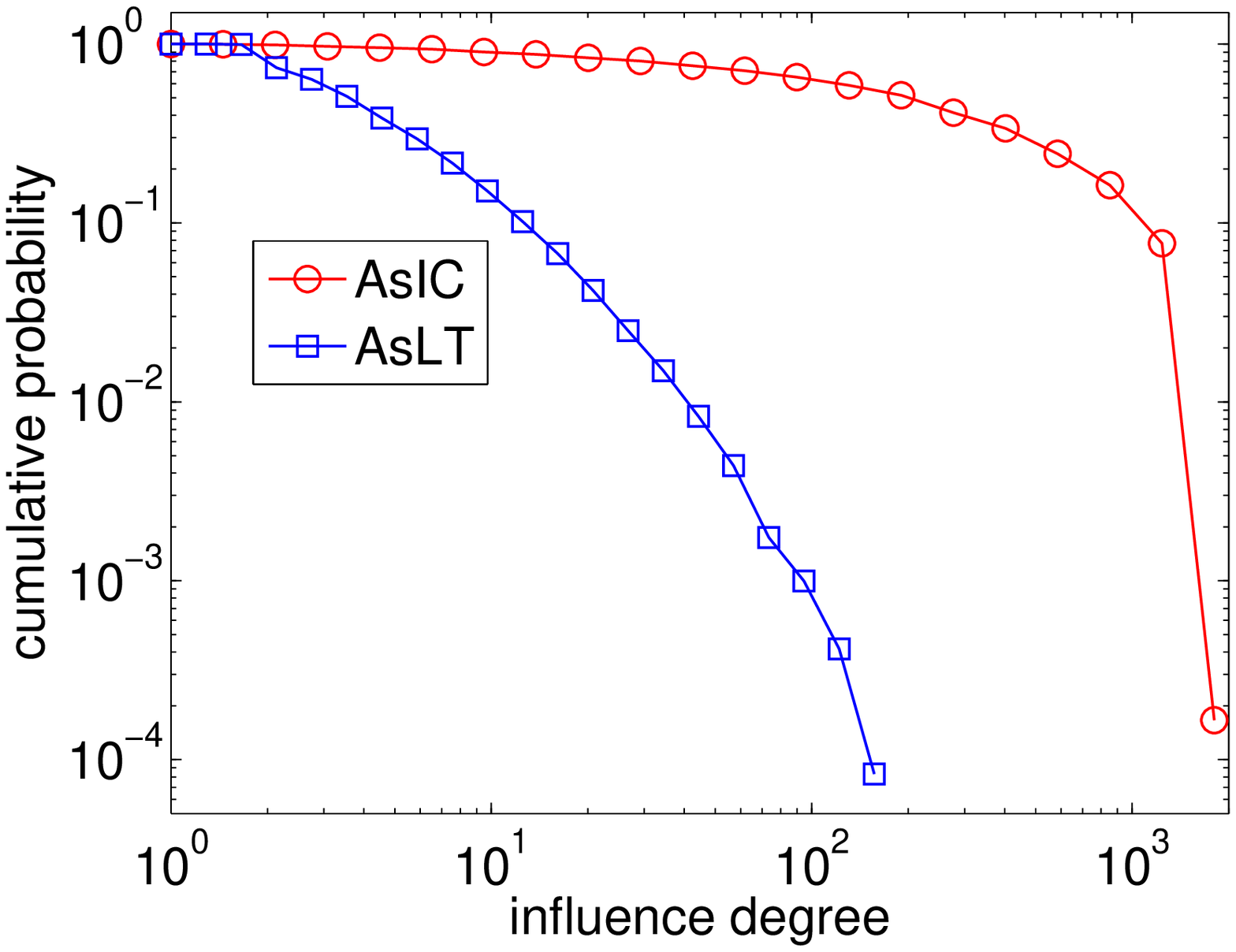}}
\hspace{3mm}
\subfloat[Wikipedia network\label{wiki_infl}]
{\includegraphics[width=7cm]{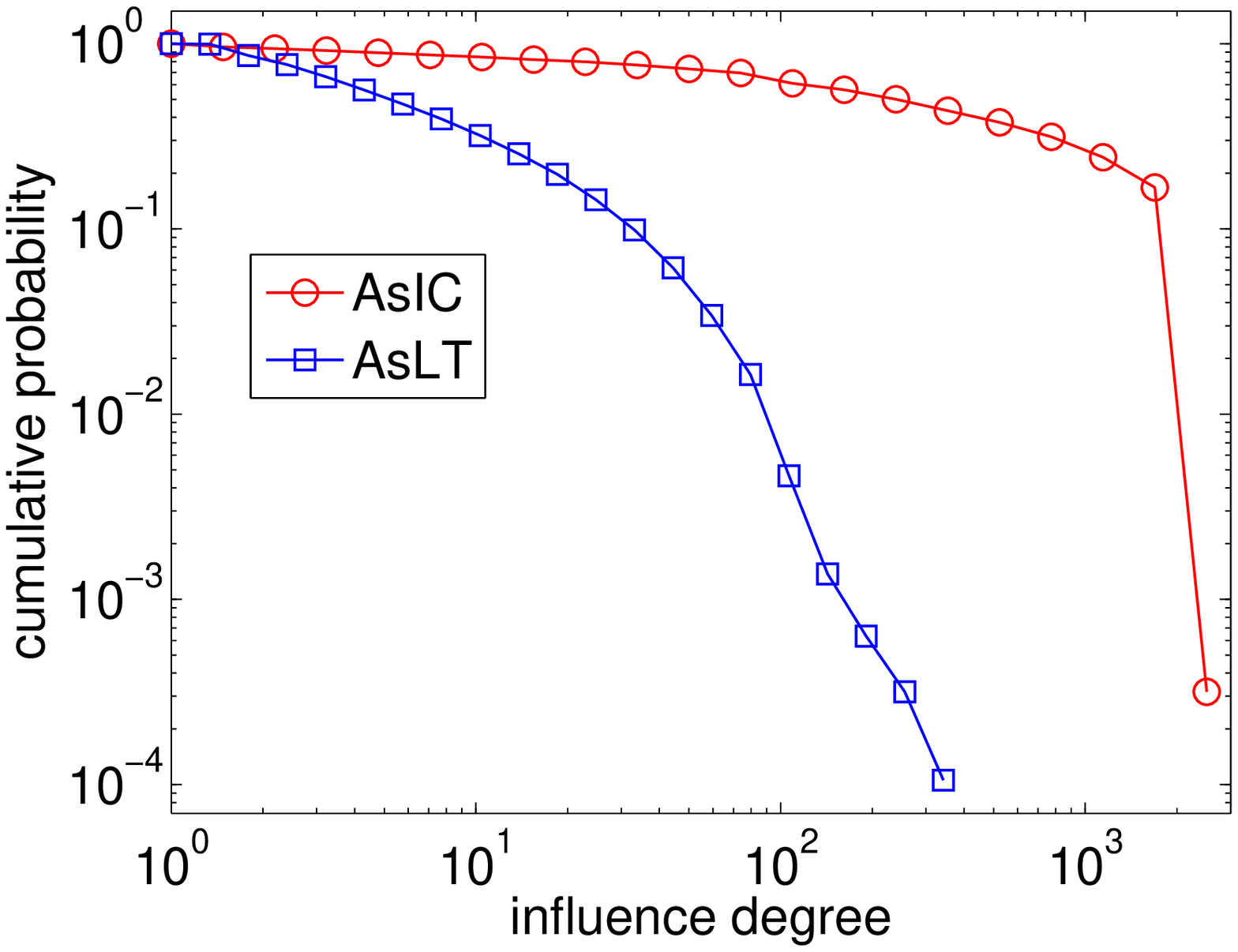}}
\hspace{1mm}
\subfloat[Enron network\label{enron_infl}]
{\includegraphics[width=7cm]{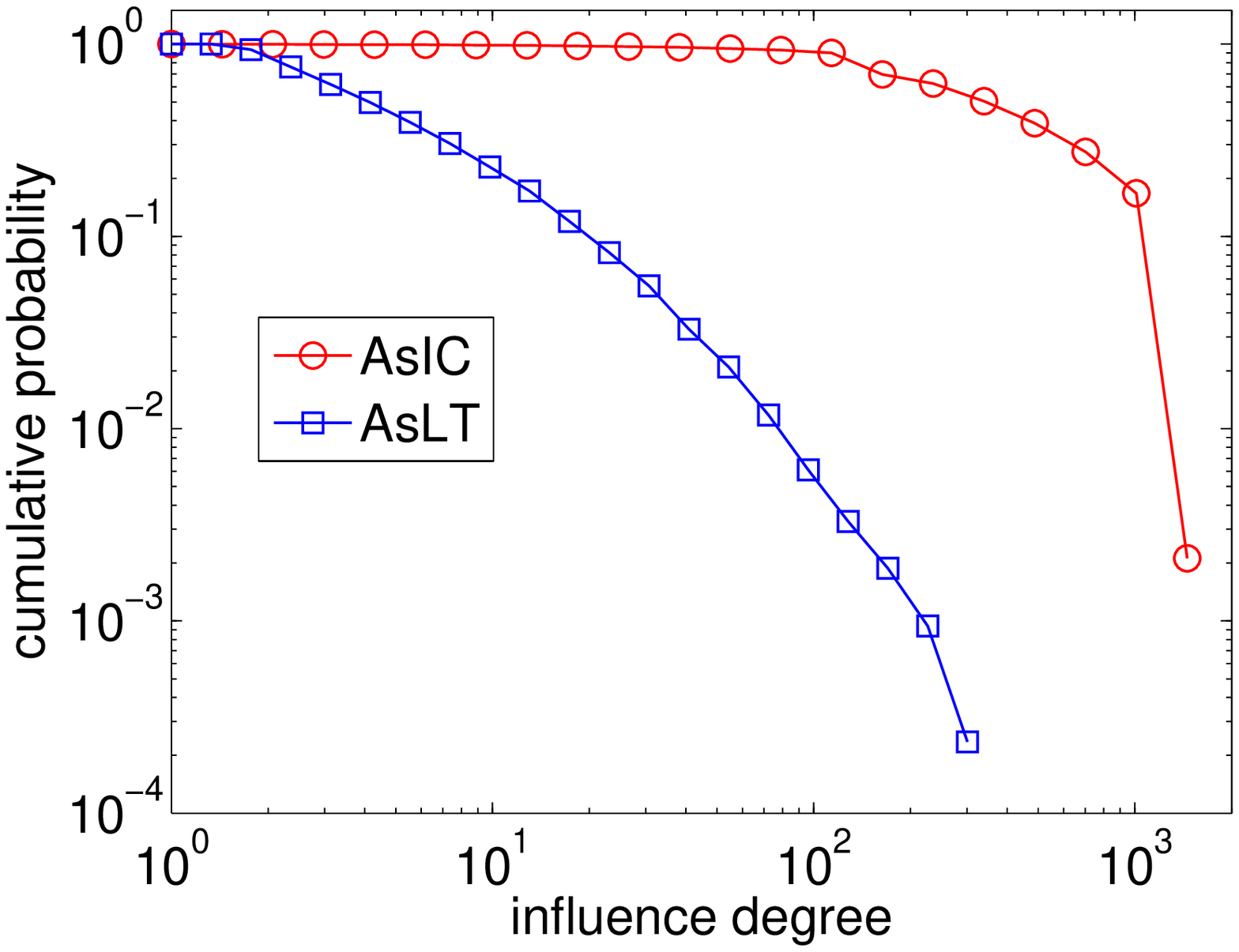}}
\hspace{3mm}
\subfloat[Coauthorship network\label{coauthor_infl}]
{\includegraphics[width=7cm]{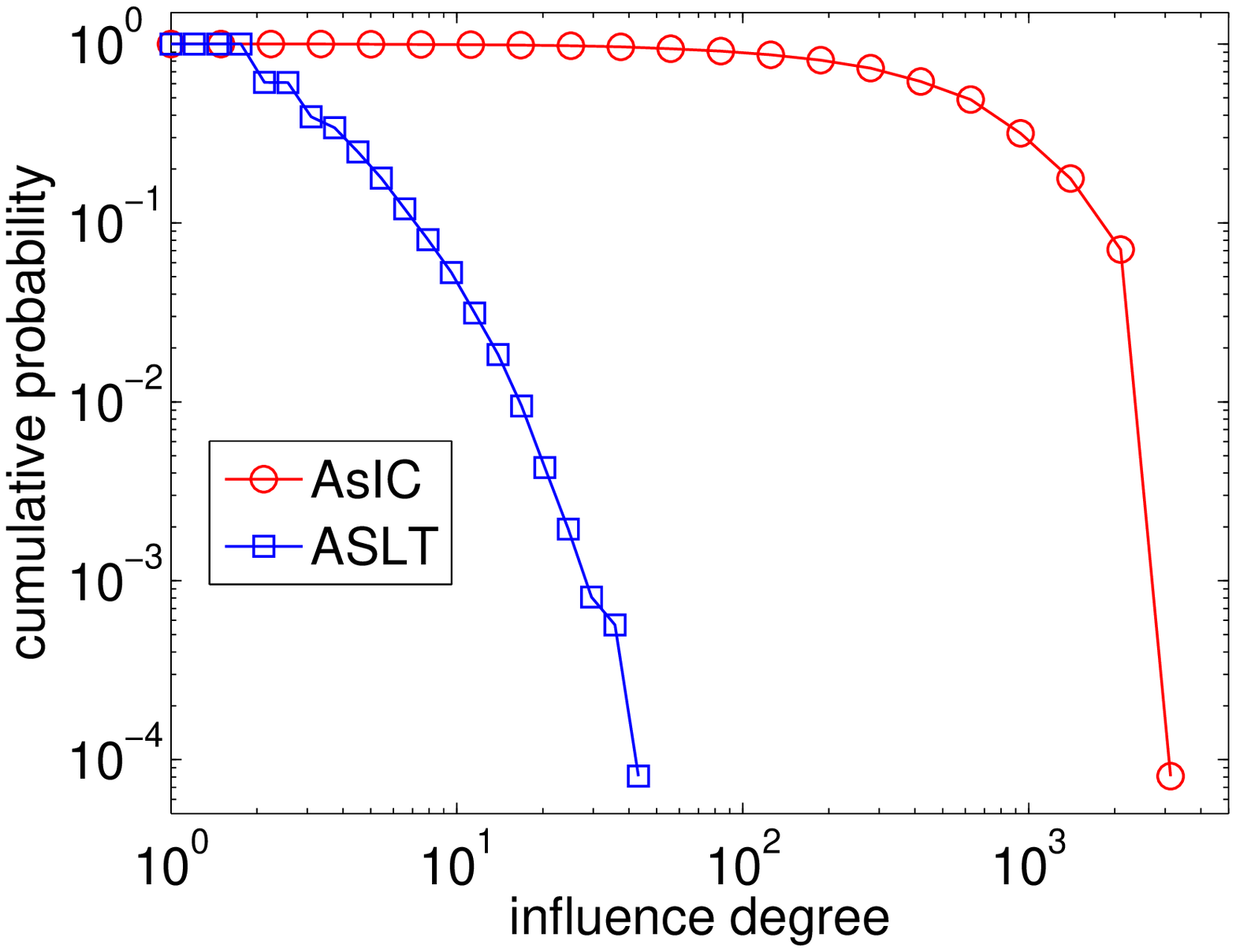}}
\caption{Comparison of influence degree between the AsIC and the AsLT models}
\label{infl_comp}
\end{figure*}

\begin{figure*}[!t]
\centering
\subfloat[Blog network\label{blog_ic_infl}]
{\includegraphics[width=7cm]{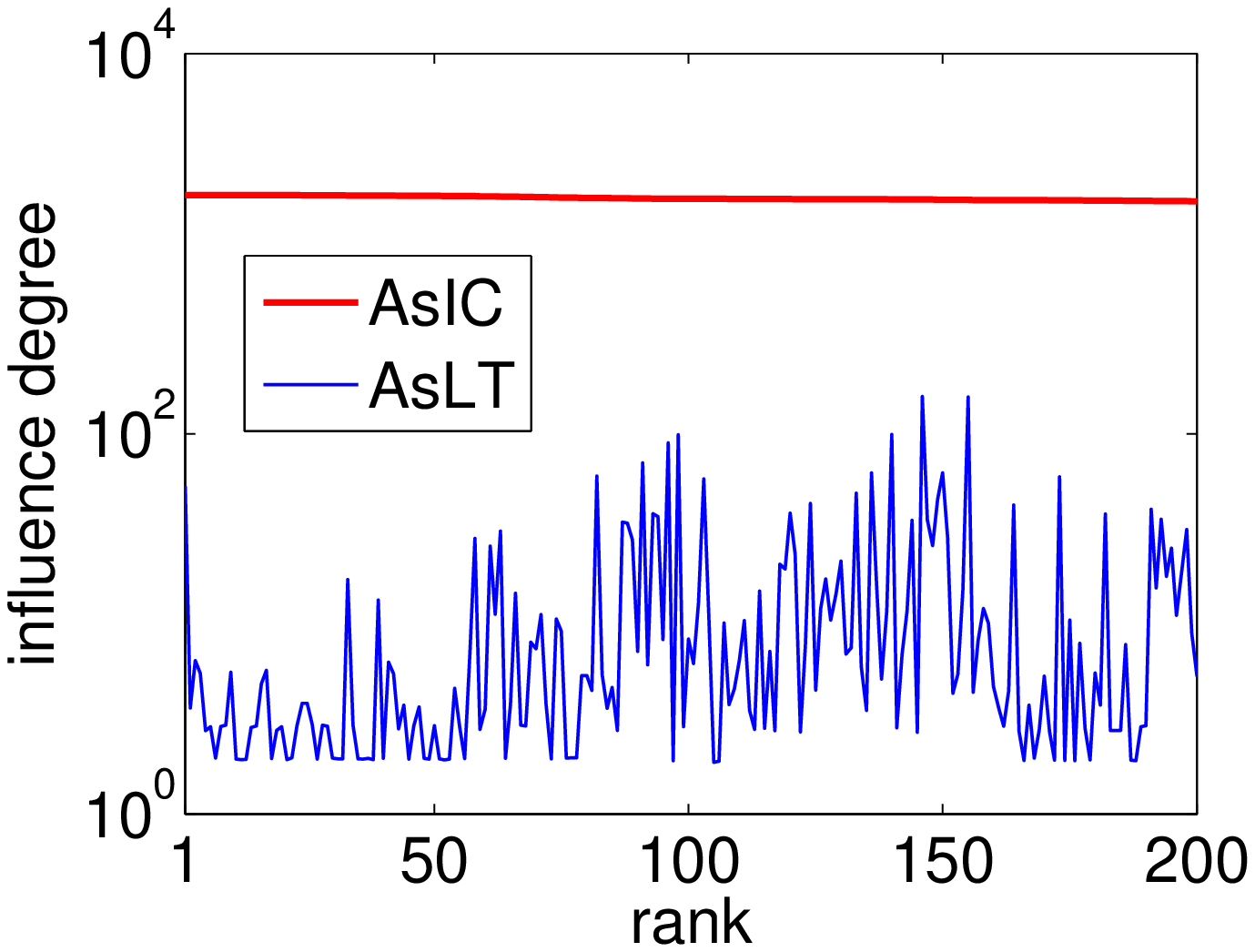}}
\hspace{3mm}
\subfloat[Wikipedia network\label{wiki_ic_infl}]
{\includegraphics[width=7cm]{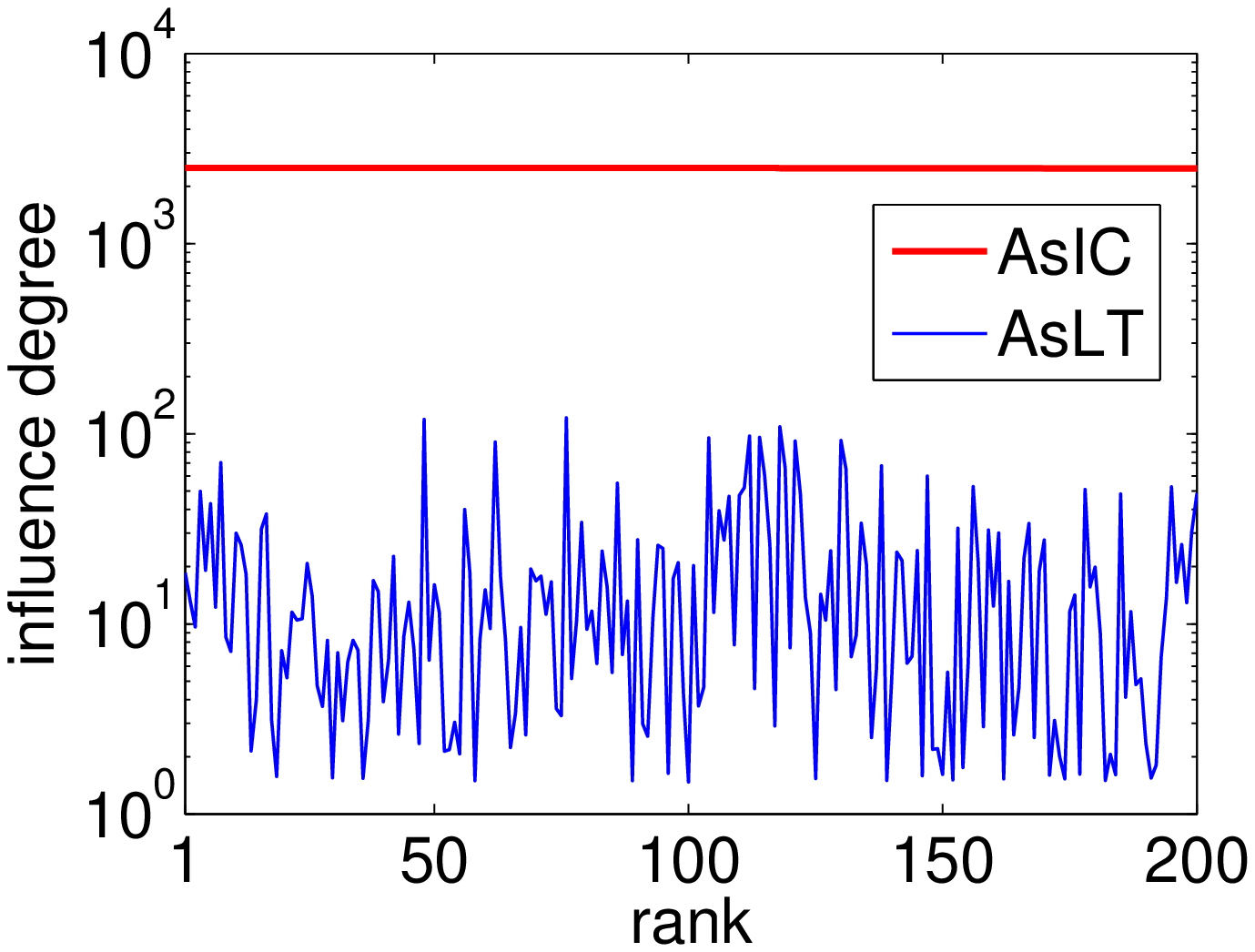}}
\hspace{1mm}
\subfloat[Enron network\label{enron_ic_infl}]
{\includegraphics[width=7cm]{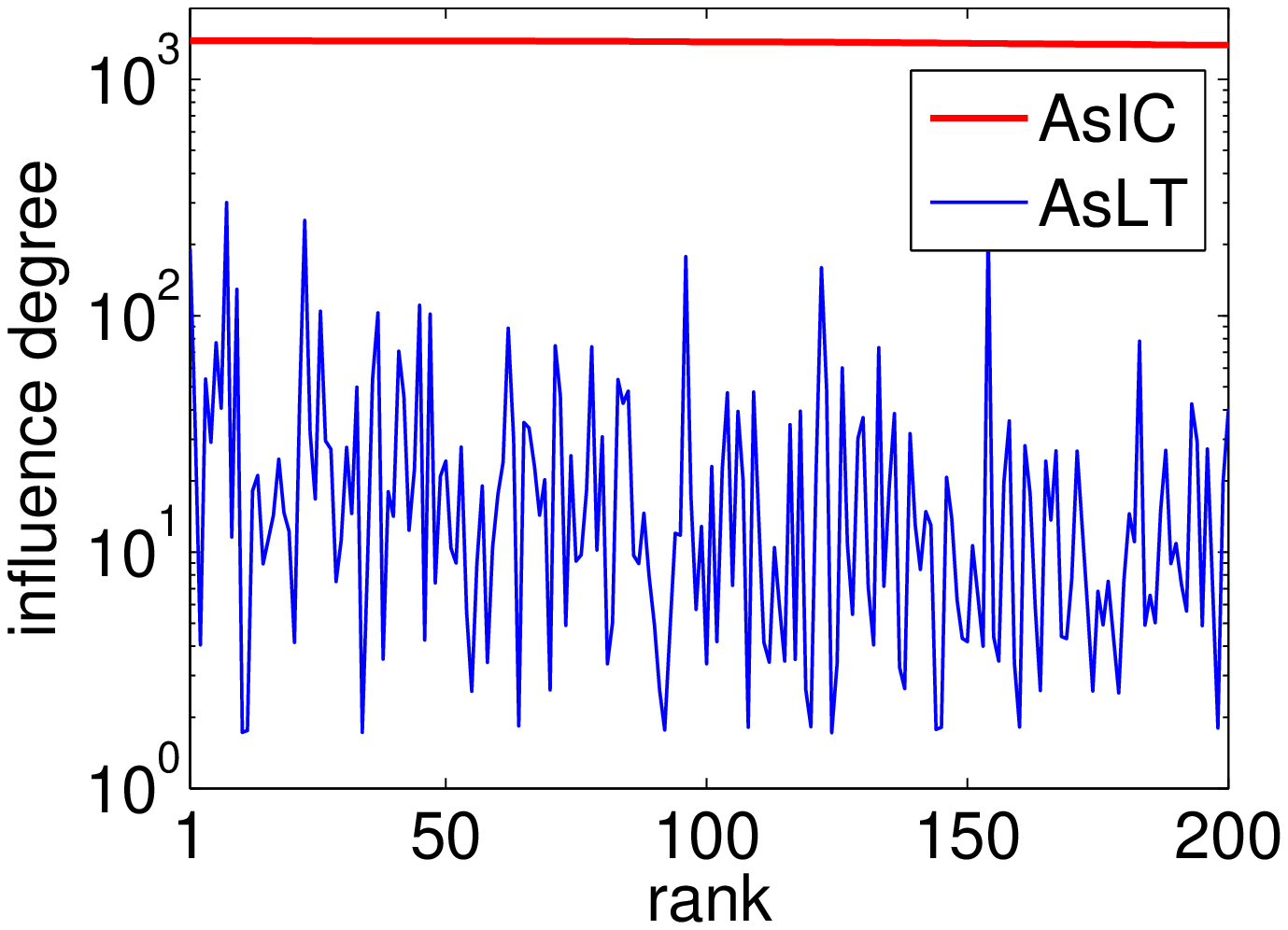}}
\hspace{3mm}
\subfloat[Coauthorship network\label{coauthor_ic_infl}]
{\includegraphics[width=7cm]{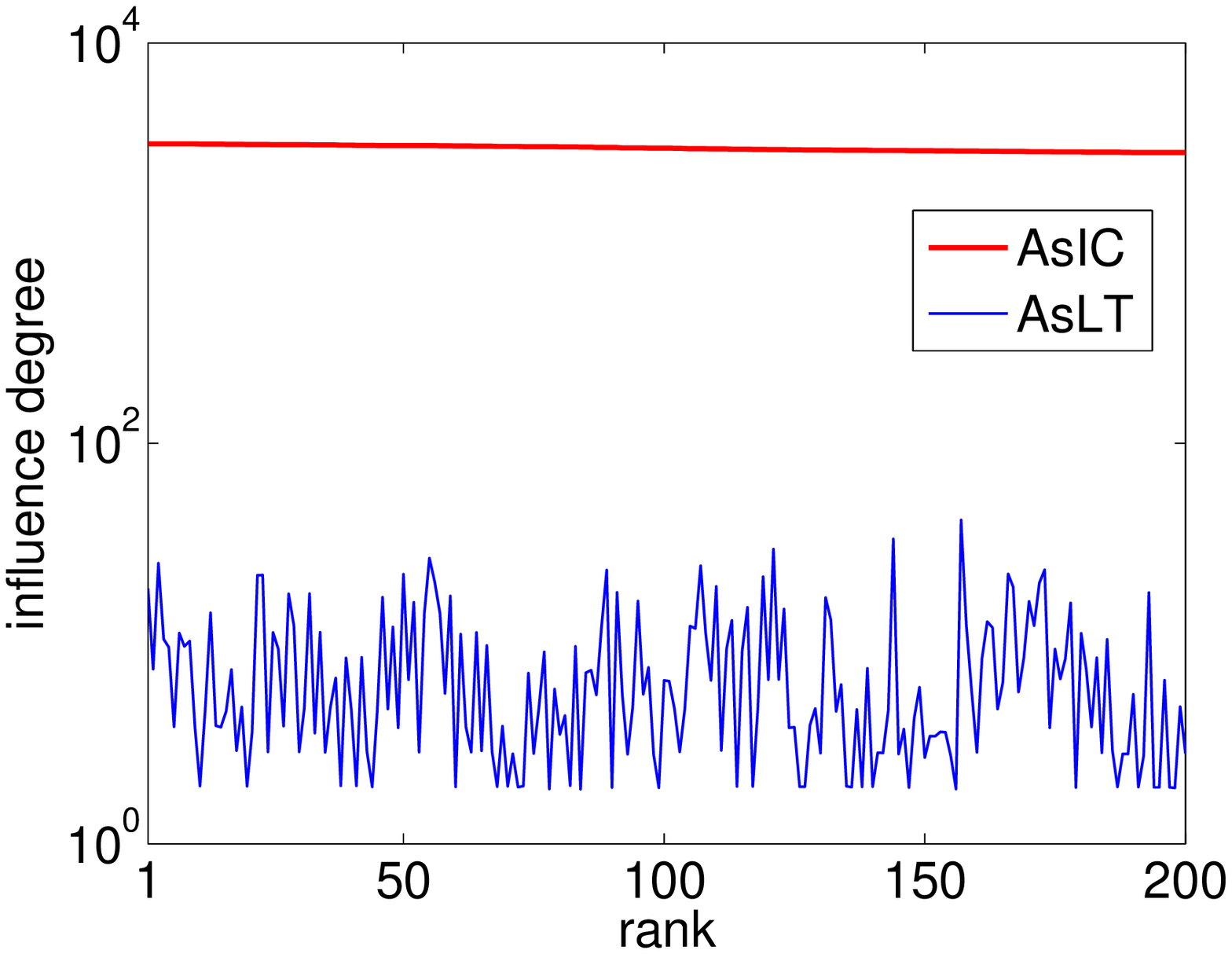}}
\caption{
Influence degree of AsIC and AsLT for the influential nodes of the AsIC model}
\label{ic_infl}
\end{figure*}

\begin{figure*}[!t]
\centering
\subfloat[Blog network\label{blog_lt_infl}]
{\includegraphics[width=7cm]{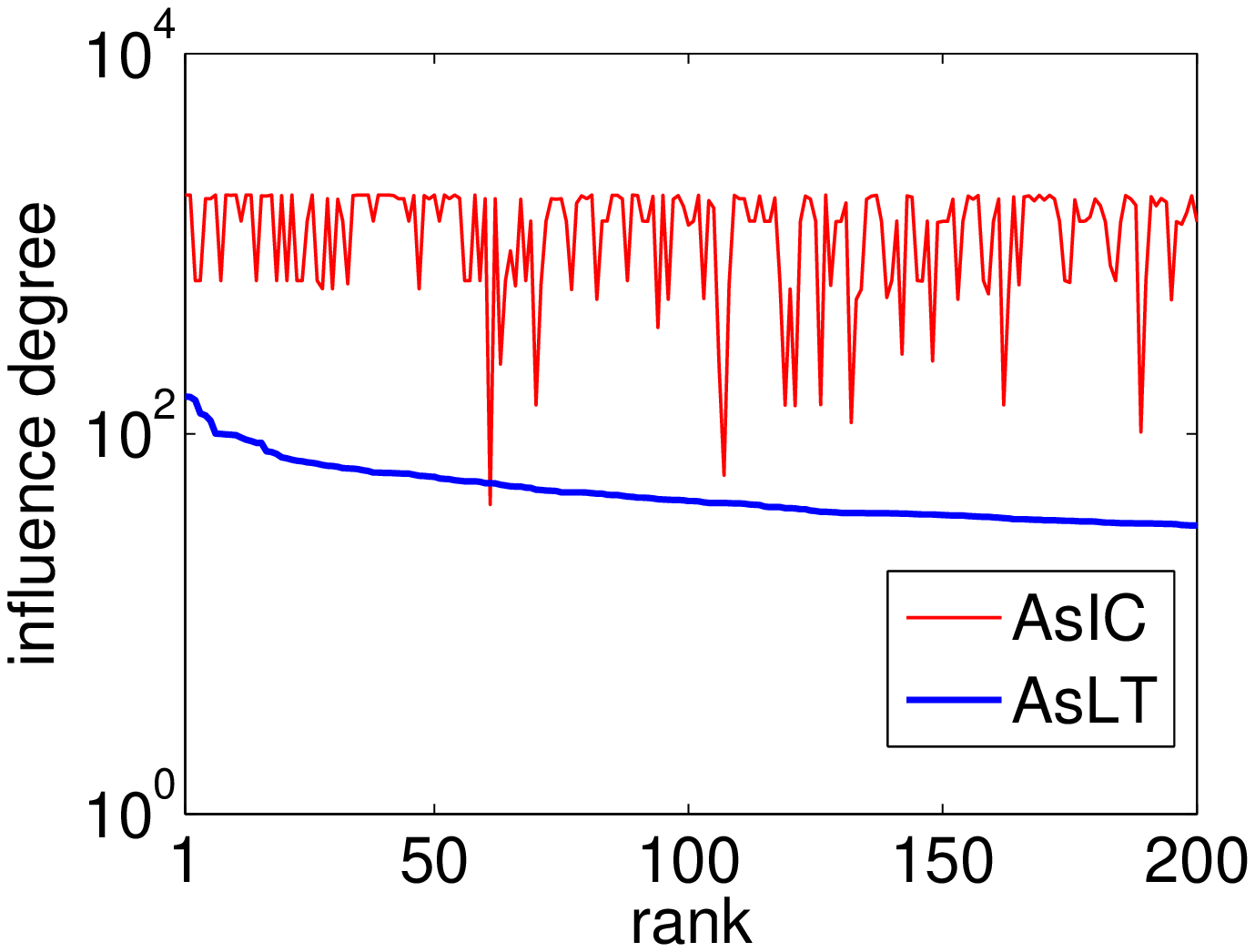}}
\hspace{3mm}
\subfloat[Wikipedia network\label{wiki_lt_infl}]
{\includegraphics[width=7cm]{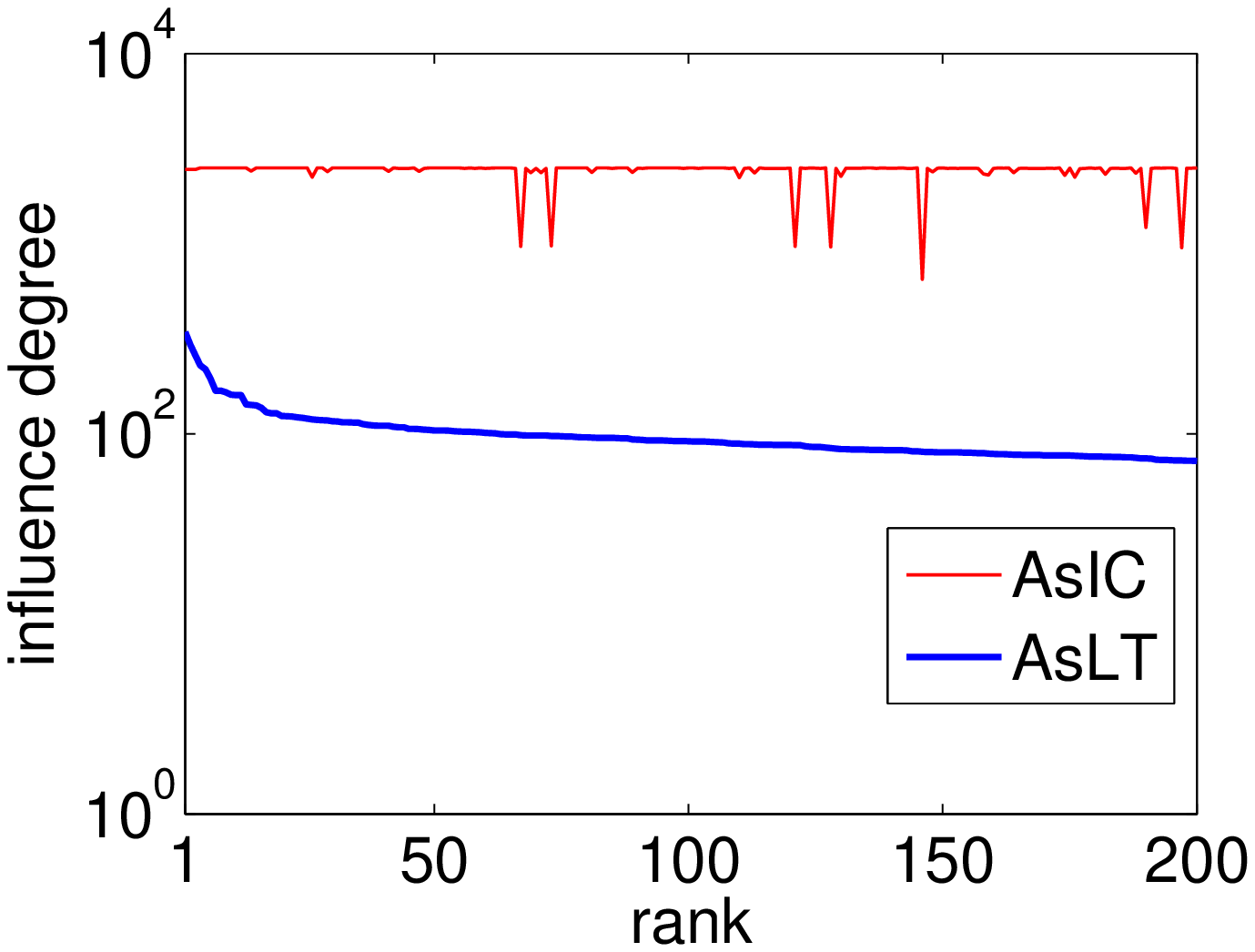}}
\hspace{1mm}
\subfloat[Enron network\label{enron_lt_infl}]
{\includegraphics[width=7cm]{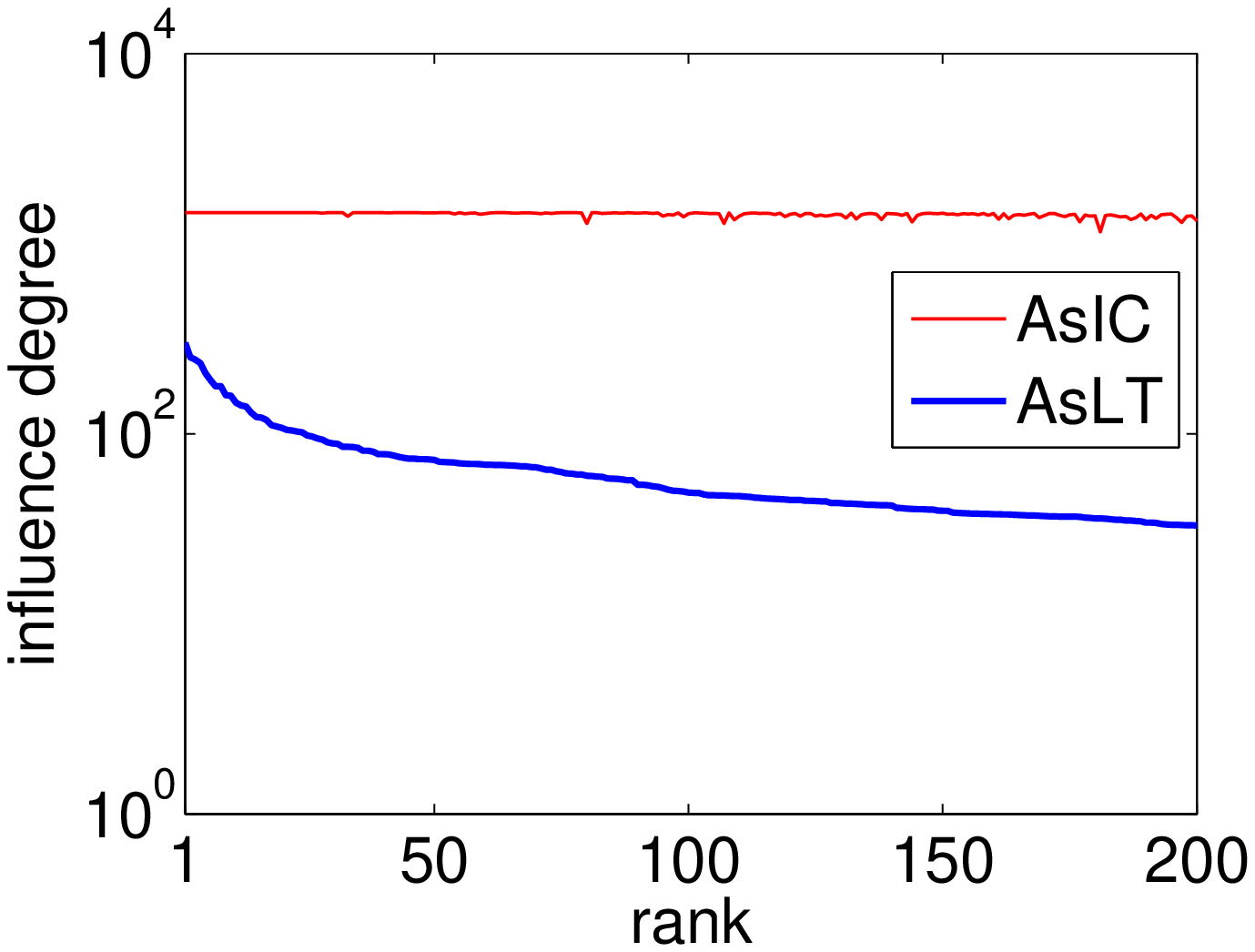}}
\hspace{3mm}
\subfloat[Coauthorship network\label{coauthor_lt_infl}]
{\includegraphics[width=7cm]{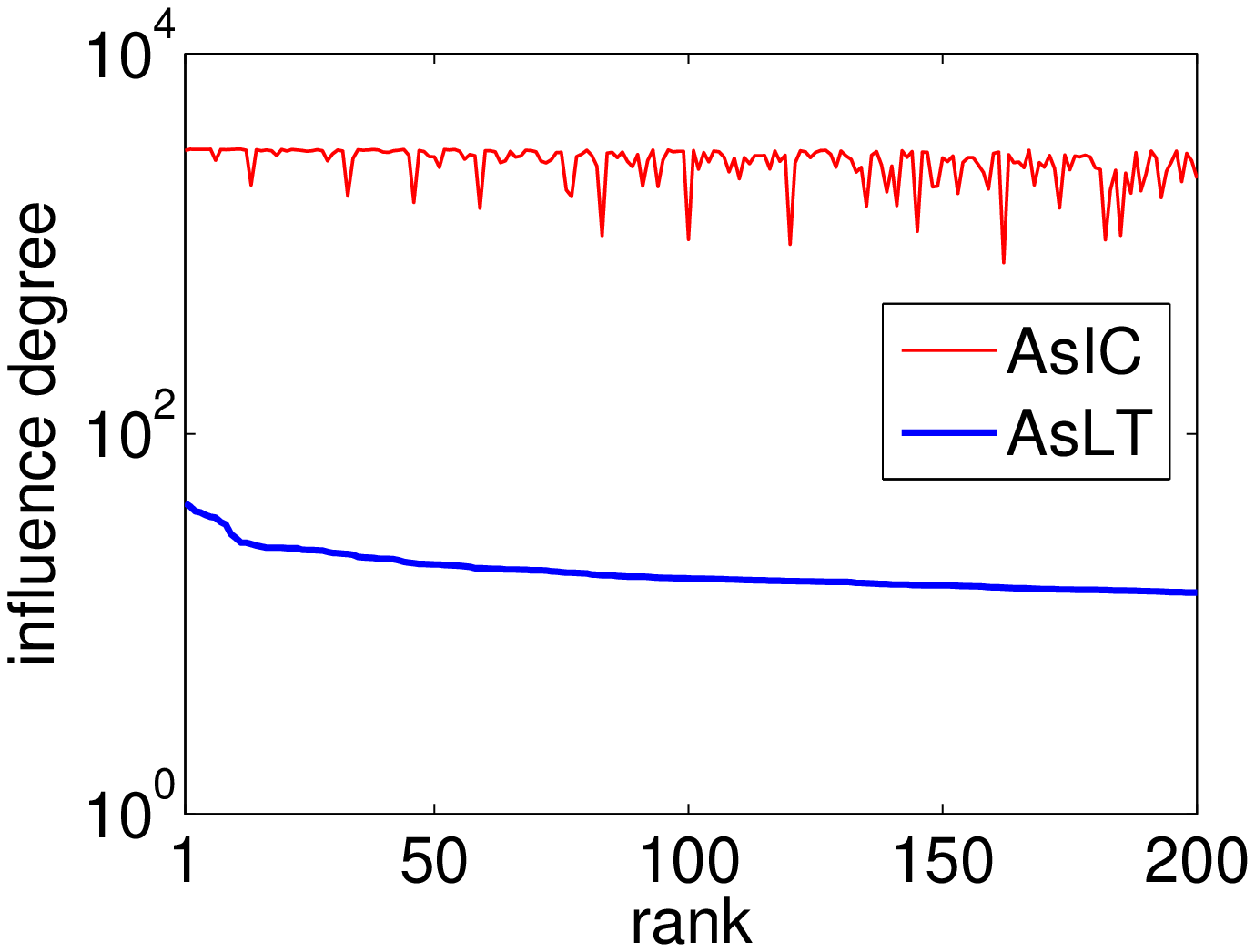}}
\caption{
Influence degree of AsIC and AsLT for the influential nodes of the AsLT model}
\label{lt_infl}
\end{figure*}

\begin{figure*}[!t]
\centering
\subfloat[Blog network\label{blog_top_ic}]
{\includegraphics[width=7cm]{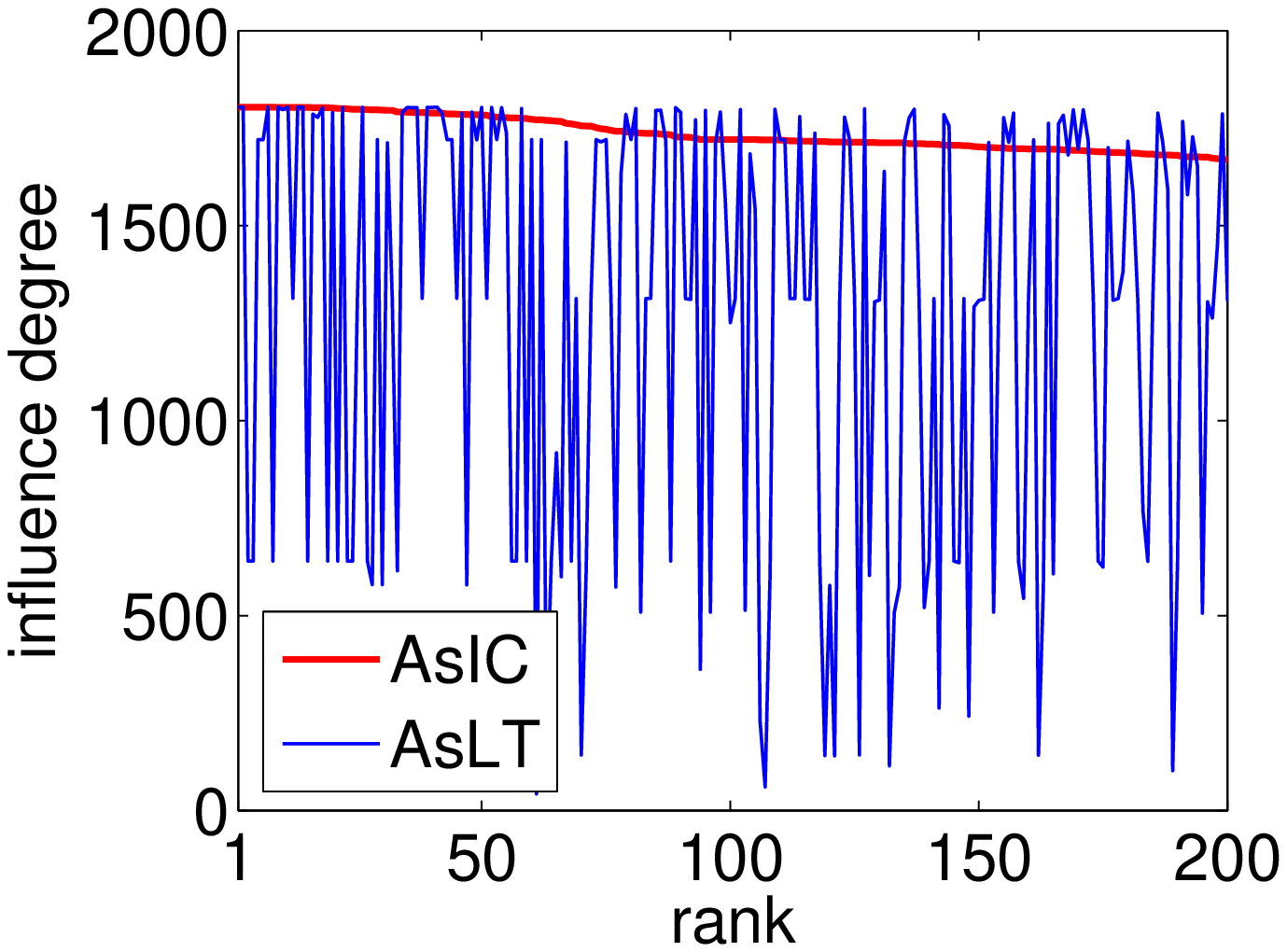}}
\hspace{3mm}
\subfloat[Wikipedia network\label{wiki_top_ic}]
{\includegraphics[width=7cm]{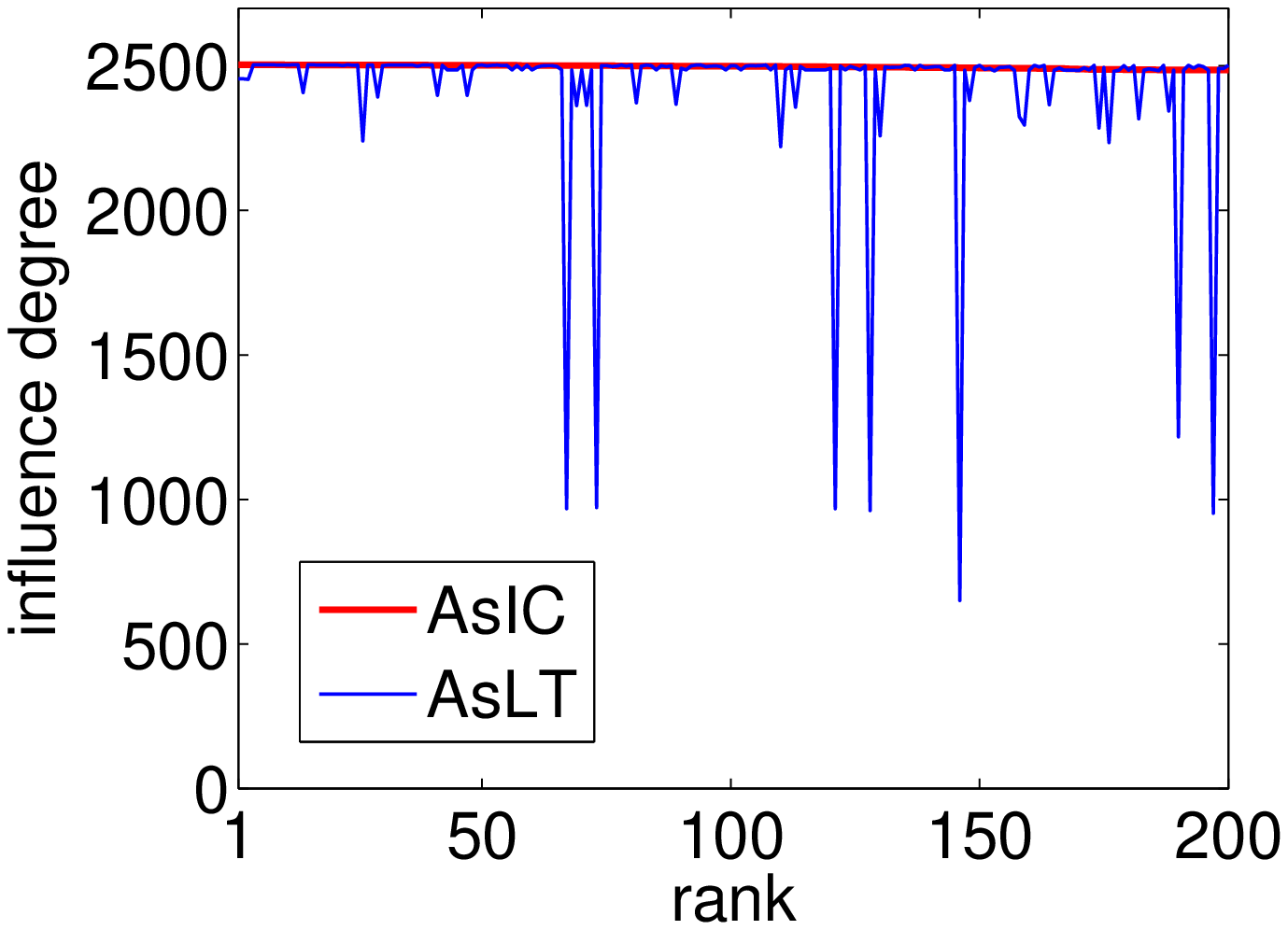}}
\hspace{1mm}
\subfloat[Enron network\label{enron_top_ic}]
{\includegraphics[width=7cm]{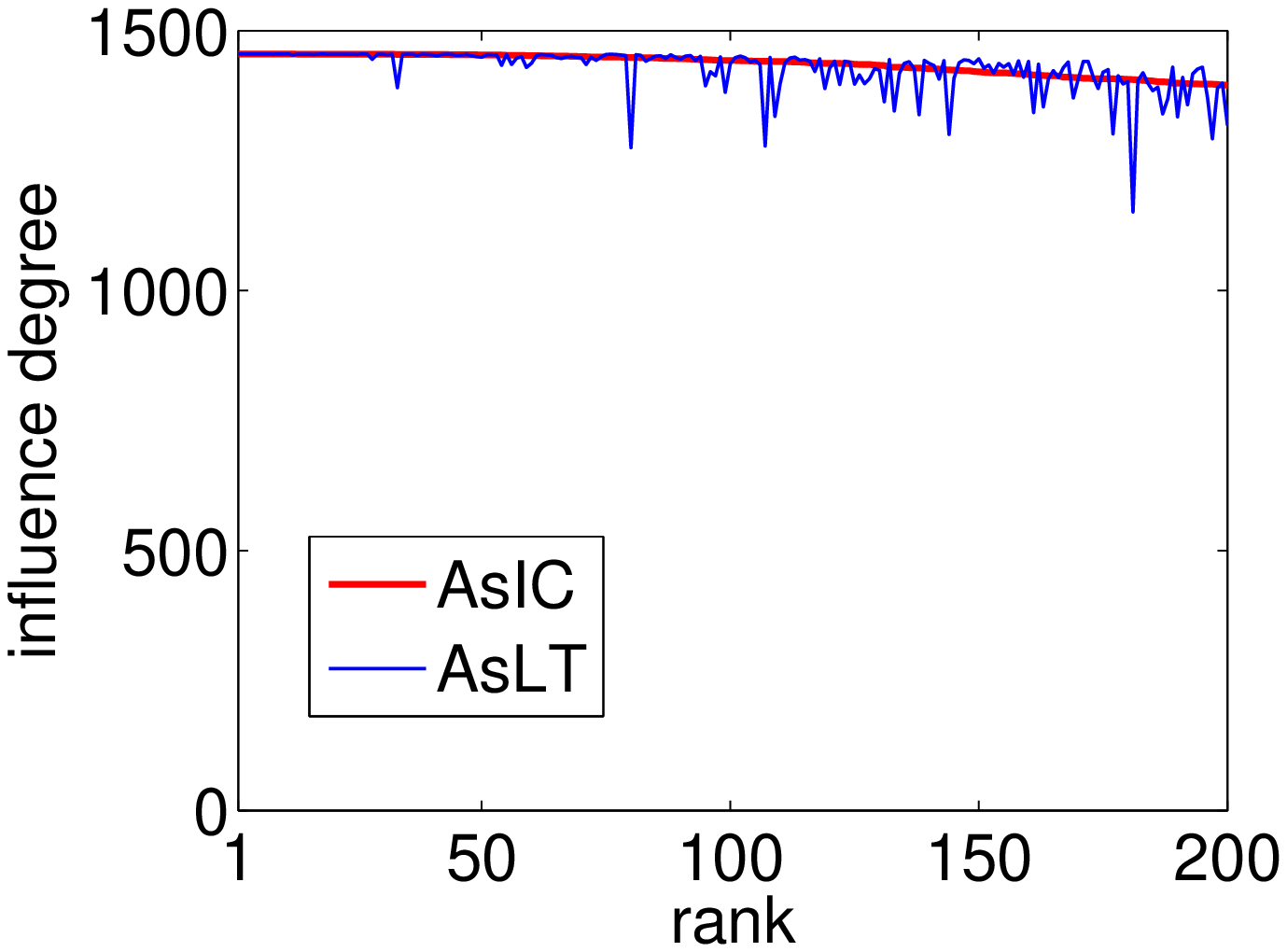}}
\hspace{3mm}
\subfloat[Coauthorship network\label{coauthor_top_ic}]
{\includegraphics[width=7cm]{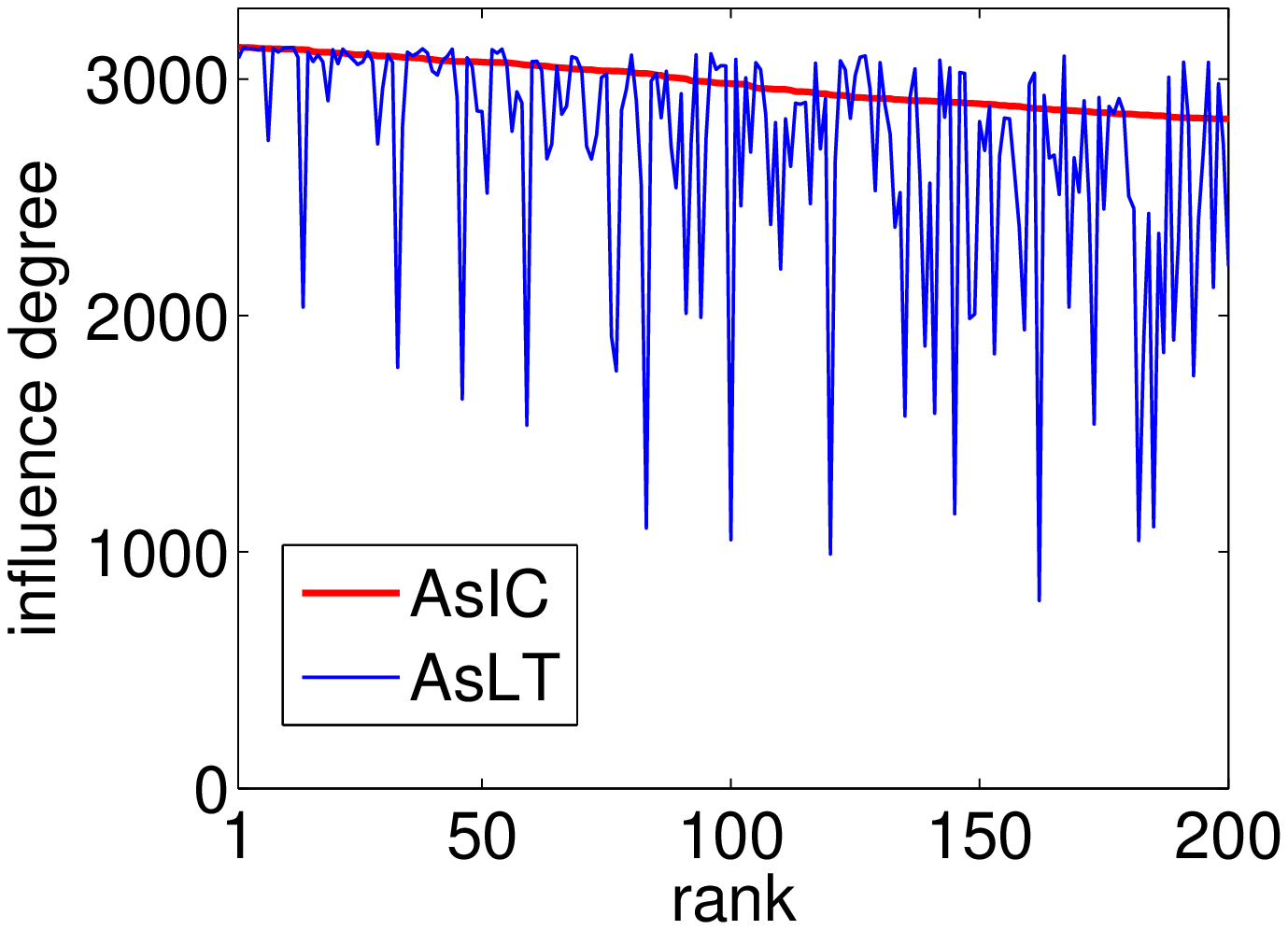}}
\caption{
Comparison of the influential nodes of AsIC
and AsLT measured in the influence degree of AsIC}
\label{top_ic}
\end{figure*}

\begin{figure*}[!t]
\centering
\subfloat[Blog network\label{blog_top_lt}]
{\includegraphics[width=7cm]{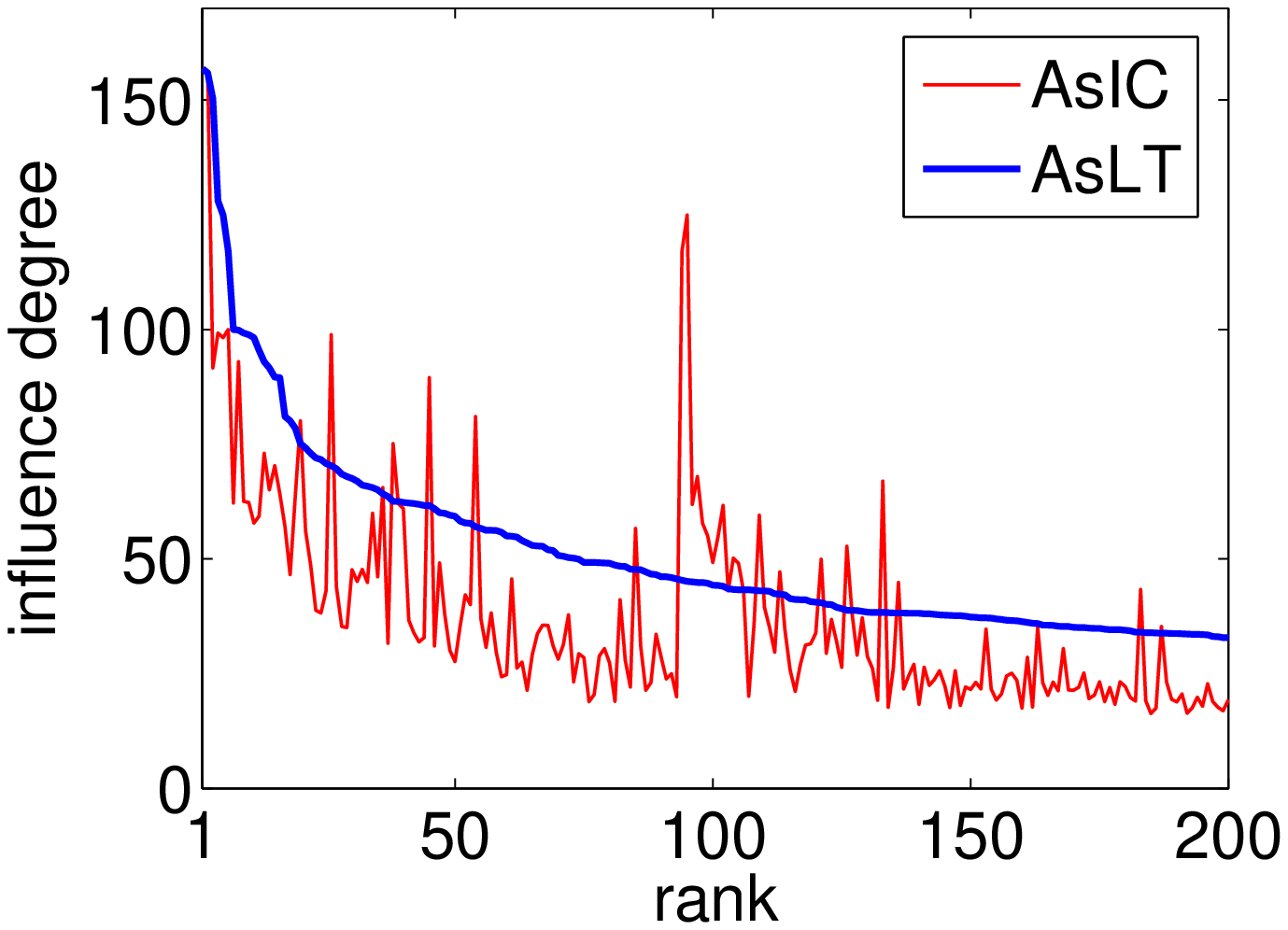}}
\hspace{3mm}
\subfloat[Wikipedia network\label{wiki_top_lt}]
{\includegraphics[width=7cm]{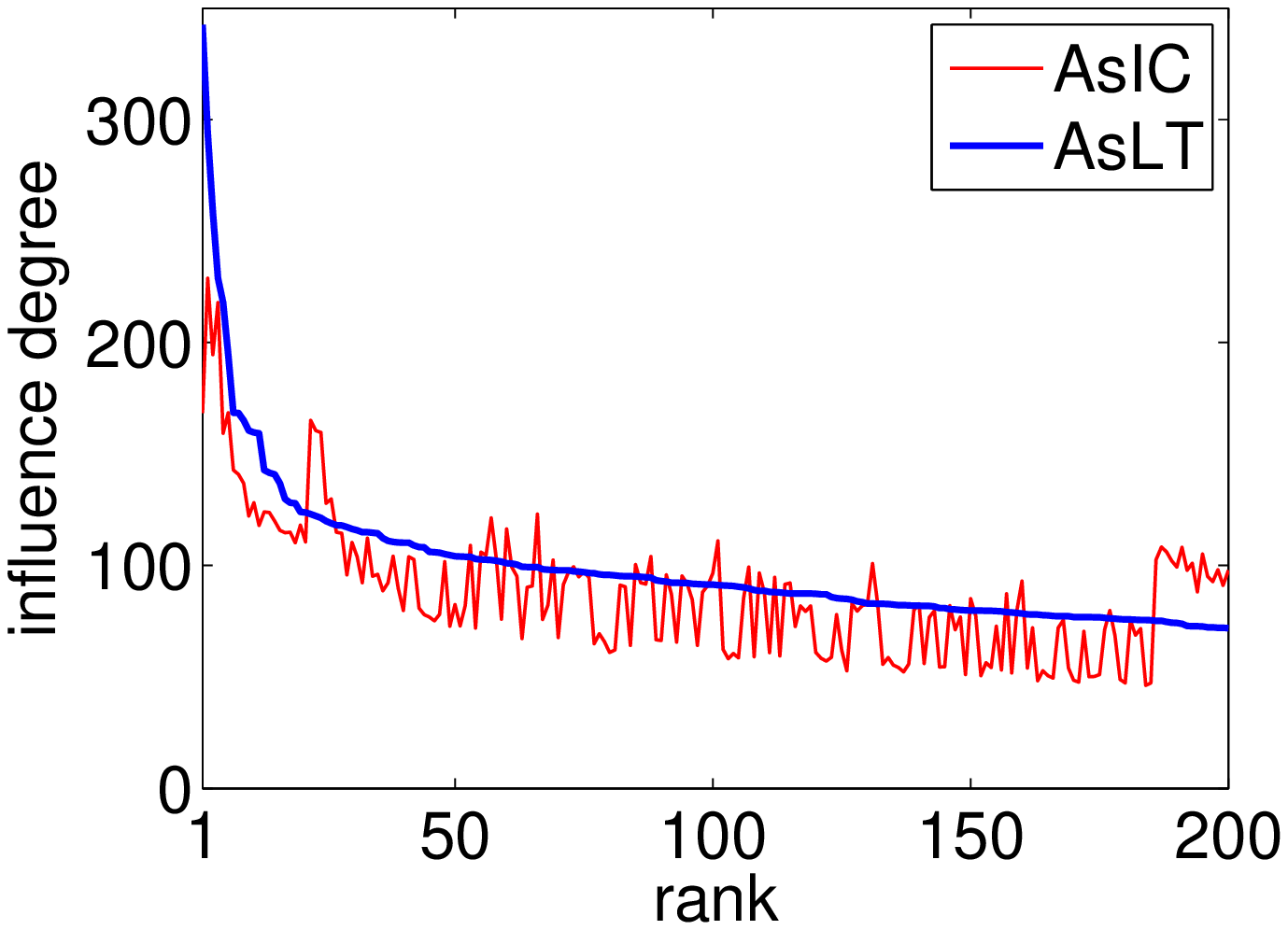}}
\hspace{1mm}
\subfloat[Enron network\label{enron_top_lt}]
{\includegraphics[width=7cm]{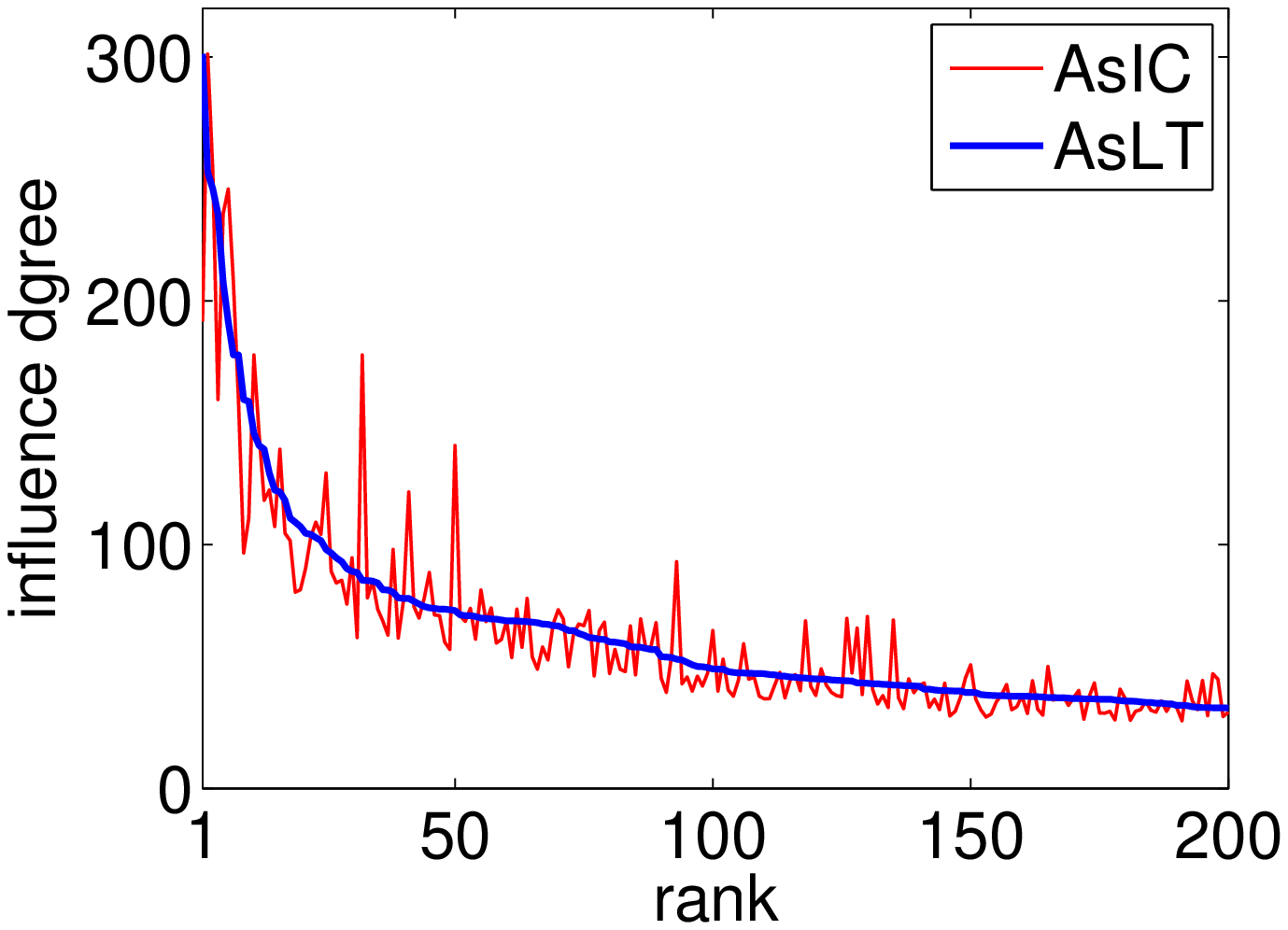}}
\hspace{3mm}
\subfloat[Coauthorship network\label{coauthor_top_lt}]
{\includegraphics[width=7cm]{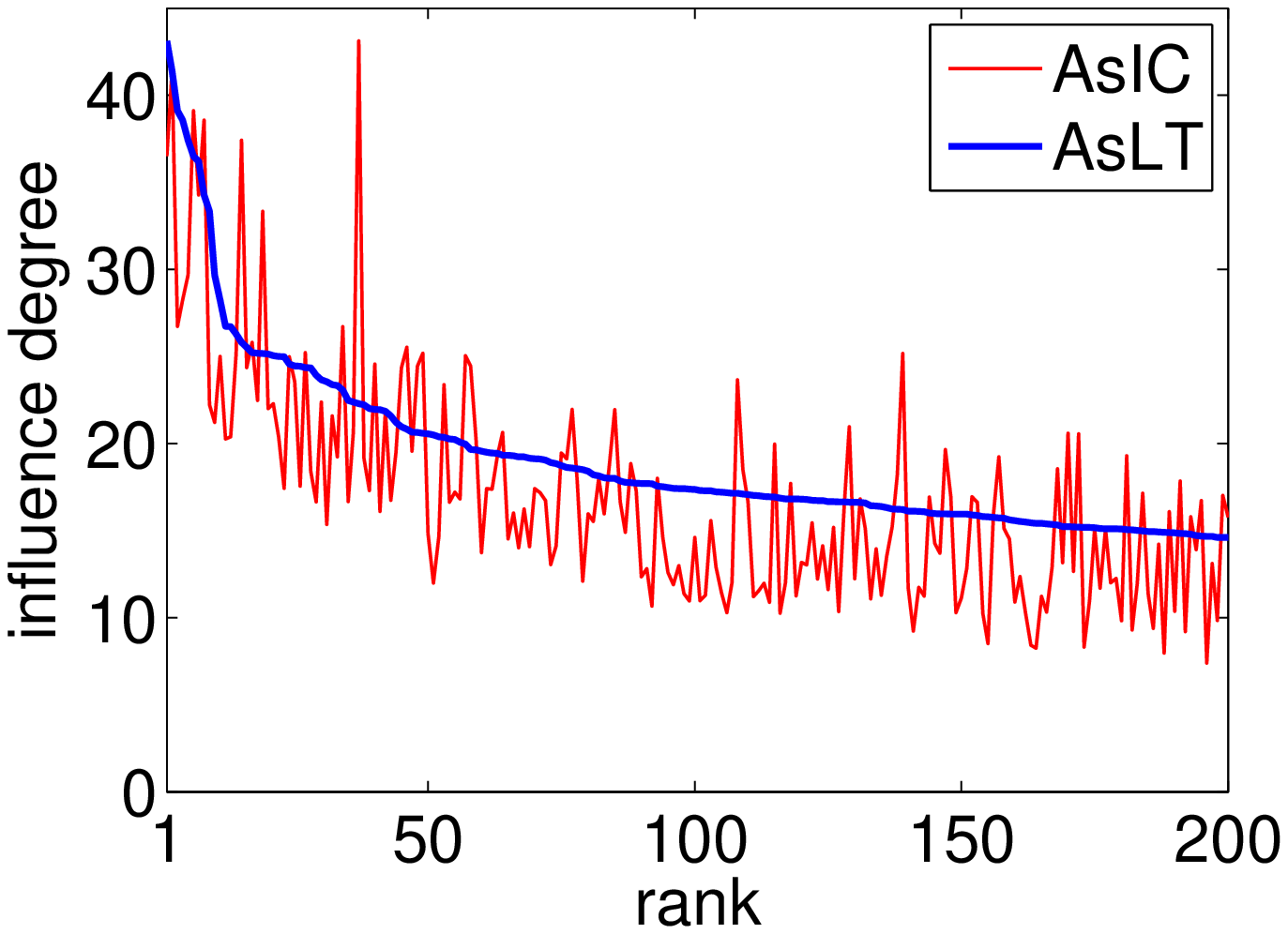}}
\caption{Comparison of the influential nodes of AsIC
and AsLT measured in the influence degree of AsLT}
\label{top_lt}
\end{figure*}

First, we investigated which of the AsIC and AsLT models can spread
information more widely.  Figure~\ref{infl_comp} shows the cumulative
probability of influence degree, $f_\sigma (x)$ $=$ $| \{ v \in V;
\sigma(v) \geq x \} | / |V|$, for the AsIC and the AsLT models. At a
glance we can see that the AsIC model has by far many more nodes of high
influence degrees than the AsLT model. Further, we examined the
difference of influence degree between the two models for the respective
influential nodes of both the AsIC and the AsLT models.  We ranked nodes
according to the influence degree of AsIC and AsLT, respectively, and
extracted the top $200$ influential nodes for each.
Figures~\ref{ic_infl} and \ref{lt_infl} display the respective influence
degree of rank $k$ node of AsIC and AsLT ($k = 1, \cdots, 200$).  Here,
the red line indicates the influence degree of AsIC, and the blue line
indicates the influence degree of AsLT. We can see that the difference
of influence degree between the two models is quite large for these
influential nodes. This clearly indicates that the information can
diffuse more widely under the AsIC model than the AsLT model. This can
be attributed to the scale-free nature (having power-law degree
distributions) of the four real networks used in the experiments.  It is
known \cite{albert:nature} that {\em hub nodes}, defined as those having
many outgoing links, play an important role for widely spreading
information in a scale-free network. By the information diffusion
mechanism of the AsIC and AsLT models, it is more difficult for the AsLT
model to transmit information to hub nodes than the AsIC model in a
scale-free network.  Therefore, the result is understandable.

Next, we compared the difference of the influential nodes between the
AsIC and the AsLT models.  The results are shown in Figures~\ref{top_ic}
and \ref{top_lt}. For both figures the horizontal axes are node ranking
($k = 1, \cdots, 200$), and the actual ranking depends which model we
are considering, {\em e.g.}, the rank $k$ node for AsIC is different
from the same rank $k$ node for AsLT. The vertical axis are influence
degree for both figures, but it is the influence degree for AsIC in
Figure~\ref{top_ic} and that for AsLT in Figure~\ref{top_lt}. The red
line corresponds to nodes for AsIC and the blue line corresponds to
nodes for AsLT. Thus, by definition of node ranking, the influence
degree of AsIC (red thick line) is non-increasing in Figure~\ref{top_ic}
and the influence degree of AsLT (blue thick line) in
Figure~\ref{top_lt} is non-increasing. However, the corresponding line
for AsLT (blue line) in Figure~\ref{top_ic} and that for AsIC (red line)
in Figure~\ref{top_lt} are very irregular. This means that almost all
the nodes that are influential for AsIC model are different from the
nodes that are influential for AsLT, and vice versa. There are small
number of influential nodes that overlap for both the models, but how
similar the influential nodes are (degree of overlapping) depends on the
characteristics of the network structure, and no general tendency can
be extracted.

\section{Learning Performance Evaluation}

\subsection{Data Sets and Parameter Setting}

We used the same four datasets that are used in Section
\ref{behavioral-difference}, and employed also the simplest
approximation for the parameter setting but with a slight difference
according to the work \citeA{saito:acml09}.

We set $p_{u, v} = p$, $r_{u, v} = r$ for AsIC and $q_{u,v} = q
|B(v)|^{-1}$, $r_{u,v} = r$ for AsLT.  Under this assumption there is no
need for the observation sequence data to pass through every link or
node at minimum once and desirably several times. This drastically
reduces the amount of data we have to generate to use as the training
data to learn the parameters. Then, our task is to estimate the values
of these parameters from the training data.  According to the work of
\citeA{kempe:kdd}, we set $p$ to a value slightly smaller than $1/{\bar
d}$. Thus, the true value of $p$ was set to $0.2$ for the coauthorship
network, $0.1$ for the blog and Enron networks, and $0.02$ for the
Wikipedia network. The true value of $q$ was set to $0.9$ for every
network to achieve reasonably long diffusion results, and the true value
of $r$ was set to $1.0$.\footnote{Note that a different value of $r$
corresponds to a different scaling of the time axis under the
assumption of uniform value.}

Using these parameter values, we generated a diffusion sequence from a
randomly selected initial active node for each of the AsIC and the AsLT
models in four networks. We then constructed a training dataset such
that each diffusion sequence has at least 10 nodes.  Parameter updating
is terminated when either the iteration number reaches its maximum (set
to 100) or the following condition is first satisfied:
$|r^{(s+1)}-r^{(s)}|+|p^{(s+1)}-p^{(s)}| \le 10^{-6}$ for AsIC and
$|r^{(s+1)}-r^{(s)}|+|q^{(s+1)}-q^{(s)}| \le 10^{-6}$ for AsLT, where
the superscript $(s)$ indicates the value for the $s$-th iteration.  In
most of the cases, the above inequality is satisfied in less than 100
iterations. The converged values are rather insensitive to the initial
parameter values, and we confirmed that the parameter updating algorithm
stably converges to the correct values which we assumed to be the true
values.

\subsection{Parameter Estimation}

\begin{table}[!t]
\centering
\caption{Parameter estimation error of the learning method 
for the AsIC model
in four networks}
\vspace{0.2cm}
\label{table:ic-learn}
\begin{tabular}{c|c|c|c}
\hline
Network & Number of active nodes & ${\cal E}_r$ & ${\cal E}_p$\\
\hline
     & 1,163 & 0.019 & 0.026\\
Blog & 5,151 & 0.018 & 0.014\\
     & 10,322 & 0.011 & 0.011\\
\hline
      & 1,275 & 0.060 & 0.032\\
Wikipedia  & 5,386 & 0.013 & 0.009\\
      & 10,543 & 0.006 & 0.007\\
\hline
      & 1,456 & 0.031 & 0.030\\
Enron & 5,946 & 0.011 & 0.011\\
      & 10,468 & 0.005 & 0.006\\
\hline
         & 1,203 & 0.028 & 0.022\\
Coauthorship & 5,193 & 0.009 & 0.007\\
         & 10,132 & 0.006 & 0.006\\
\hline  
\end{tabular}
\end{table}

\begin{table}[!t]
\centering
\caption{Parameter estimation error of the learning method 
for the AsLT model
in four networks}
\vspace{0.2cm}
\label{table:lt-learn}
\begin{tabular}{c|c|c|c}
\hline
Network & Number of active nodes & ${\cal E}_r$ & ${\cal E}_q$\\
\hline
     & 1,023 & 0.020 & 0.020\\
Blog & 5,018 & 0.012 & 0.020\\
     & 10,037 & 0.012 & 0.020\\
\hline
      & 1018 & 0.032 & 0.024\\
Wikipedia  & 5,038 & 0.015 & 0.020\\
      & 10,025 & 0.006 & 0.017\\
\hline
      & 1,017 & 0.023 & 0.014\\
Enron & 5,054 & 0.013 & 0.011\\
      & 10,024 & 0.007 & 0.010\\
\hline
         & 1,014 & 0.017 & 0.034\\
Coauthorship & 5,023 & 0.017 & 0.029\\
         & 10,023 &  0.006 & 0.027\\
\hline  
\end{tabular}
\end{table}

We generated the training set for each of the AsIC and the AsLT models
as follows to evaluate the proposed learning methods as a function of
the number of observed active nodes, {\em i.e.}, amount of the training
data. First we specified the target number $K$ of the active nodes we
want to have, and the training set is generated by increasing the
sequence one by one such that the total number of active nodes reaches
$K$ with each sequence starting from a randomly chosen initial active
node, skipping very short ones (those in which the number of nodes is
less than 10). In the experiments, we investigated the cases of $K =
1,000, 5,000, 10,000$.  Let $r^*$, $p^*$ and $q^*$ denote the true
values of $r$, $p$ and $q$, respectively, and $\hat{r}$, $\hat{p}$ and
$\hat{q}$ the estimated values of $r$, $p$ and $q$, respectively.  We
define the parameter estimation
errors ${\cal E}_r$, ${\cal E}_p$ and ${\cal E}_q$ by
$$
{\cal E}_r = \frac{|\hat{r} - r^*|}{r^*}, \ \ \ 
{\cal E}_p = \frac{|\hat{p} - p^*|}{p^*}, \ \ \ 
{\cal E}_q = \frac{|\hat{q} - q^*|}{q^*}. 
$$
Tables~\ref{table:ic-learn} and \ref{table:lt-learn}
show the parameter estimation errors of the proposed learning methods
for the AsIC model and the AsLT model in four networks
as a function of the number of observed active nodes, respectively.
Here, the results are averaged over five independent experiments.
As can be expected, the error is progressively reduced as the number of
active nodes becomes larger. The algorithm guarantees to converge but
does not guarantee the global optimal solution. In most of the cases,
the number of iterations is less than 100. These results indicate that
it converges to the correct solution in practice for all the parameters
and for all the networks, which demonstrate the effectiveness of the
proposed methods.

\begin{table}[!t]
\centering
\caption{Parameter estimation error of the learning method
from a single observed sequence
for four networks (Values in parentheses are standard deviations.)}
\vspace{0.2cm}
\label{table:lm_result}
{
\begin{tabular}{c|c|c|c|c|c}
\hline
\multicolumn{2}{c|}{Network}
                   & ~~Blog~~    & ~~Wikipedia~~ & ~~Enron~ & Coauthorship \\
\hline
\hline
AsIC
       & ${\cal E}_r$       & 0.091 (0.121) & 0.088 (0.132) & 0.029
(0.020)& 0.119 (0.173) \\
       & ${\cal E}_p$  & 0.064 (0.085) & 0.043 (0.056) & 0.022 (0.019)&
0.121 (0.255) \\
\hline
AsLT
       & ${\cal E}_r$      & 0.188 (0.219) & 0.192 (0.272) & 0.143
(0.140) & 0.214 (0.194) \\
       & ${\cal E}_q$       & 0.078 (0.049) & 0.069 (0.043) & 0.077
(0.053) & 0.086 (0.054) \\

\hline  
\end{tabular}
}
\end{table}

\begin{figure*}[!t]
\centering
\subfloat[Blog network\label{blog_single_icm}]
{\includegraphics[width=3.7cm]{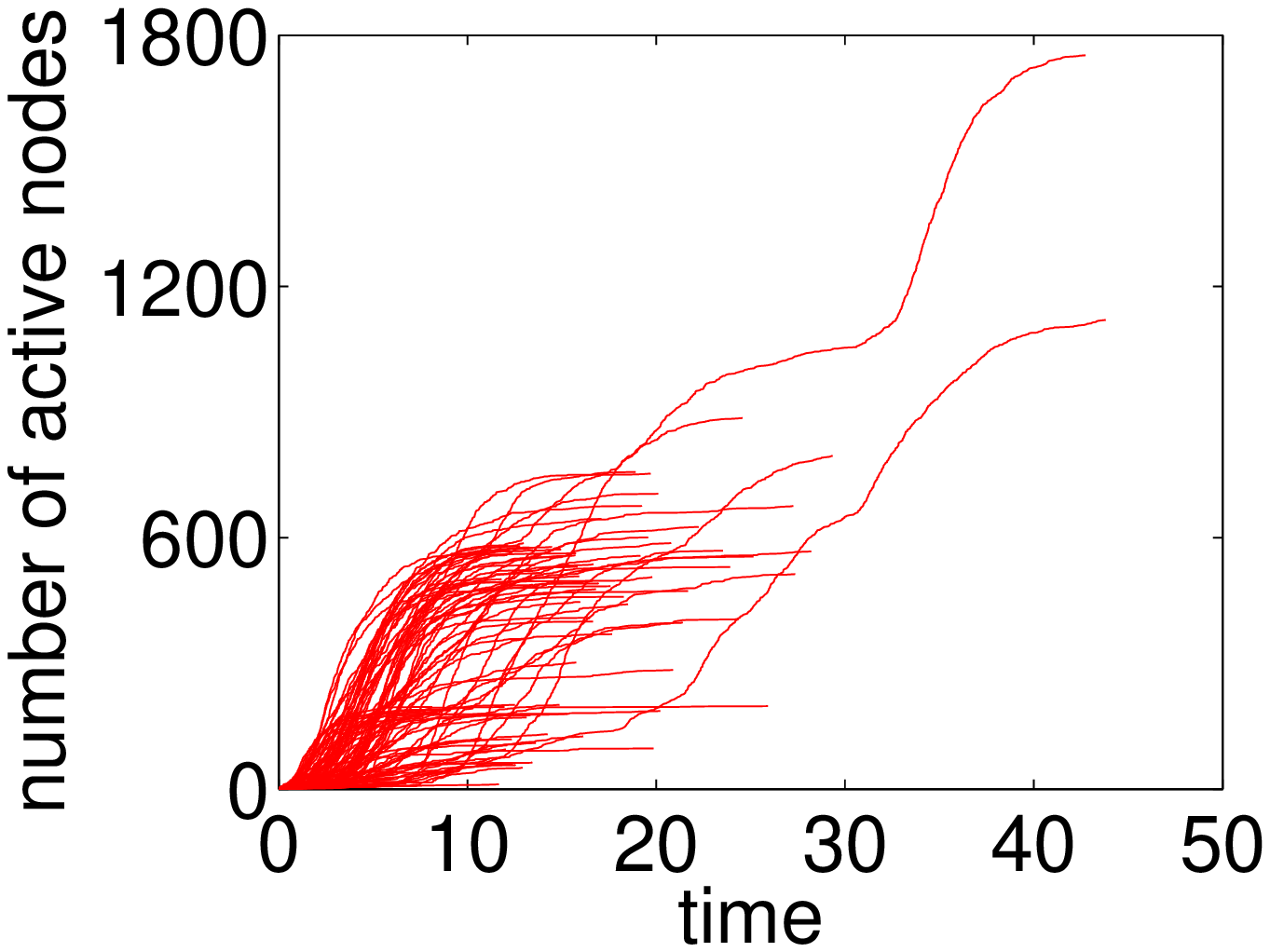}
\includegraphics[width=3.7cm]{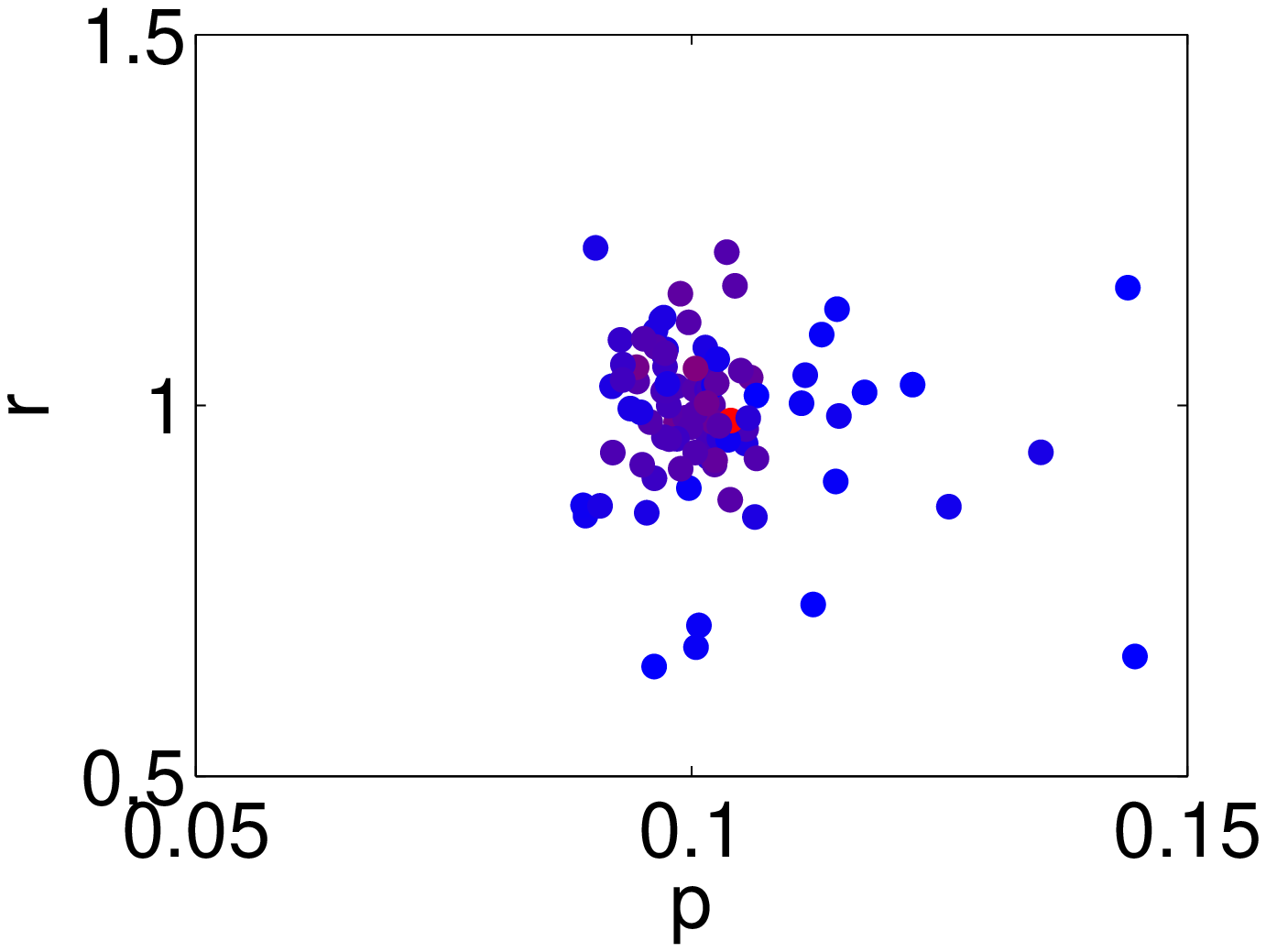}}
\hspace{1mm}
\subfloat[Wikipedia network\label{wiki_single_icm}]
{\includegraphics[width=3.7cm]{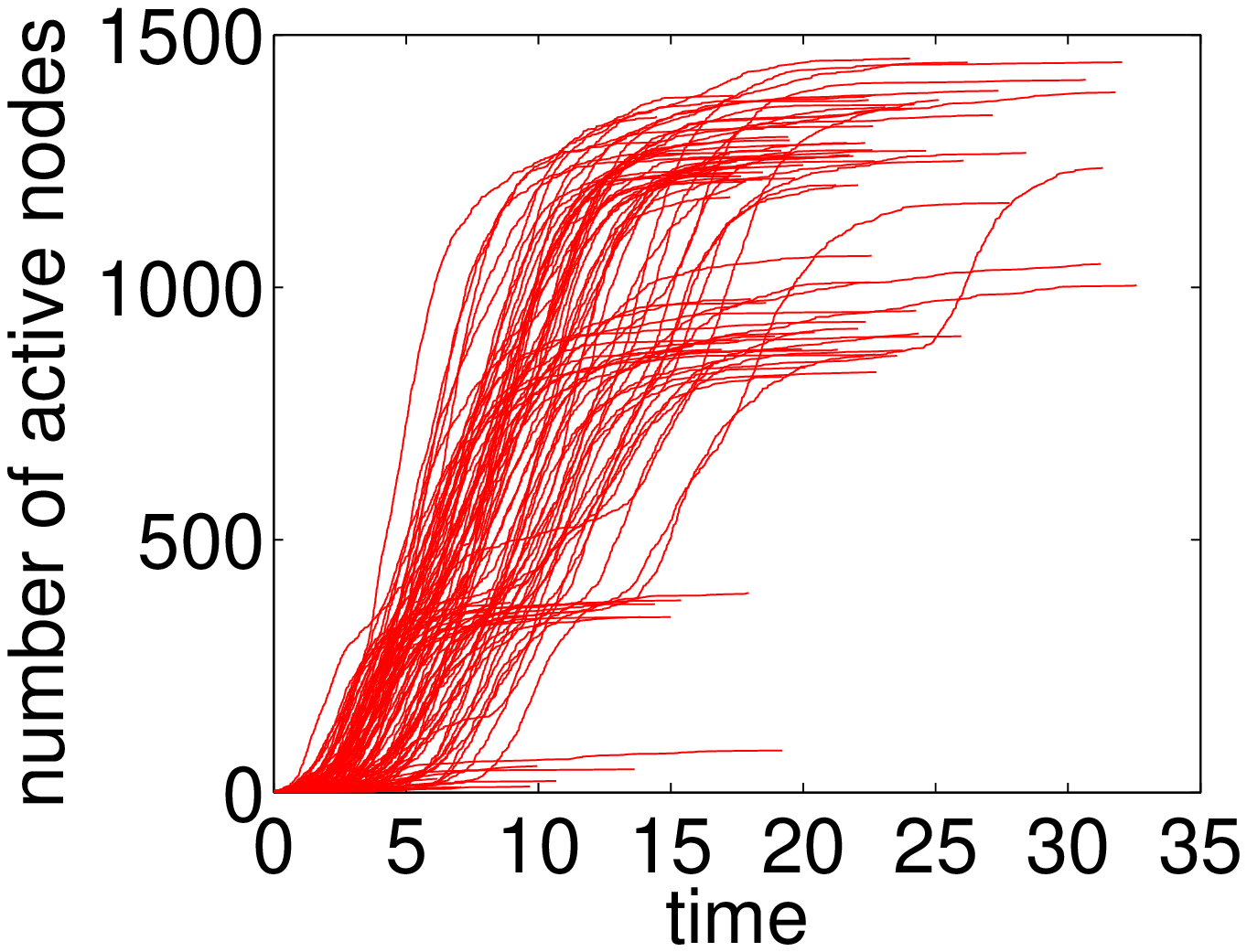}
\includegraphics[width=3.7cm]{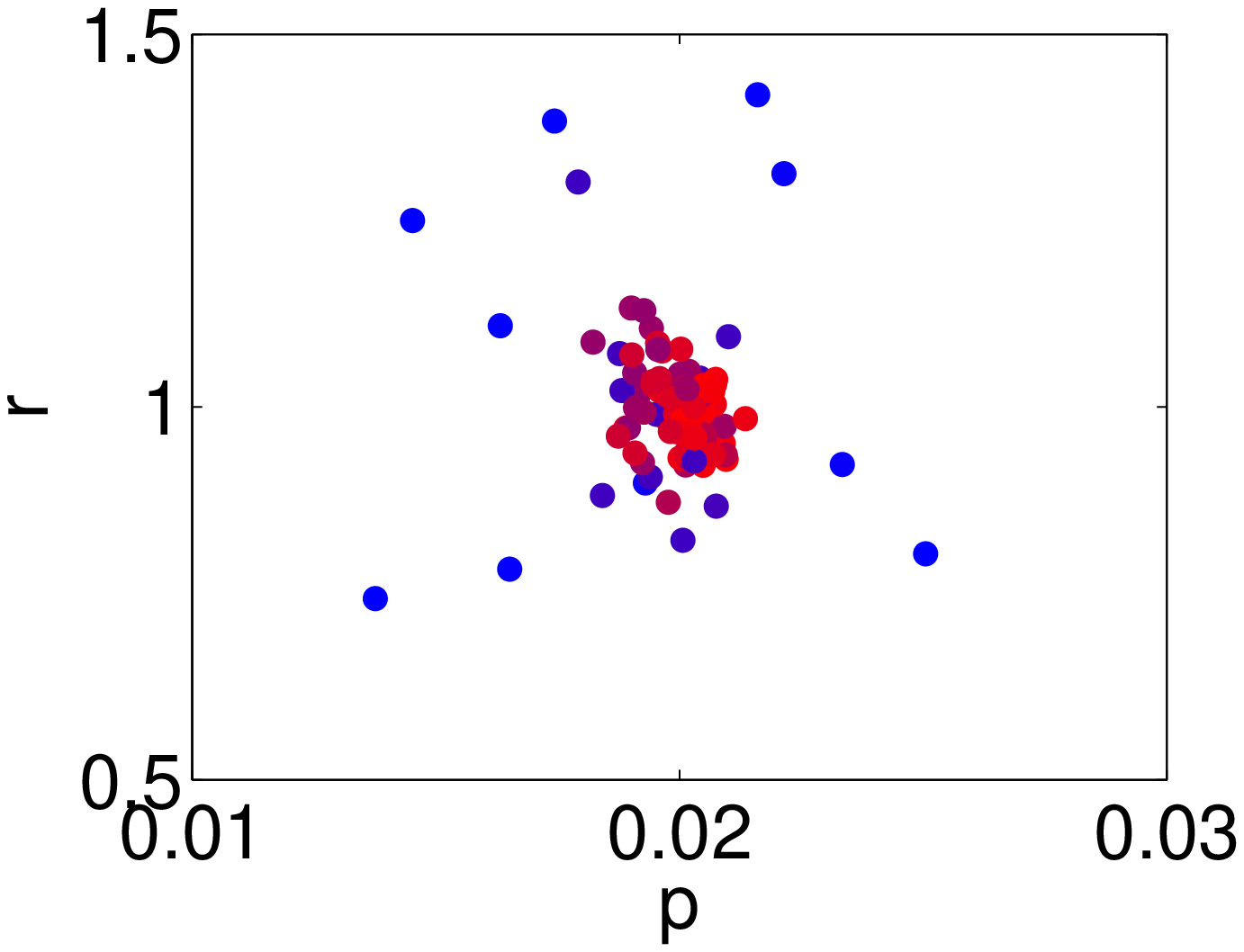}}

\subfloat[Enron network\label{enron_single_icm}]
{\includegraphics[width=3.7cm]{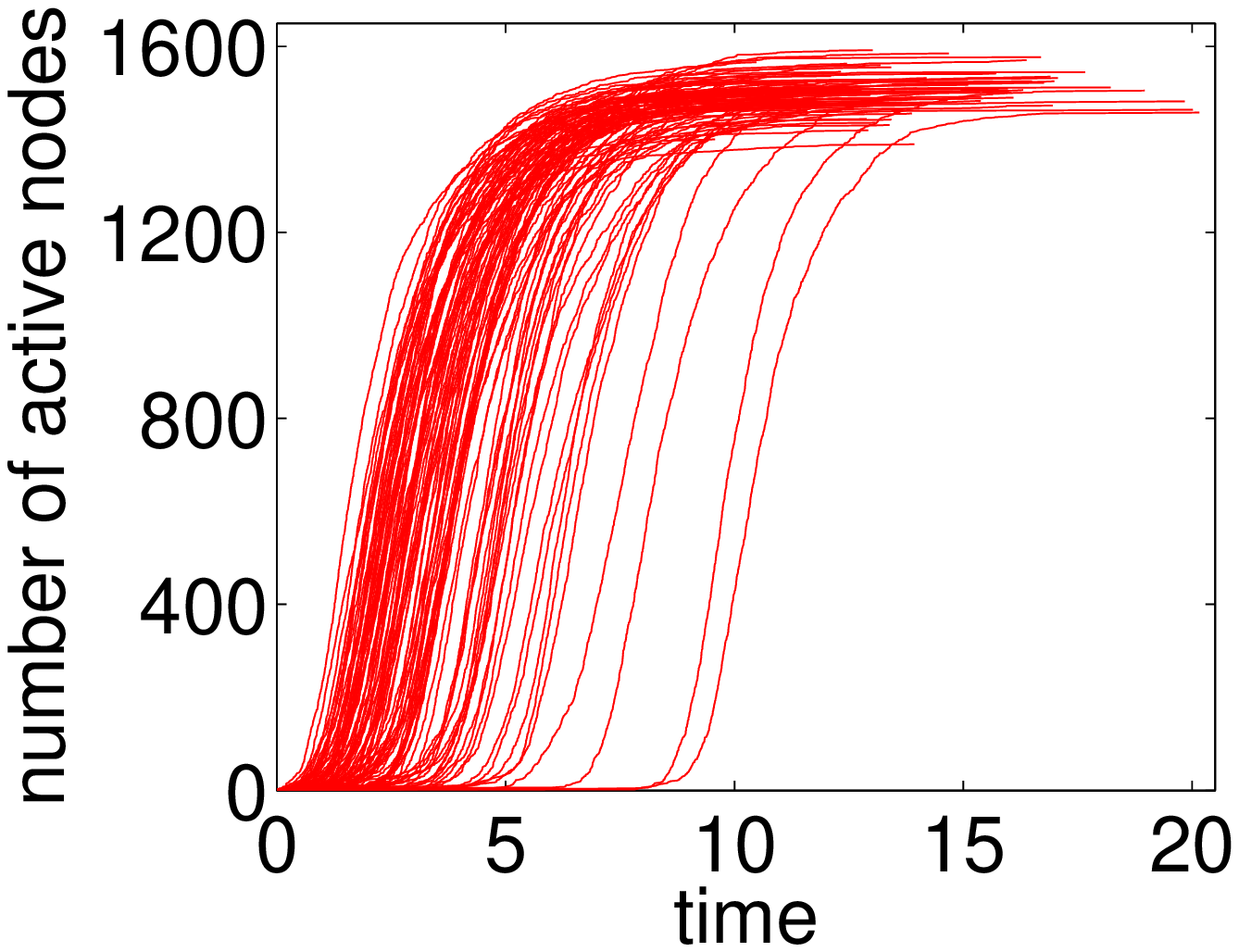}
\includegraphics[width=3.7cm]{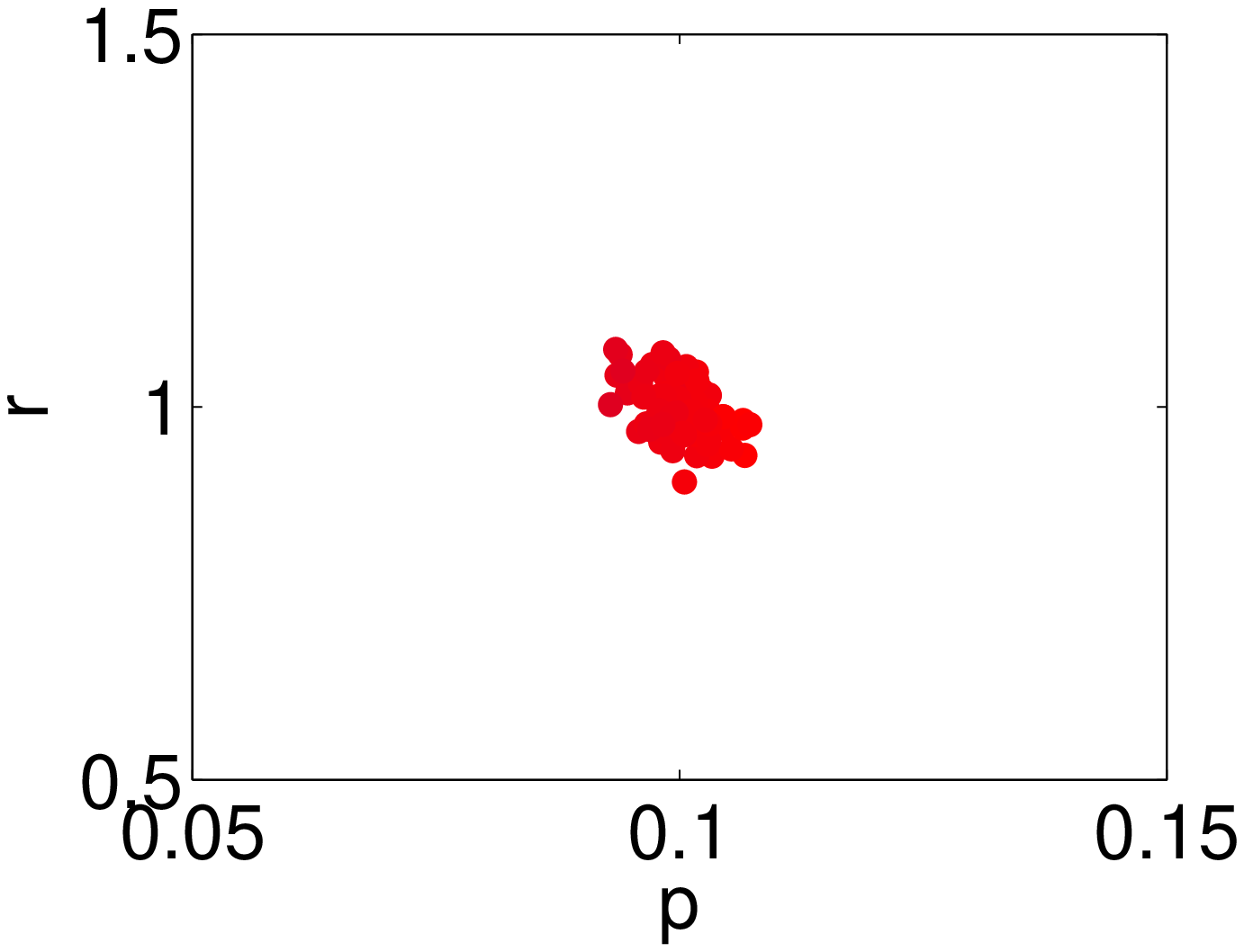}}
\hspace{1mm}
\subfloat[Coauthorship network\label{auth_single_icm}]
{\includegraphics[width=3.7cm]{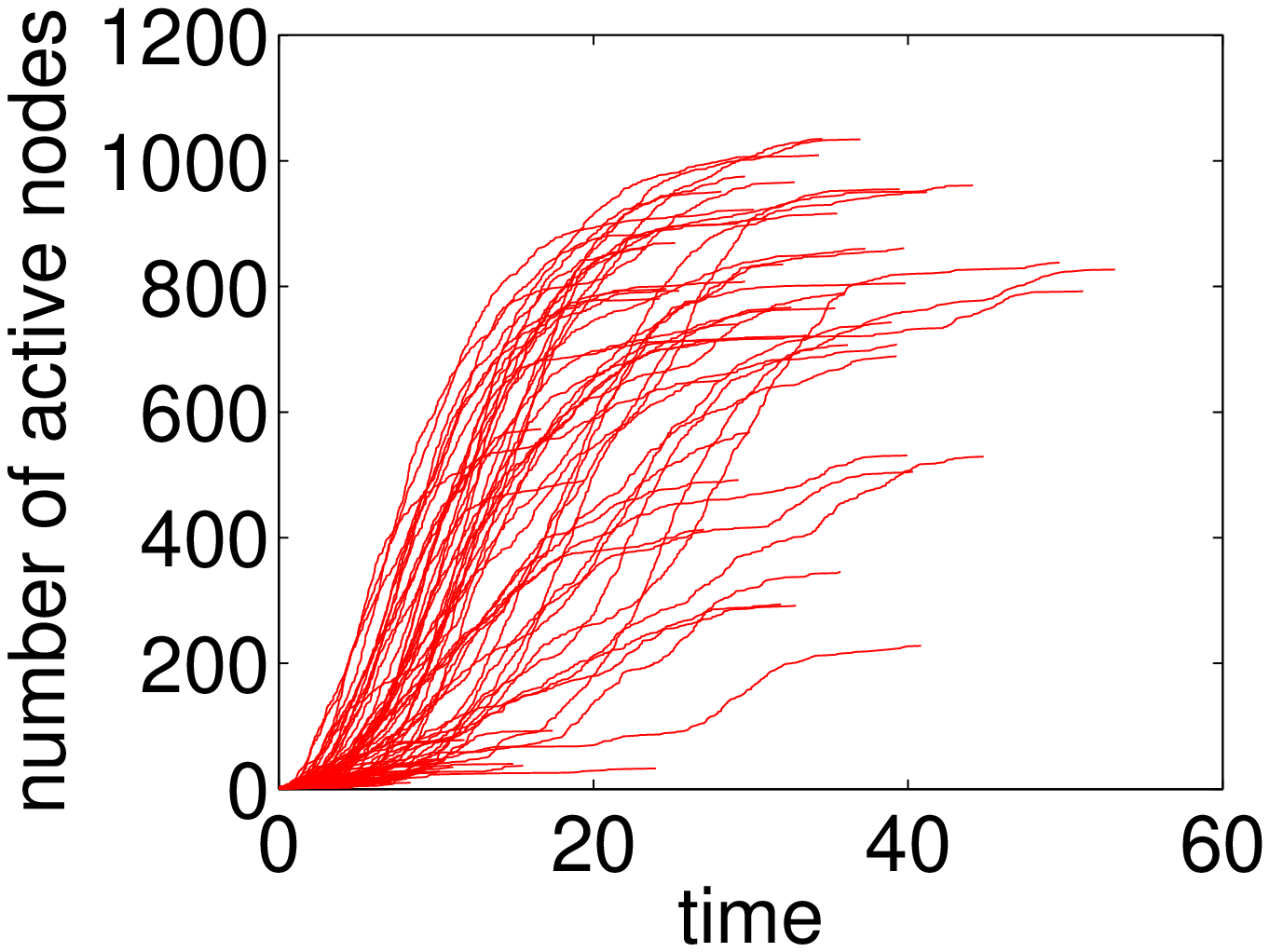}
\includegraphics[width=3.7cm]{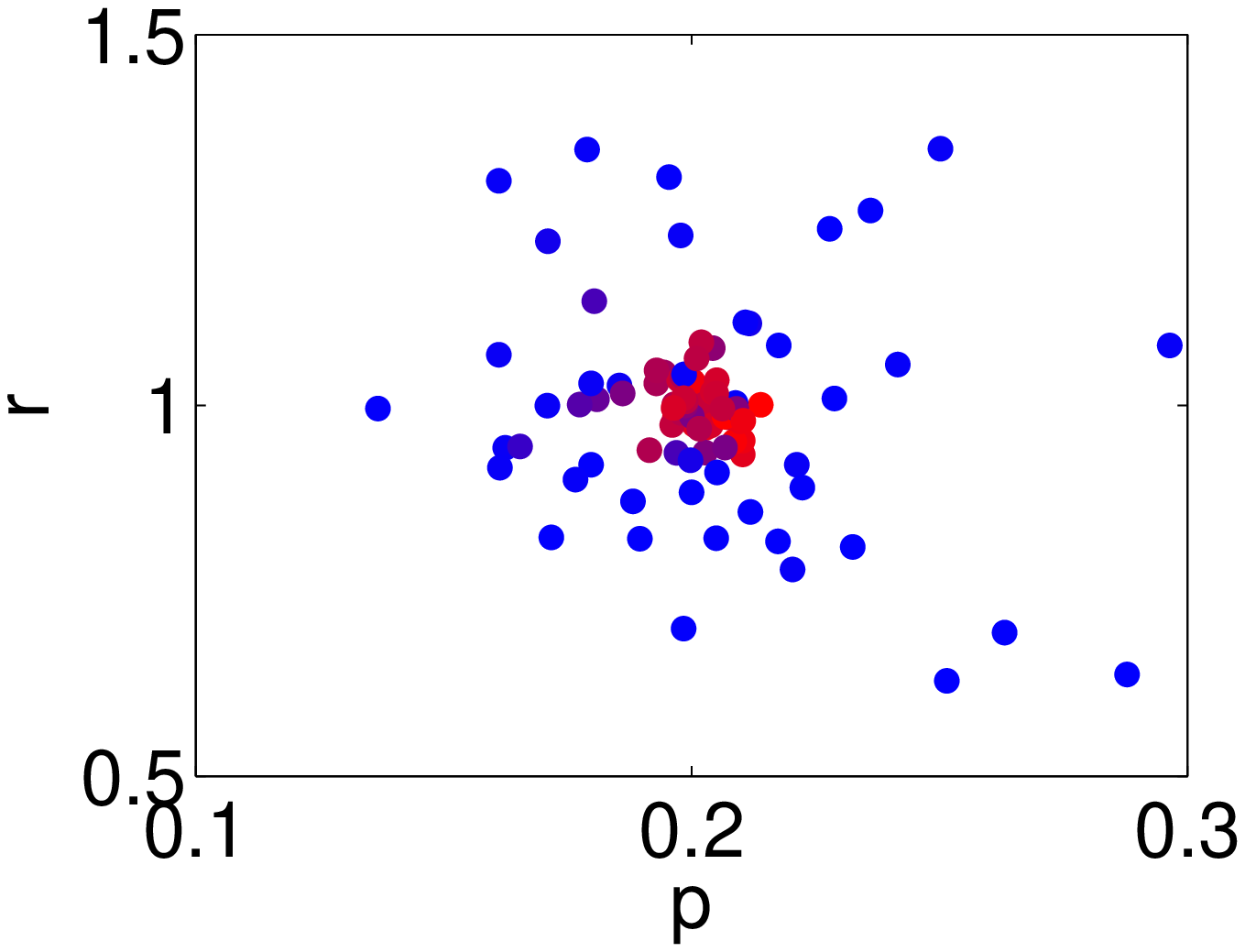}}
\caption{Influence curve and the learned parameter values from a single
observed sequence in case of AsIC (There are 100 sequences and 100
points in each figure.)}
\label{fig:singleIC}
\end{figure*}

\begin{figure*}[!t]
\centering
\subfloat[Blog network\label{blog_single_ltm}]
{\includegraphics[width=3.7cm]{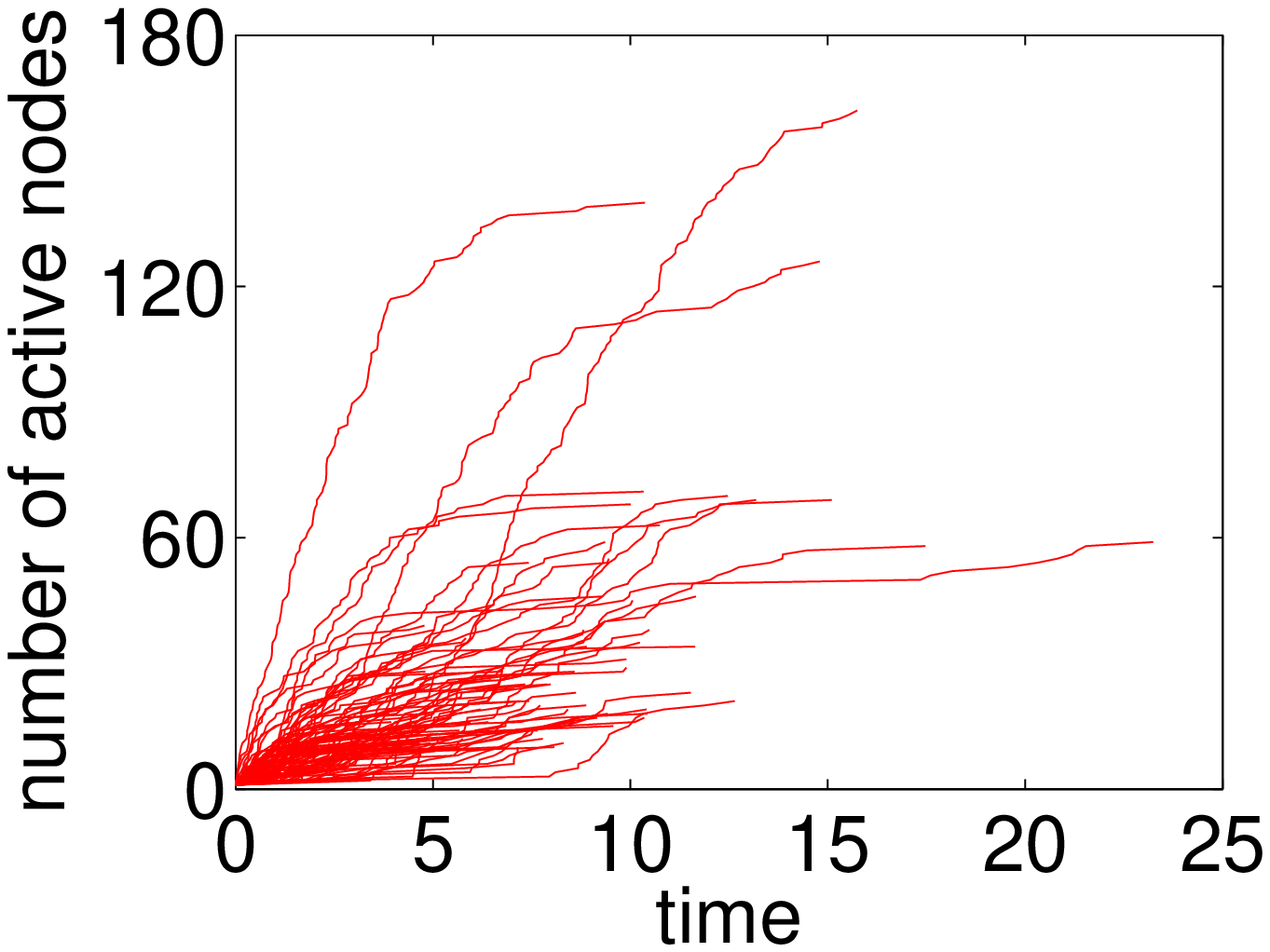}
\includegraphics[width=3.7cm]{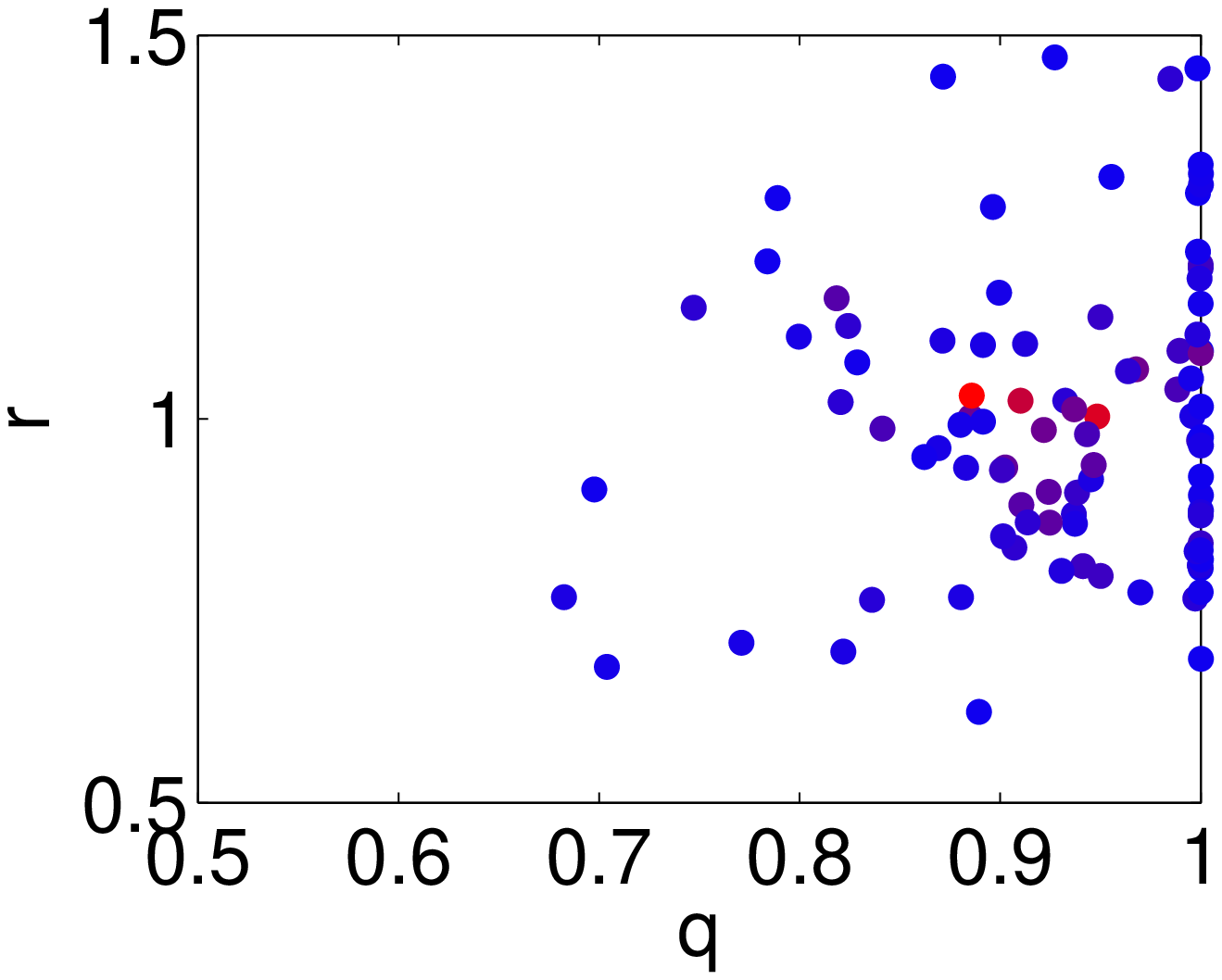}}
\hspace{1mm}
\subfloat[Wikipedia network\label{wiki_single_ltm}]
{\includegraphics[width=3.7cm]{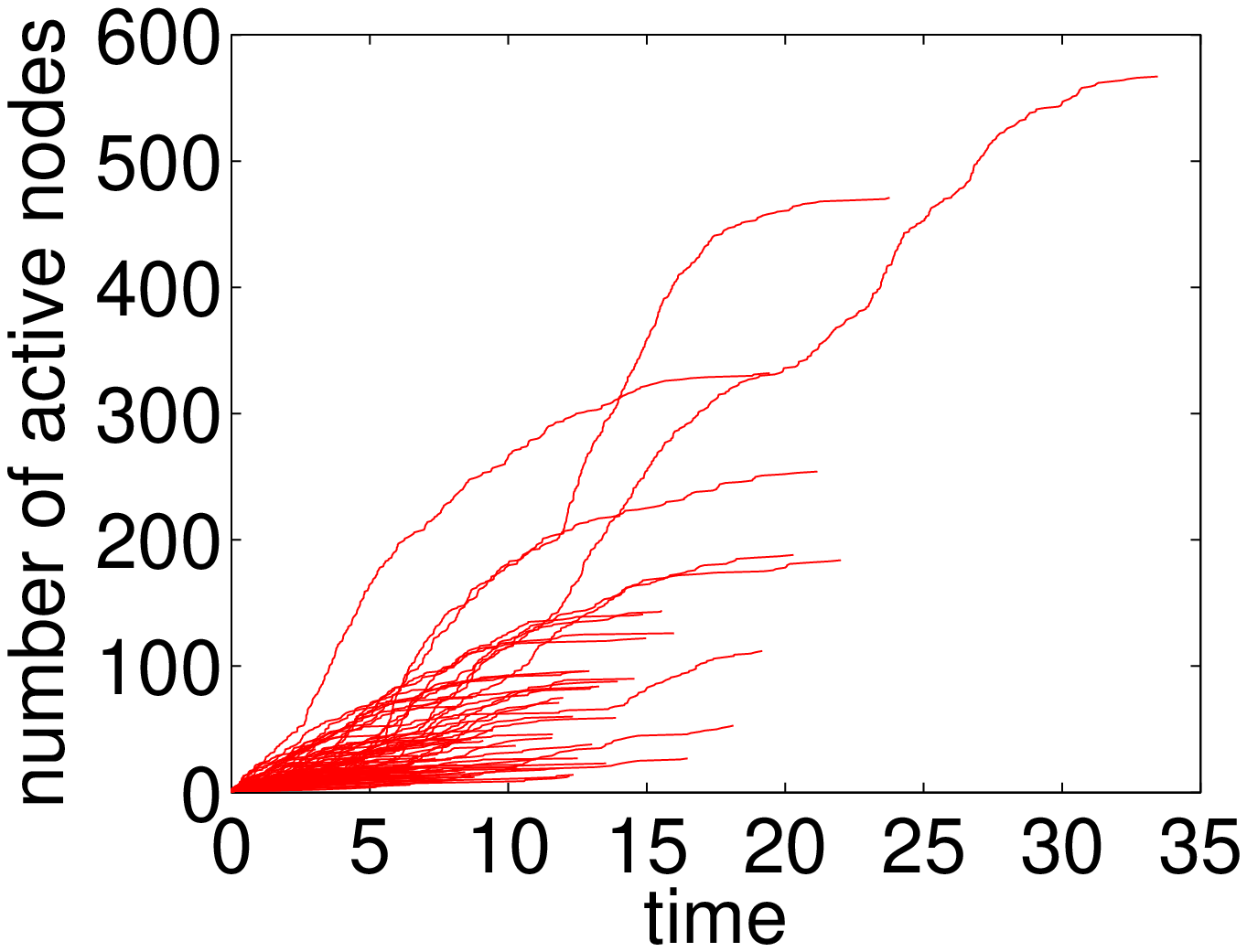}
\includegraphics[width=3.7cm]{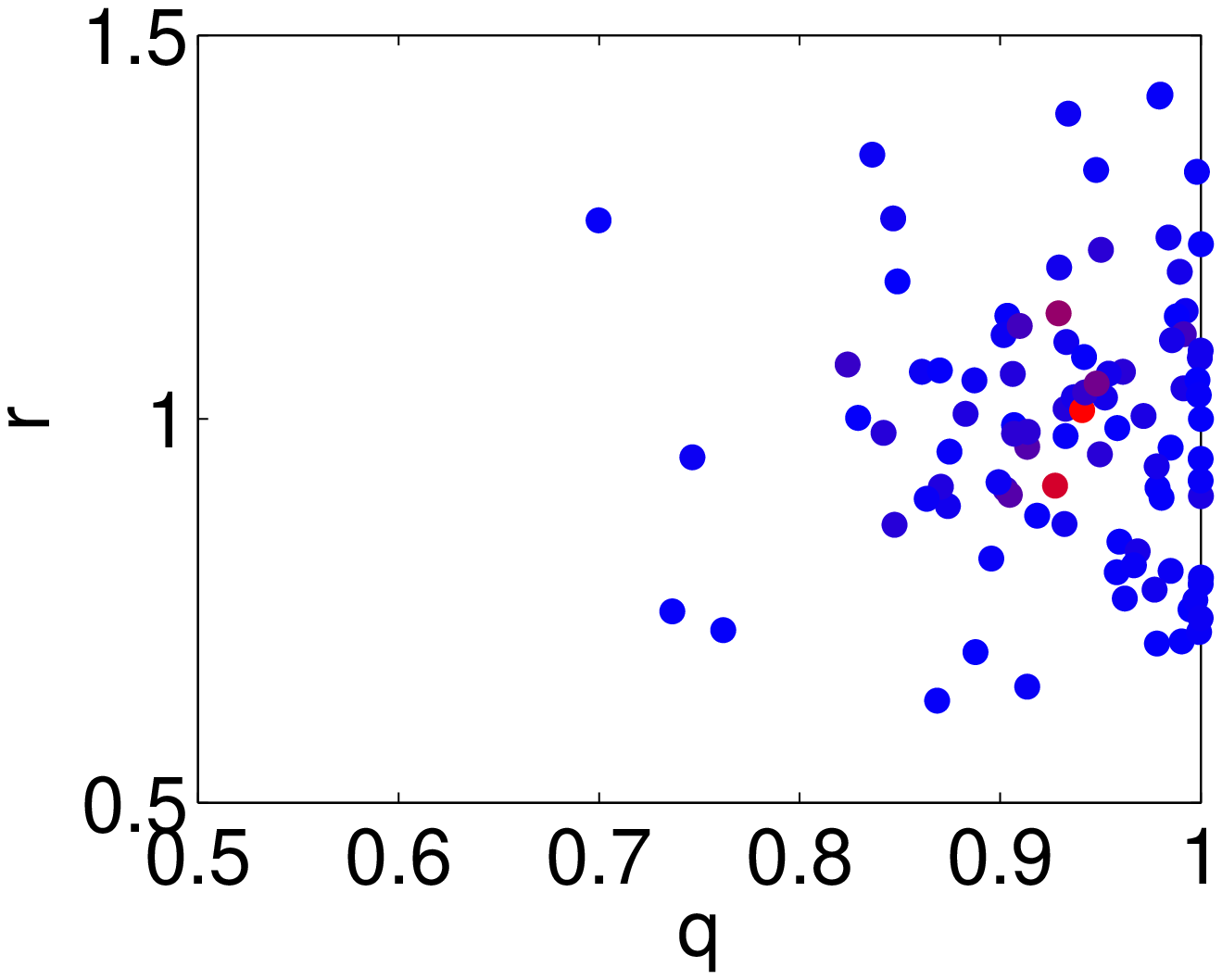}}

\subfloat[Enron network\label{enron_single_ltm}]
{\includegraphics[width=3.7cm]{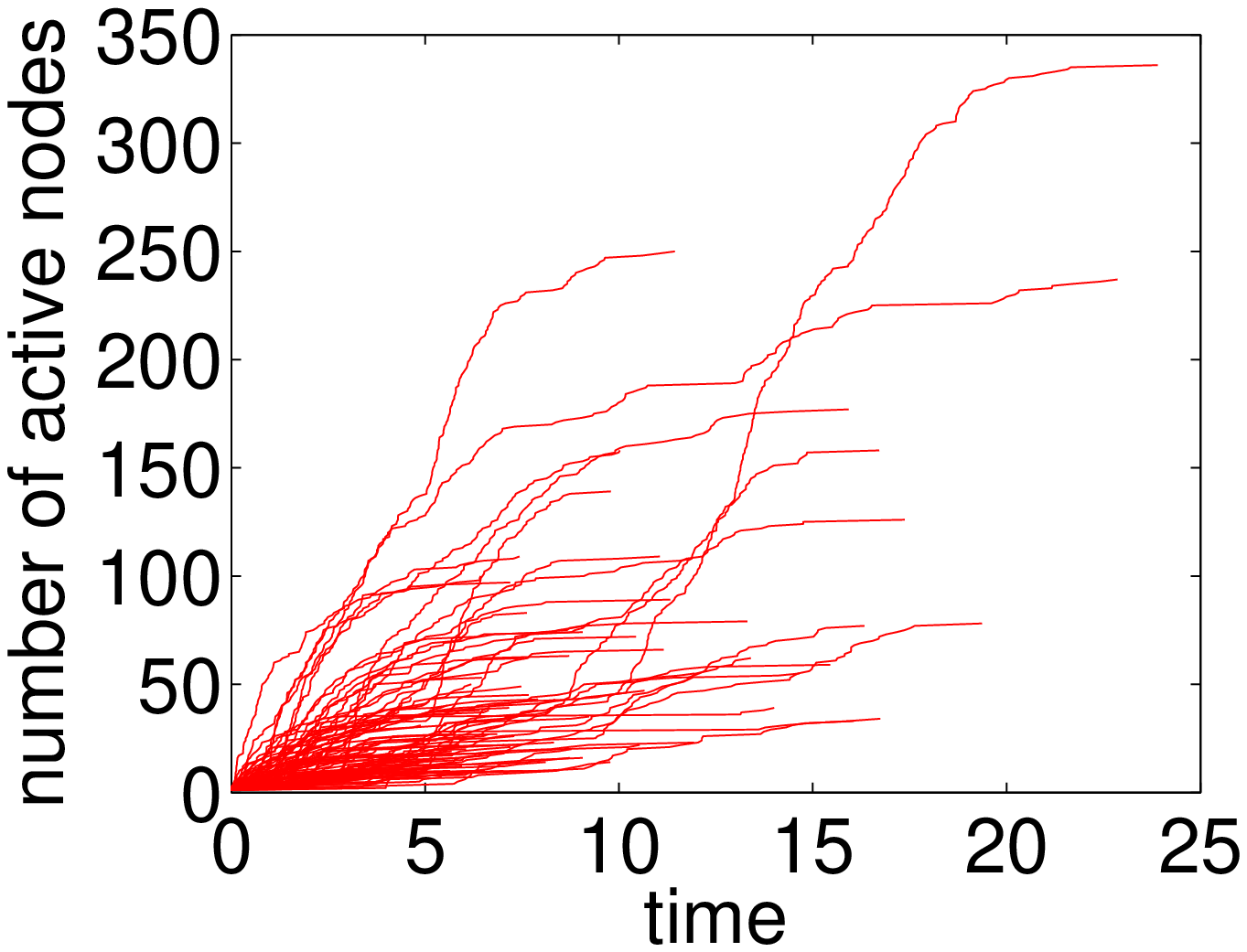}
\includegraphics[width=3.7cm]{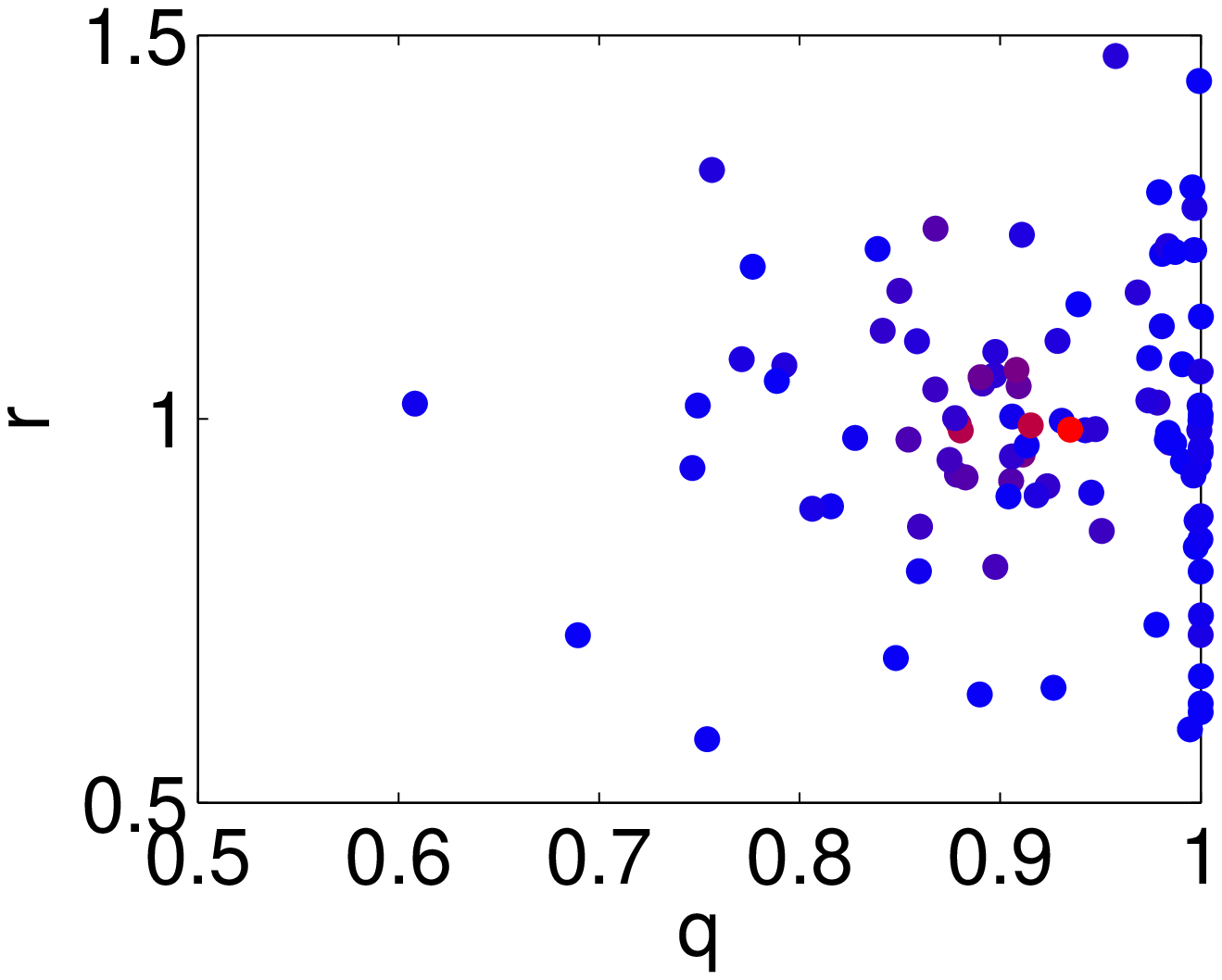}}
\hspace{1mm}
\subfloat[Coauthorship network\label{auth_single_ltm}]
{\includegraphics[width=3.7cm]{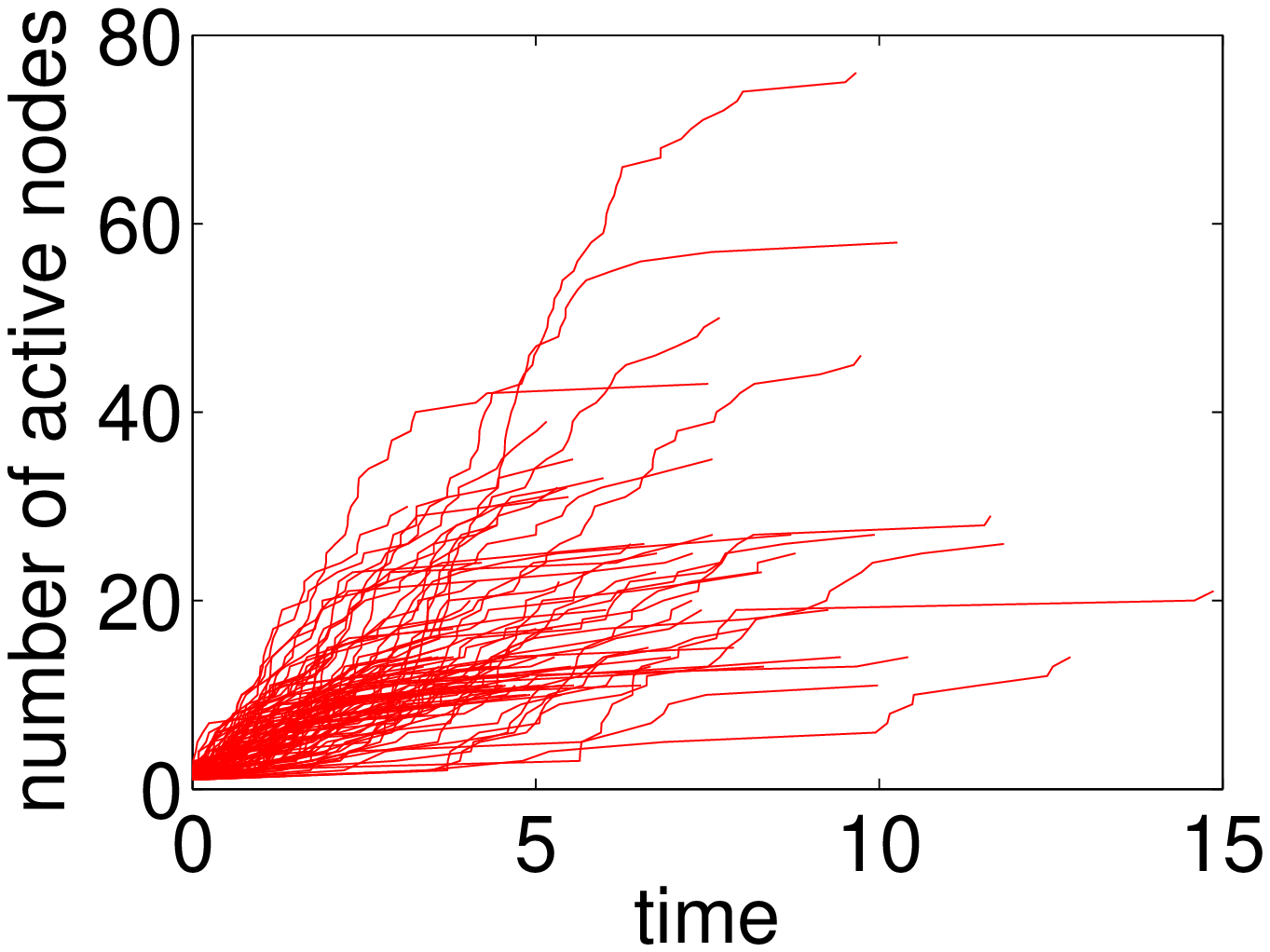}
\includegraphics[width=3.7cm]{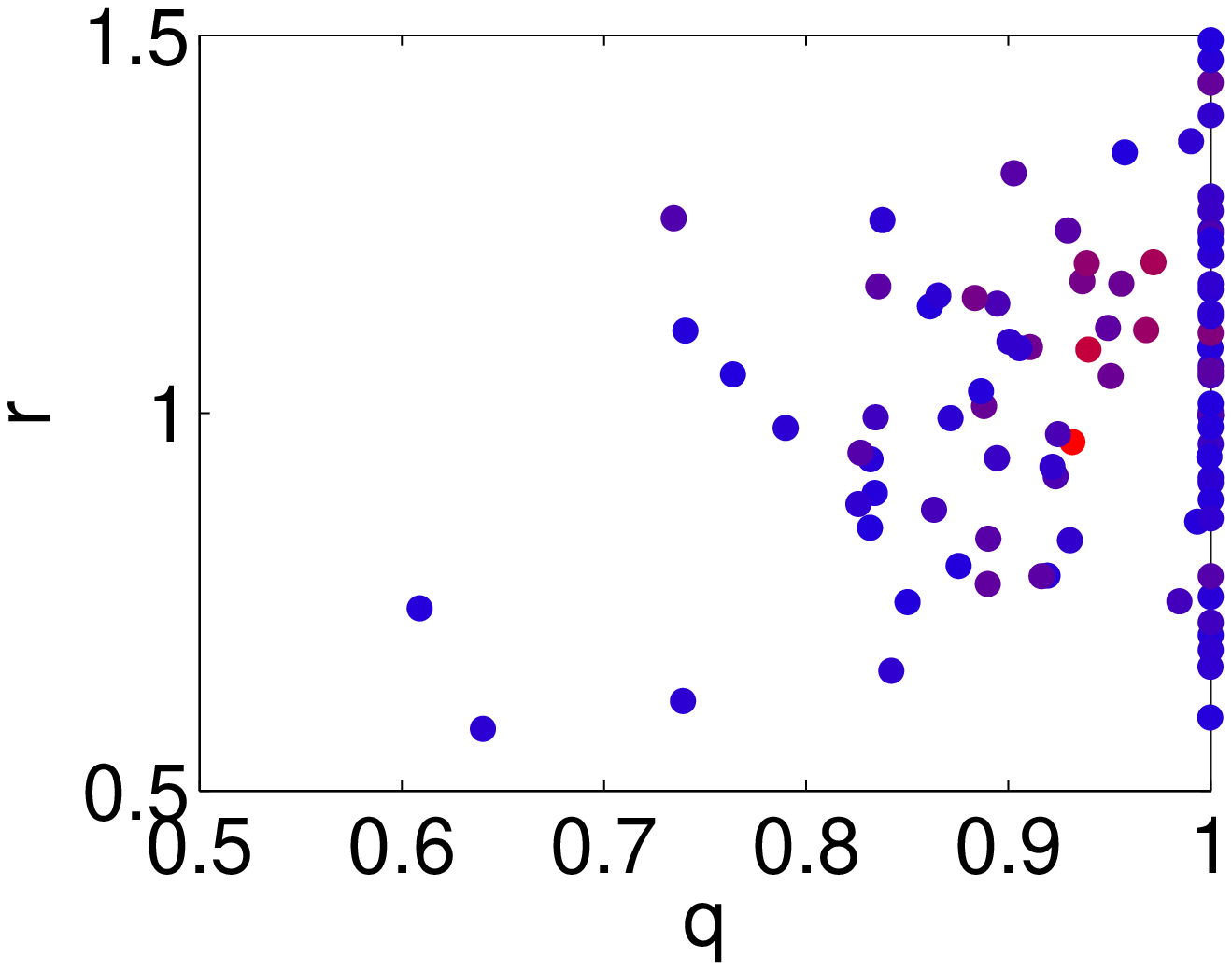}}
\caption{Influence curve and the learned parameter values from a single
observed sequence in case of AsLT (There are 100 sequences and 100
points in each figure.)}
\label{fig:singleLT}
\end{figure*}

Next, we investigated the performance of the proposed learning method
when the training set is a single diffusion sequence.
Table~\ref{table:lm_result} shows the results for four networks, where
the results are averaged over $100$ independent experiments.  Compared
with Tables~\ref{table:ic-learn} and \ref{table:lt-learn}, the errors
become larger. The average error of $p$ and $r$ for AsIC is 6\% and 8\%,
and the average error of $q$ and $r$ for AsLT is 8\% and 18\%,
respectively. The best results for AsIC is Enron network (2\% for $p$
and 3\% for $r$), and the best results for AsLT is Wikipedia network
(7\% for $q$) and Enron network (14\% for $r$). The worst results for
AsIC is Coauthorship network (12\% for $p$ and 11\% for $r$), and the
worst results for AsLT is Coauthorship network (9\% for $q$ and 21\% for
$r$). In general the accuracy is better for AsIC than for AsLT. This is
because the lengths of the sequences are larger for AsIC. Further, $r$
is more difficult to correctly estimate than $p$ and $q$. In order to
see the difference in the learning result for each sequence in more
depth, we plotted the number of active nodes as a function of time (the
influence curve),\footnote{This is different from the influence degree
$\sigma$ described in Section \ref{dataset1} which is the expected
value of the number of active nodes at the final time.} and the
values of the parameters learned, $(p, r)$ for AsIC and $(q, r)$ for
AsLT, in Figures~\ref{fig:singleIC} and \ref{fig:singleLT}. The length
of each sequence varies considerably. Some sequences are short and some
others are long. The color of the dots for the learned parameters is
determined in such a way that it goes from true blue to true red in
proportion to the sequence length, i.e., the shortest sequence is true
blue and the longest sequence is true red. From these results we can see
the algorithm learns the parameter values within 10\% of the correct
values if the length is reasonably long. For example, Enron network
generates long sequences from all the randomly chosen initial active
nodes in case of AsIC and the learning accuracy is very good.  We draw a
conclusion that although it is not desirable we can still estimate the
parameter values from a single observation sequence if this is the only
choice available.

\subsection{Node Ranking}\label{ranking}

We measure the influence of node $v$ by the influence degree $\sigma
(v)$ for the diffusion model that has generated ${\cal D}_M$. We
compared the result of the high ranked influential nodes for the true
model that uses the assumed true parameter values with 1) the proposed
method that uses the learned parameter values, 2) four heuristics widely
used in social network analysis (all computed by the network topology
alone) and 3) the same proposed method in which an incorrect diffusion
model is assumed, {\em i.e.}, data generated by AsIC but learning
assumed AsLT and vice versa. Here again the influence degree is
estimated by the bond percolation method~\cite{kimura:aaai07,kimura:dmkd},
where we used $10,000$ bond percolation processes
according to \citeA{kimura:tkdd} and \citeA{kimura:dmkd}.

We call the proposed method the model based method. We call it the AsIC
model based method if it employs the AsIC model as the information
diffusion model. We then learn the parameters of the AsIC model from the
observed data ${\cal D}_M$, and rank nodes according to the influence
degrees based on the learned model. The AsLT model based method is
defined in the same way.  Among the four heuristics we used, the first
three are ``degree centrality'', ``closeness centrality'', and
``betweenness centrality''. These are commonly used as influence measure
in sociology \cite{wasserman:book94}, where the out-degree of node $v$
is defined as the number of links going out from $v$, the closeness of
node $v$ is defined as the reciprocal of the average distance between
$v$ and other nodes in the network, and the betweenness of node $v$ is
defined as the total number of shortest paths between pairs of nodes
that pass through $v$. The fourth is ``authoritativeness'' obtained by
the ``PageRank'' method \cite{brin:www7}. We considered this measure as
one alternative since this is a well known method for identifying
authoritative or influential pages in a hyperlink network of web pages.
This method has a parameter $\varepsilon$; when we view it as a model of
a random web surfer, $\varepsilon$ corresponds to the probability with
which a surfer jumps to a page picked uniformly at random
\cite{ng:ijcai01}.  In our experiments, we used a typical setting of
$\varepsilon = 0.15$.

\begin{figure*}[!t]
\centering
\subfloat[Blog network\label{blog_icm}]
{\includegraphics[width=7cm]{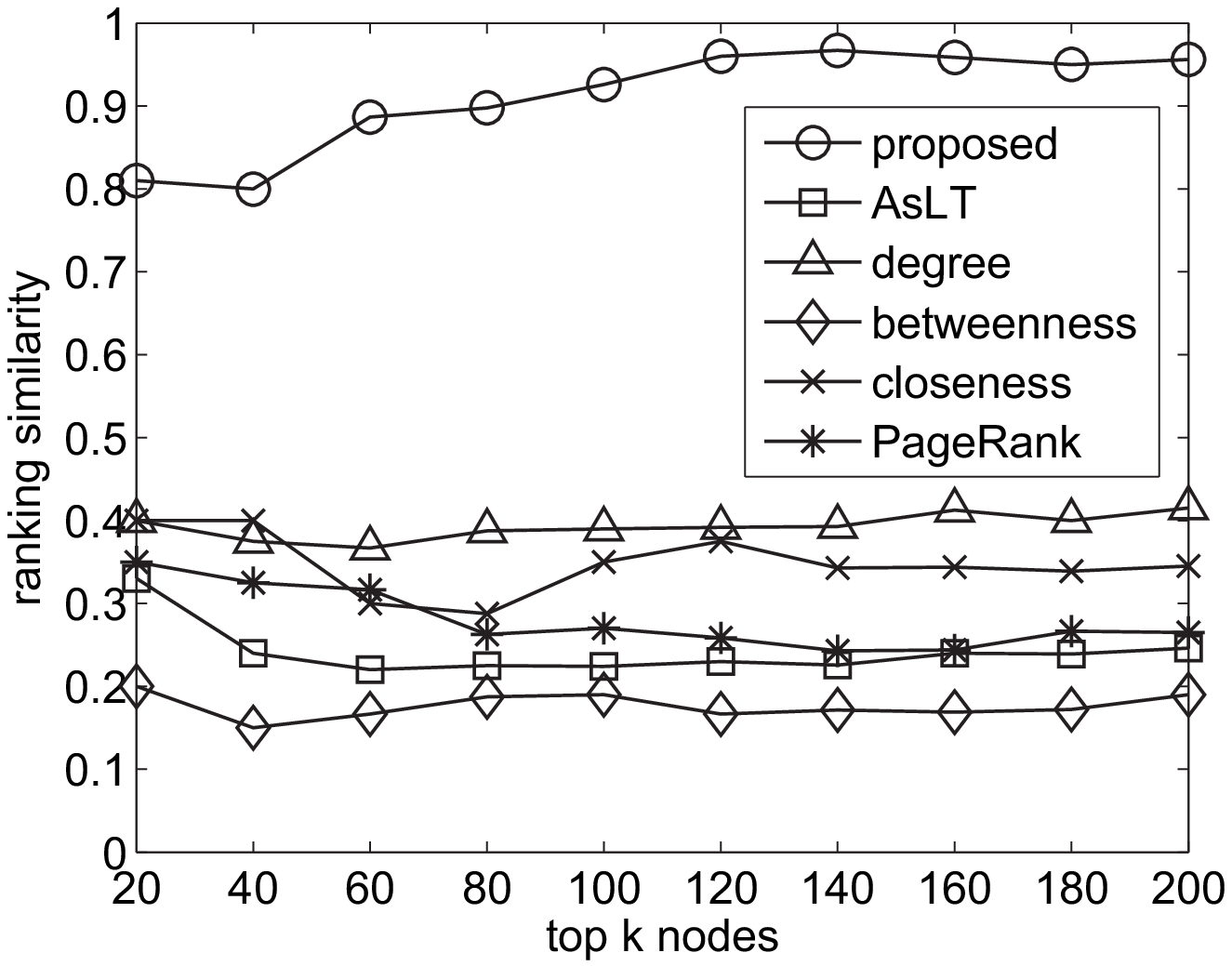}}
\hspace{3mm}
\subfloat[Wikipedia network\label{wiki_icm}]
{\includegraphics[width=7cm]{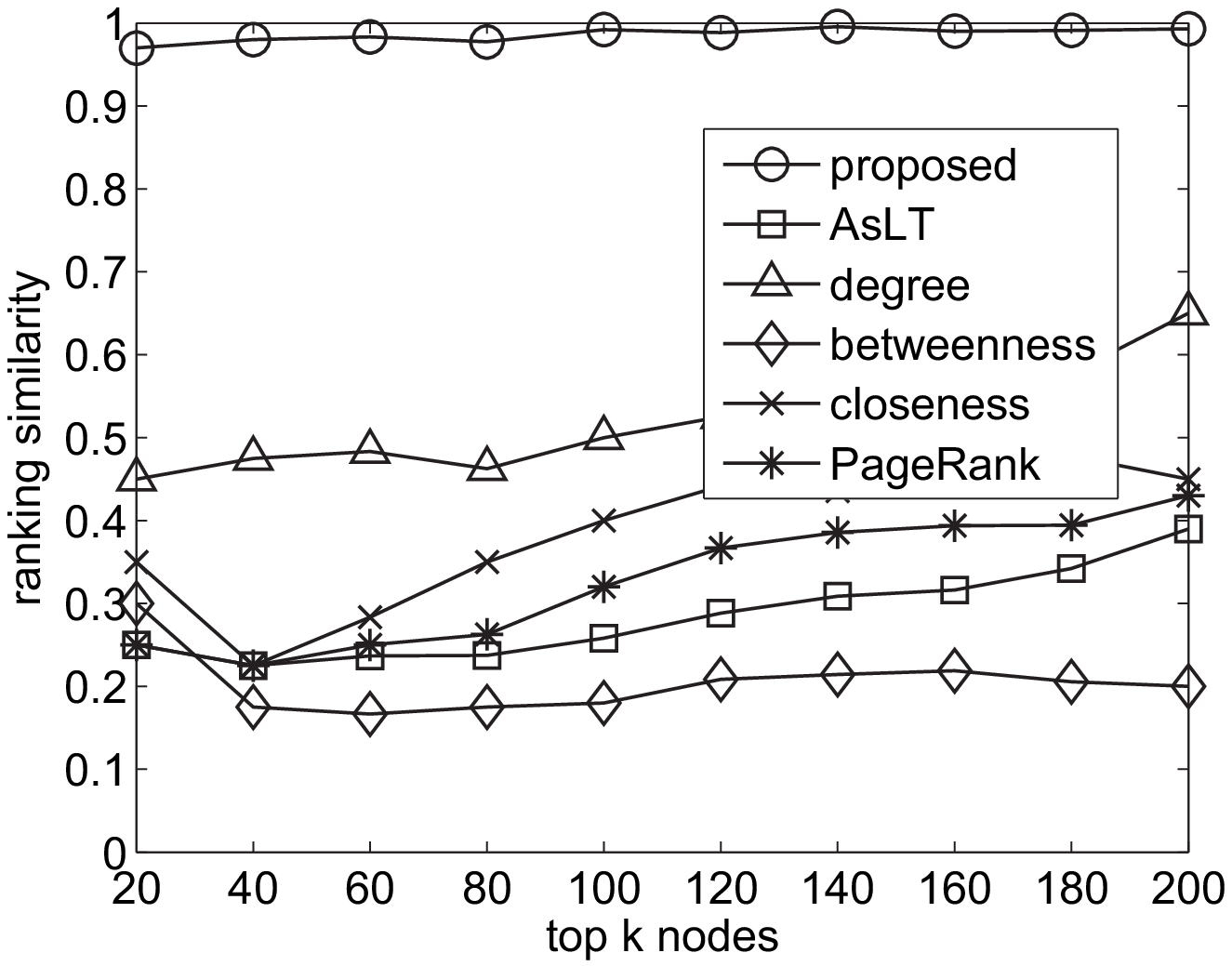}}
\hspace{1mm}
\subfloat[Enron network\label{enron_icm}]
{\includegraphics[width=7cm]{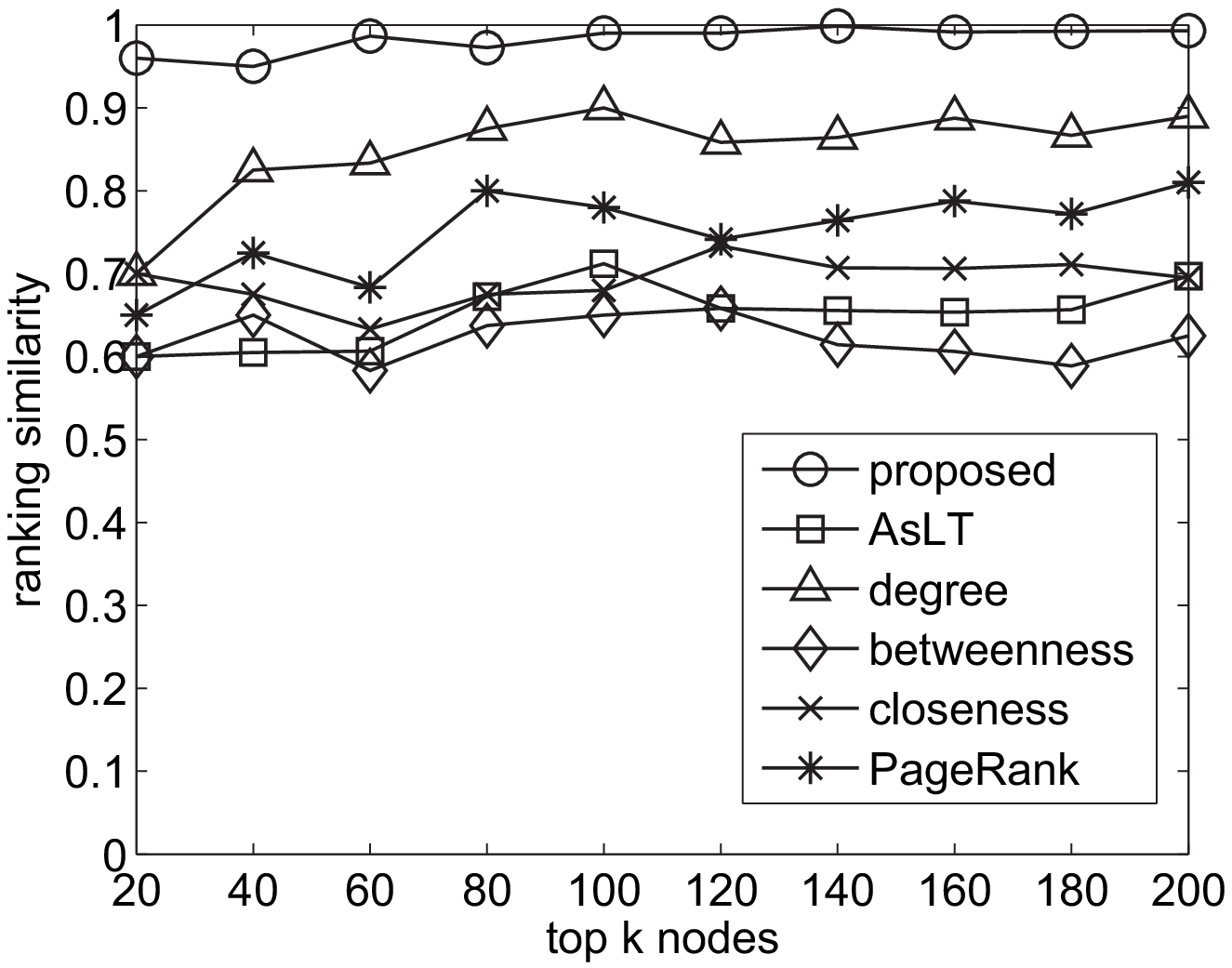}}
\hspace{3mm}
\subfloat[Coauthorship network\label{coauthor_icm}]
{\includegraphics[width=7cm]{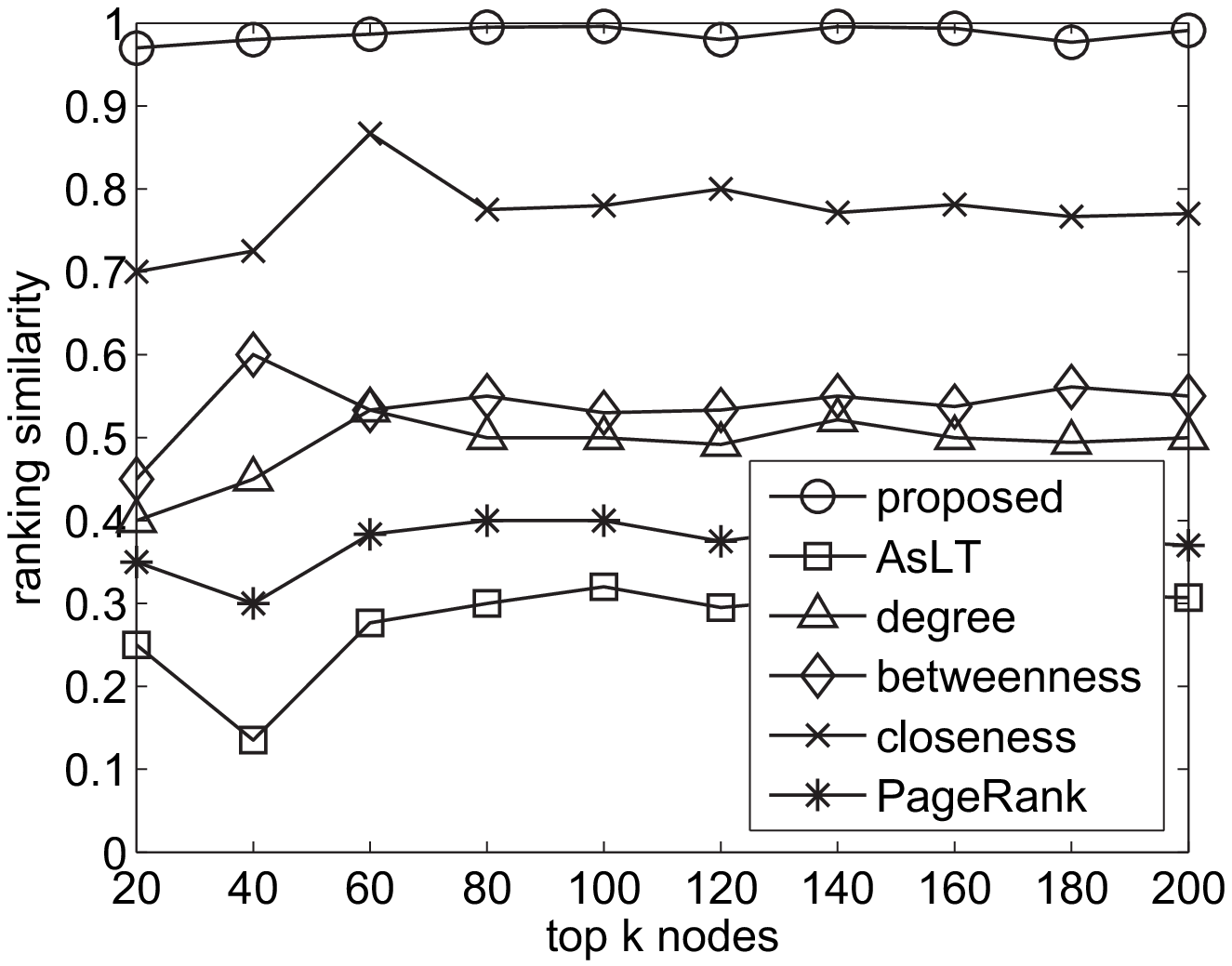}}
\caption{Performance comparison in extracting influential nodes
for the AsIC model}
\label{AsIC:ranking}
\end{figure*}

\begin{figure*}[!t]
\centering
\subfloat[Blog network\label{blog_ltm}]
{\includegraphics[width=7cm]{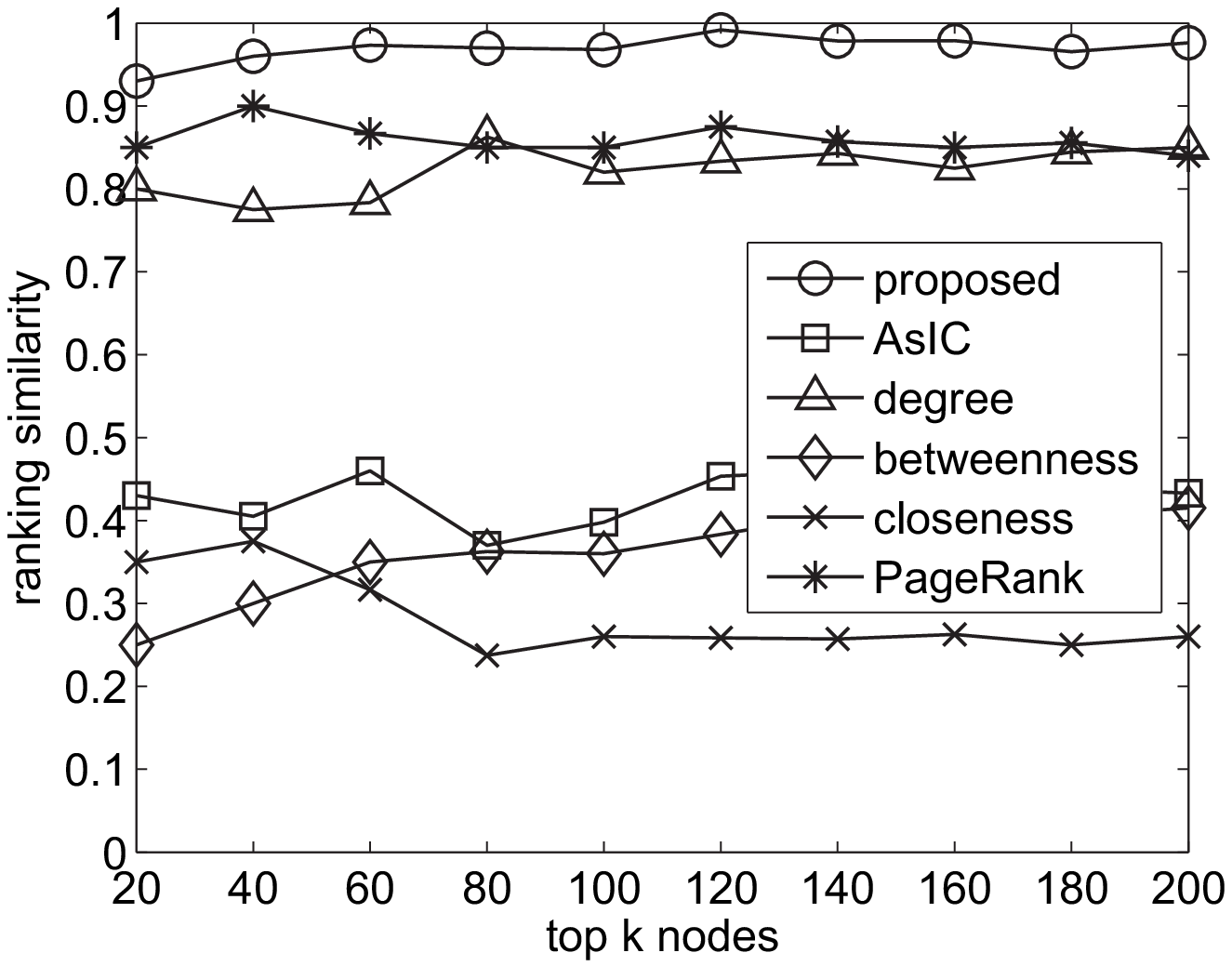}}
\hspace{3mm}
\subfloat[Wikipedia network\label{wiki_ltm}]
{\includegraphics[width=7cm]{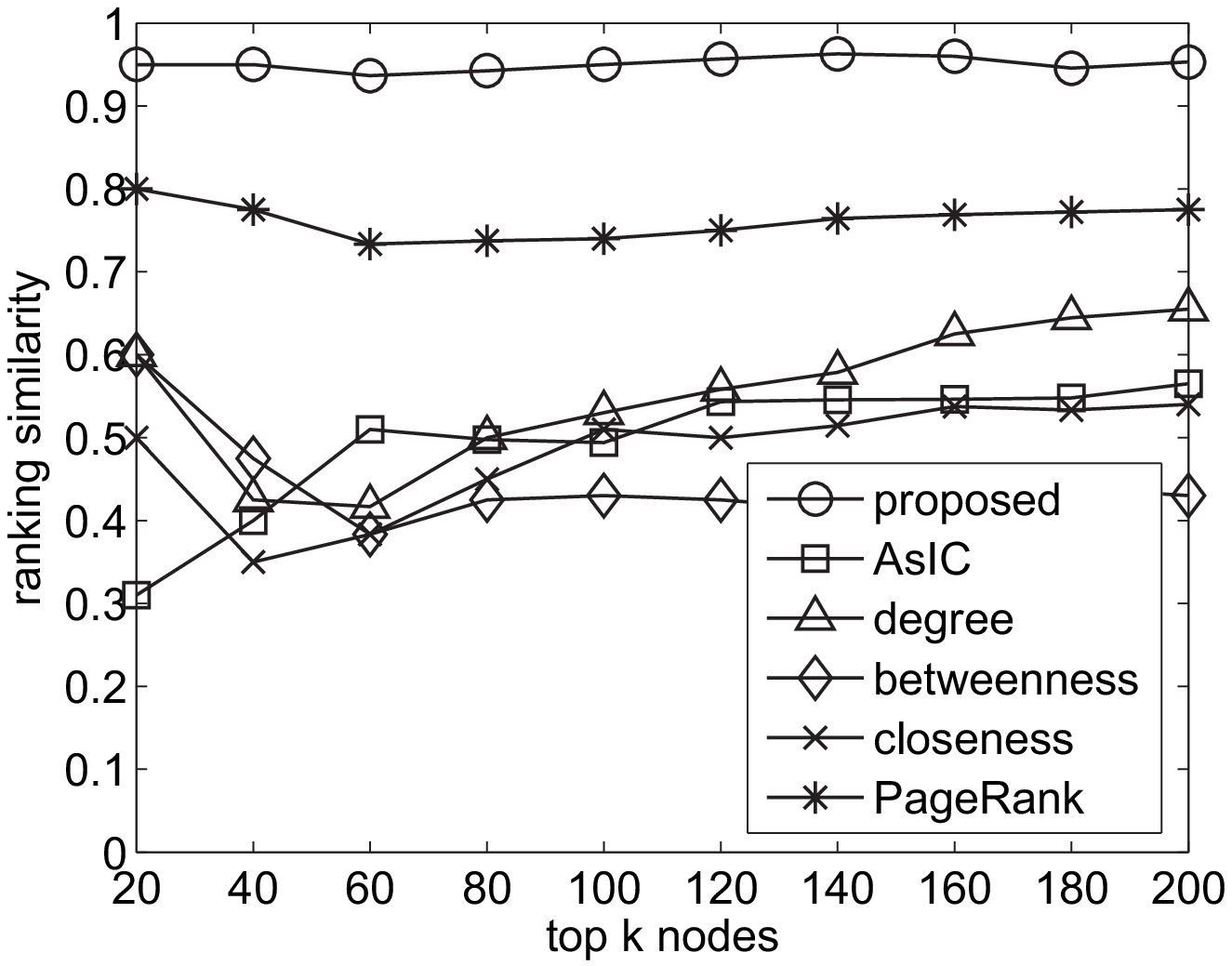}}
\hspace{1mm}
\subfloat[Enron network\label{enron_ltm}]
{\includegraphics[width=7cm]{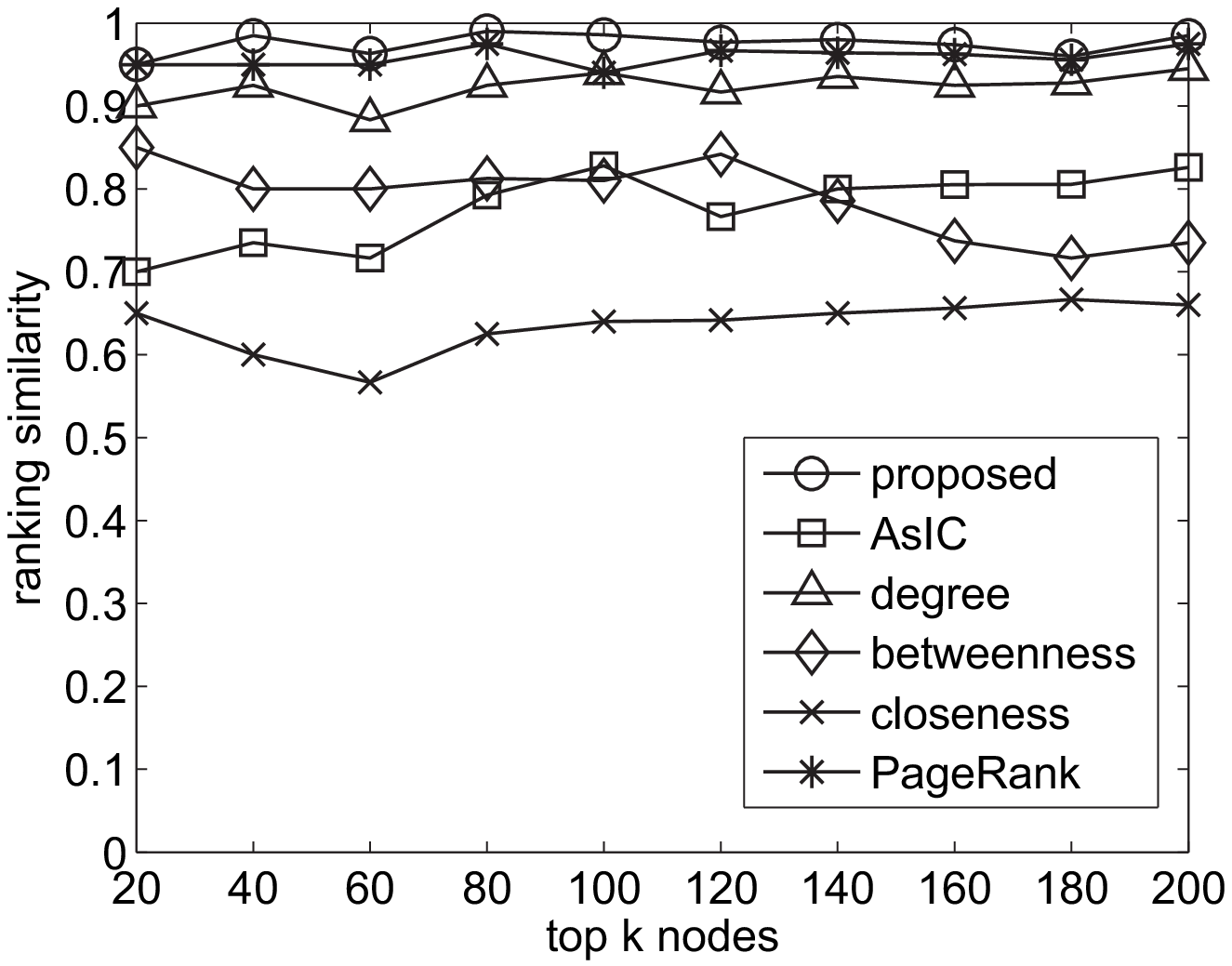}}
\hspace{3mm}
\subfloat[Coauthorship network\label{coauthor_ltm}]
{\includegraphics[width=7cm]{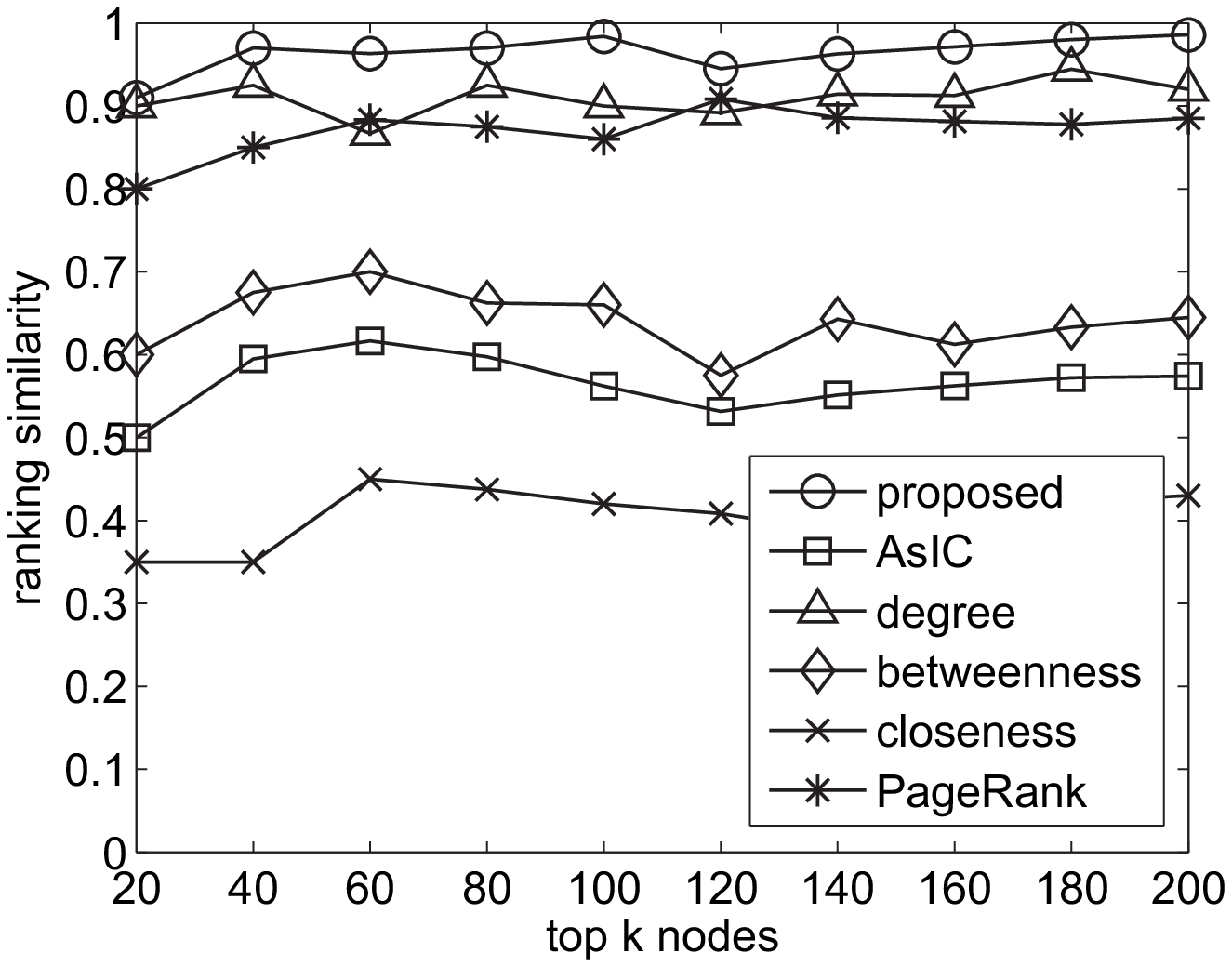}}
\caption{Performance comparison in extracting influential nodes
for the AsLT model}
\label{AsLT:ranking}
\end{figure*}

In terms of extracting influential nodes from the network $G = (V,E)$,
we evaluated the performance of the ranking methods mentioned above by
the {\em ranking similarity} ${\cal F}(k) = {|L^* (k) \cap L (k)|}/{k}$
within the rank $k (>0)$, where $L^*(k)$ and $L(k)$ are the true set of
top $k$ nodes and the set of top $k$ nodes for a given ranking method,
respectively.  We focused on the performance for high ranked nodes since
we are interested in extracting influential nodes.
Figures~\ref{AsIC:ranking} and \ref{AsLT:ranking} show the results for
the AsIC and the AsLT models, respectively.  For the diffusion model
based methods, we plotted the average value of ${\cal F}(k)$ at $k$ for
five independent experimental results.  We see that the proposed method
gives better results than the other methods for these networks,
demonstrating the effectiveness of our proposed learning
method. It is interesting to note that the model based method in which
an incorrect diffusion model is used is as bad as and in general worse
than the heuristic methods. The results imply that it is important to
consider the information diffusion process explicitly in discussing
influential nodes and also to identify the correct model of information
diffusion for the task in hand, same observation as in Section
\ref{behavioral-difference}.

\section{Model Selection}

Now we have a method to estimate the parameter values from the
observation for each of the assumed models. In this section we discuss
whether the proposed learning method can correctly identify which of the
two models: AsIC and AsLT the observed data come from, {\em i.e.}, {\em
  Model Selection} problem. We assume that the topic is the decisive
factor in determining the parameter values and place a constraint that
the parameters depend only on topics but not on nodes and links of the
network $G$, and differentiate different topics by assigning an index
$l$ to topic $l$.

Therefore, we set $r_{l, u, v} = r_l$ and $p_{l, u, v} = p_l$ for any
link $(u, v) \in E$ in case of the AsIC model and $r_{l, u, v} = r_l$
and $q_{l, u, v} = q_l |B(v)|^{-1}$ for any node $v \in V$ and link
$(u,v) \in E$ in case of the AsLT model. Note that $0 < q_l < 1$ and
$q_{v,v} = 1 - q_l$. Since we normally have a very small number of
observation for each $(l, u, v)$, often only one, without this
constraint, there is no way to learn the parameters.

\subsection{Model Selection based on Predictive Accuracy}
\label{model-selection}

We have to select a model which is more appropriate to the model for the
observed diffusion sequence. We decided to use predictive accuracy as
the criterion for selection. We cannot use an information
theoretic criterion such as AIC (Akaike Information
Criterion)\cite{akaike} or MDL (Minimum Description
Length)\cite{rissanen} because we need to select the one from models
with completely different probability distributions.  Moreover, for both
models, it is quite difficult to efficiently calculate the exact
activation probability of each node for more than two information
diffusion cascading steps ahead.  In order to avoid these difficulties,
we propose a method based on a hold-out strategy, which attempts to
predict the activation probabilities at one step ahead and repeat this
multiple times.

We now group the observed data sequences $D_m$ into topics. Assume that
each topic $l$ has $M_l$ sequences of observation, i.e., $D_l =
\{D_{l,m},\ m=1, \cdots, M_l\}$, where each $D_{l,m}$ is a set of pairs
of active node and its activation time in the $m$-th diffusion result in
the $l$-th topic. Accordingly we add a subscript $l$ to other variables,
{\em e.g.}, we denote $t_{l,m,v}$ to indicate the time $t$ that a node $v$ is
activated in the $m$-th sequence of the $l$-th topic.

We learn the model parameters for each topic separately. This does not
exclude treating each sequence in a topic separately and learn from
each, i.e., $M_l=1$, which would help investigating if the same topic
propagate similarly or not. For simplicity, we assume that for each
$D_{l,m}$, the initial observation time $t_{l,m}$ is zero, i.e.,
$t_{l,m} = 0$ for $m = 1, \cdots, M_l$.  Then, we introduce a set of
observation periods
$$
{\cal I}_l = \{ [0, \tau_{l,n}); \ n = 1, \cdots, N_l \},
$$ where $N_l$ is the number of observation data we want to predict
sequentially and each $\tau_{l,n}$ has the following property: There
exists some $(v, t_{l,m,v}) \in D_{l,m}$ such that $0 < \tau_{l,n} <
t_{l,m,v}$.  Let $D_{l,m; \tau_{l,n}}$ denote the observation data in
the period $[0, \tau_{l,n})$ for the $m$-th diffusion result in the $l$th
topic, i.e.,
$$
D_{l,m; \tau_{l,n}} =
\{ (v, t_{l,m,v}) \in D_{l,m}; \ t_{l,m,v} < \tau_{l,n} \}.
$$ We also set ${\cal D}_{M_l; \tau_{l,n}}$ $=$ $\{ (D_{l,m;
  \tau_{l,n}}, \tau_{l,n})$; $m$ $=$ $1$, $\cdots$, $M_l \}$.  Let
$\bbtheta$ denote the set of parameters for either the AsIC or the AsLT
models, i.e., $\bbtheta = (\bbr, \bbp)$ or $\bbtheta = (\bbr, \bbq)$.
We can estimate the values of $\bbtheta$ from the observation data
${\cal D}_{M_l; \tau_{l,n}}$ by using the learning algorithms in
Sections \ref{learning AsIC} (Appendix A.)  and \ref{learning AsLT}
(Appendix B.).  Let $\widehat{\bbtheta}_{\tau_{l,n}}$ denote the
estimated values of $\bbtheta$.  Then, we can calculate the activation
probability $q^{}_{\tau_{l,n}} (v, t)$ of node $v$ at time $t$ ($\geq
\tau_{l,n}$) using $\widehat{\bbtheta}_{\tau_{l,n}}$.

For each $\tau_{l,n}$, we select the node $v(\tau_{l,n})$ and the time
$t_{l,m(\tau_{l,n}), v(\tau_{l,n})}$ by
\begin{eqnarray*}
t_{l,m(\tau_{l,n}), v(\tau_{l,n})} = \min \left \{ t_{l,m,v}; \ (v,
t_{l,m,v}) \in \bigcup_{m=1}^{M_l} ( D_{l,m} \setminus D_{l,m;
\tau_{l,n}} ) \right \}.
\end{eqnarray*}
Note that $v(\tau_{l,n})$ is the first active node in $t \geq
\tau_{l,n}$. We evaluate the predictive performance for the node
$v(\tau_{l,n})$ at time $t_{l,m(\tau_{l,n}),
v(\tau_{l,n})}$. Approximating the empirical distribution by
$$ p^{}_{\tau_{l,n}} (v, t) \ = \ \delta_{v, v(\tau_{l,n})} \ \delta(t -
t_{l,m(\tau_{l,n}), v(\tau_{l,n})})
$$
with respect to $(v(\tau_n), t_{l,m(\tau_{l,n}), v(\tau_{l,n})})$,
we employ the Kullback-Leibler (KL) divergence
$$
KL (p^{}_{\tau_{l,n}} \, || \, q^{}_{\tau_{l,n}}) \ = \
- \sum_{v \in V} \int_{\tau_{l,n}}^{\infty} 
p^{}_{\tau_{l,n}}(v, t) 
\log \frac{q^{}_{\tau_{l,n}}(v, t)}{p^{}_{\tau_{l,n}} (v, t)} \, dt,
$$
where
$\delta_{v,w}$ and $\delta (t)$ stand for Kronecker's delta and
Dirac's delta function, respectively.
Then, we can easily show
\begin{eqnarray}
KL(p^{}_{\tau_{l,n}} \, || \, q^{}_{\tau_{l,n}}) \ = \
- \log h_{m(\tau_{l,n}), v(\tau_{l,n})}.
\label{KL}
\end{eqnarray}
By averaging the above KL divergence with respect to ${\cal I}_l$,
we propose the following model selection criterion $\cal E$
(see Equation~(\ref{KL})):
\begin{eqnarray}
{\cal E}({\cal A}; D_{l,1} \cup \cdots \cup D_{l,M_l}) & = & -
\frac{1}{N_l} \sum_{n=1}^{N_l} \log h_{m(\tau_{l,n}), v(\tau_{l,n})},
\label{criterion}
\end{eqnarray}
where ${\cal A}$ expresses the information diffusion model (i.e., the
AsIC or the AsLT models).
In our experiments, we adopted
$$ {\cal I}_l \ = \ \{ [0, t_{l,m,v}); \ (v, t_{l,m,v}) \in D_{l,1} \cup
\cdots \cup D_{l,M_l}, \ t_{l, m, v} \geq \tau_0 \},
$$
where $\tau_0$ is the median time of all the observed activation time points.

\subsection{Evaluation by Synthetic Data}
\label{ms_eval}

Our goal here is to evaluate the model selection method to see how
accurately it can detect the true model that generated the data, using
topological structure of four large real networks described in Section
\ref{dataset1}. We assumed the true model by which the data are
generated to be either AsLT or AsIC. We have to repeatedly estimate the
parameters using the proposed parameter update algorithms. In actual
computation the learned values for observation period $[0, \tau_{l,n}]$
are used as the initial values for observation period $[0,
 \tau_{l,n+1}]$ for efficiency purpose.

The average KL divergence given by Equation~(\ref{criterion}) is the
measure for the goodness of the model ${\cal A}$ for a training set
$D_l$ of $M_l$ sequences with respect to topic $l$.  The smaller its
value is, the better the model explains the data in terms of
predictability.  Thus, we can estimate the true model from which $D_l$
is generated to be AsIC if ${\cal E}(AsIC; D_l) < {\cal E}(AsLT; D_l)$,
and vice versa.  Using each of the AsIC and the AsLT models as the true
model, we generated a training set $D_l$. Here we set $M_l=1$, i.e., we
generated a single diffusion sequence, learned a model and performed the
model selection. We repeated this 100 times independently for the four
networks mentioned before. We could have set $M_l=100$ and learned a
single parameter set. This is more reliable, but we wanted to know
whether the model selection algorithm works well or not using only a
single sequence of data.

\begin{table}[!t]
\centering
\caption{Accuracy of the model selection method for four
networks}
\vspace{0.2cm}
\label{table:ms_result}
{
\begin{tabular}{c|c|c|c|c}
\hline
Network            & Blog    & Wikipedia &    Enron & Coauthorship \\
\hline
\hline
AsIC 
 &     92  &      100   &      100 &        93 \\
                   & (370.2) &   (920.8)  & (1500.6) &   (383.5) \\
\hline
AsLT 
&     79 &         86 &     99 &        76 \\
                   & (28.2) &     (54.0) & (47.7) &    (19.0) \\
\hline  
\end{tabular}
}
\end{table}

\begin{figure*}[!t]
\centering
 \subfloat[Blog network \label{icmBlog}]
{\includegraphics[width=4.8cm]{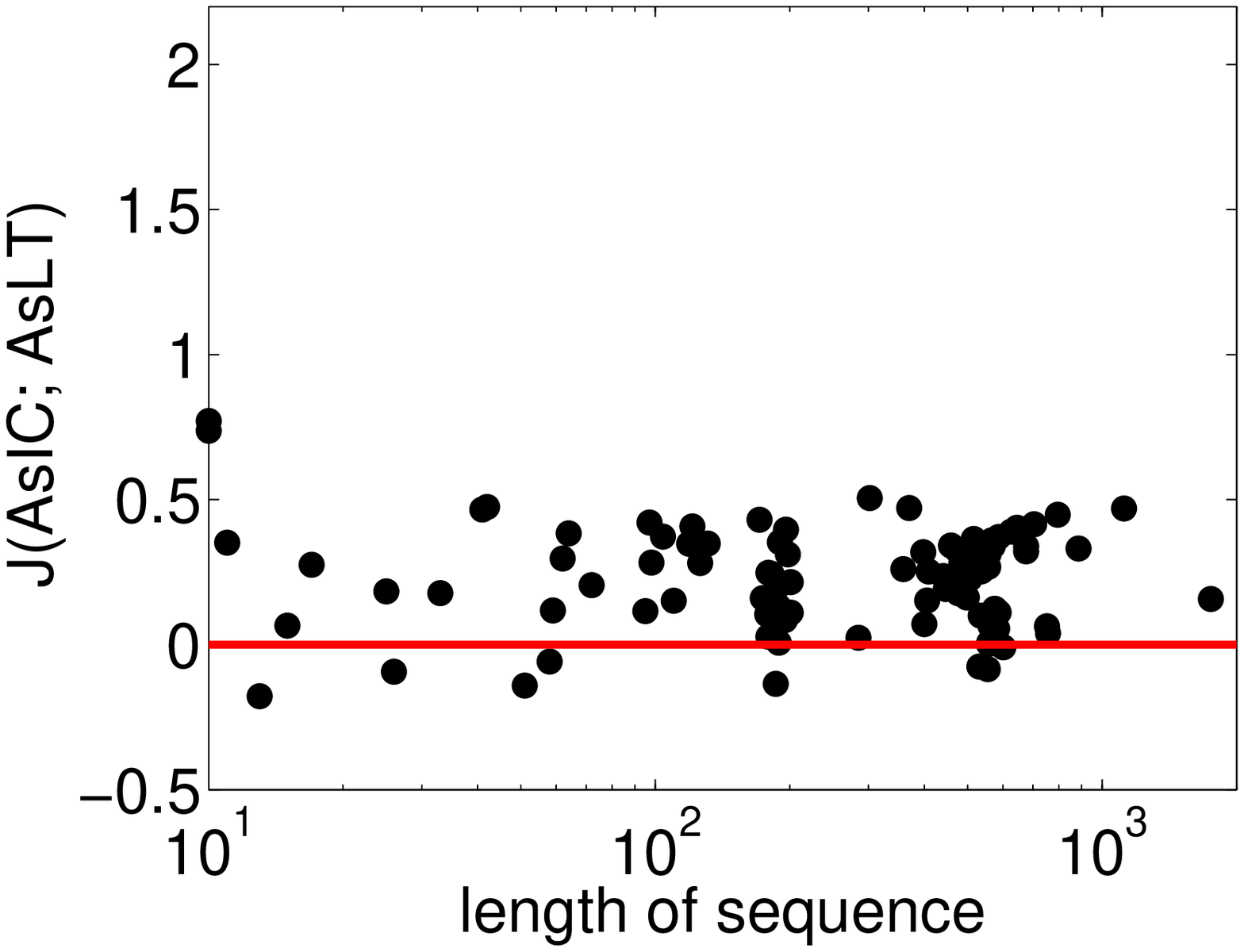}}
\hspace{10mm}
\subfloat[Wikipedia network \label{icmWiki}]
{\includegraphics[width=4.8cm]{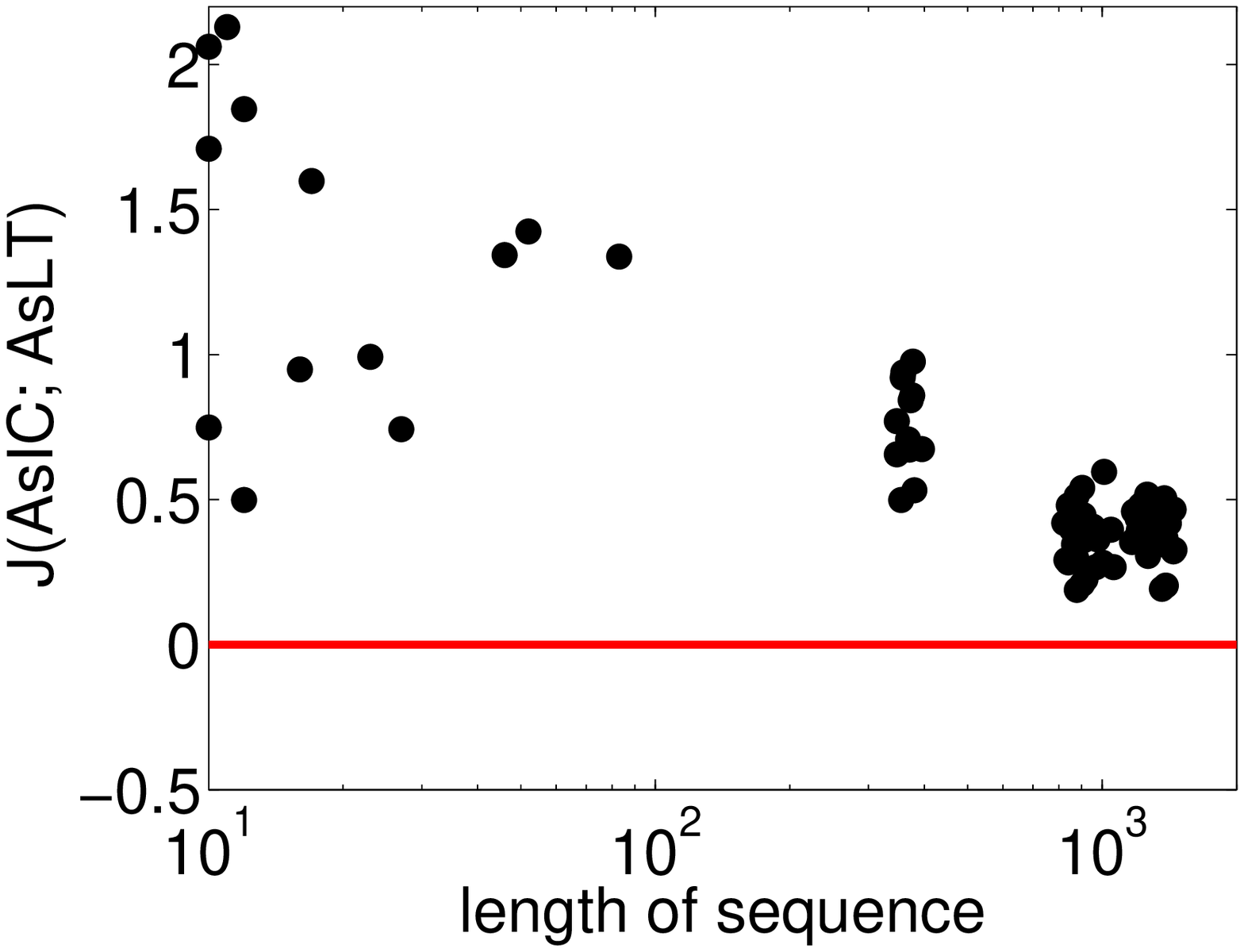}}
\hspace{10mm}
\subfloat[Enron network\label{icmEnron}]
{\includegraphics[width=4.8cm]{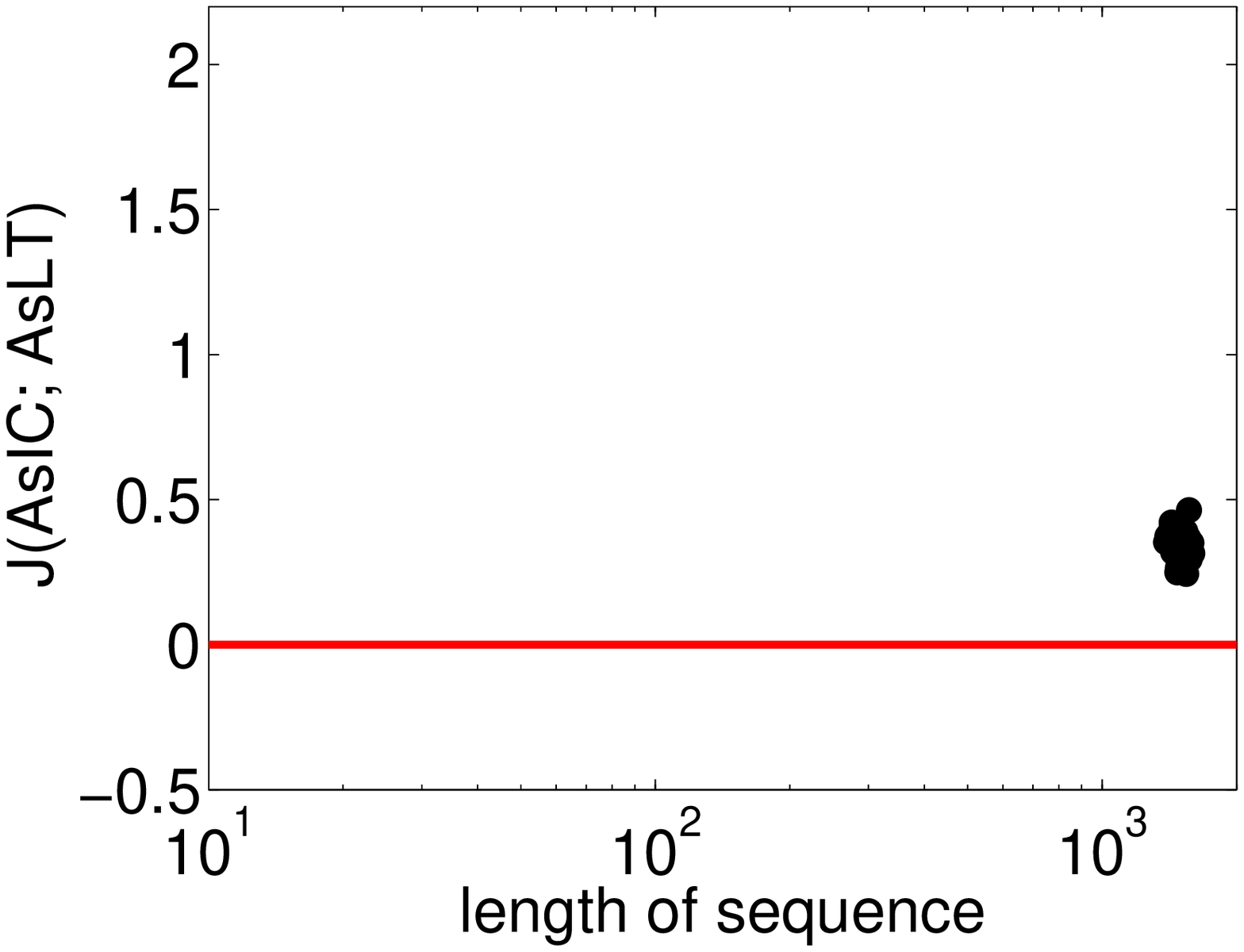}}
\hspace{10mm}
\subfloat[Coauthorship network \label{icmAuthor}]
{\includegraphics[width=4.8cm]{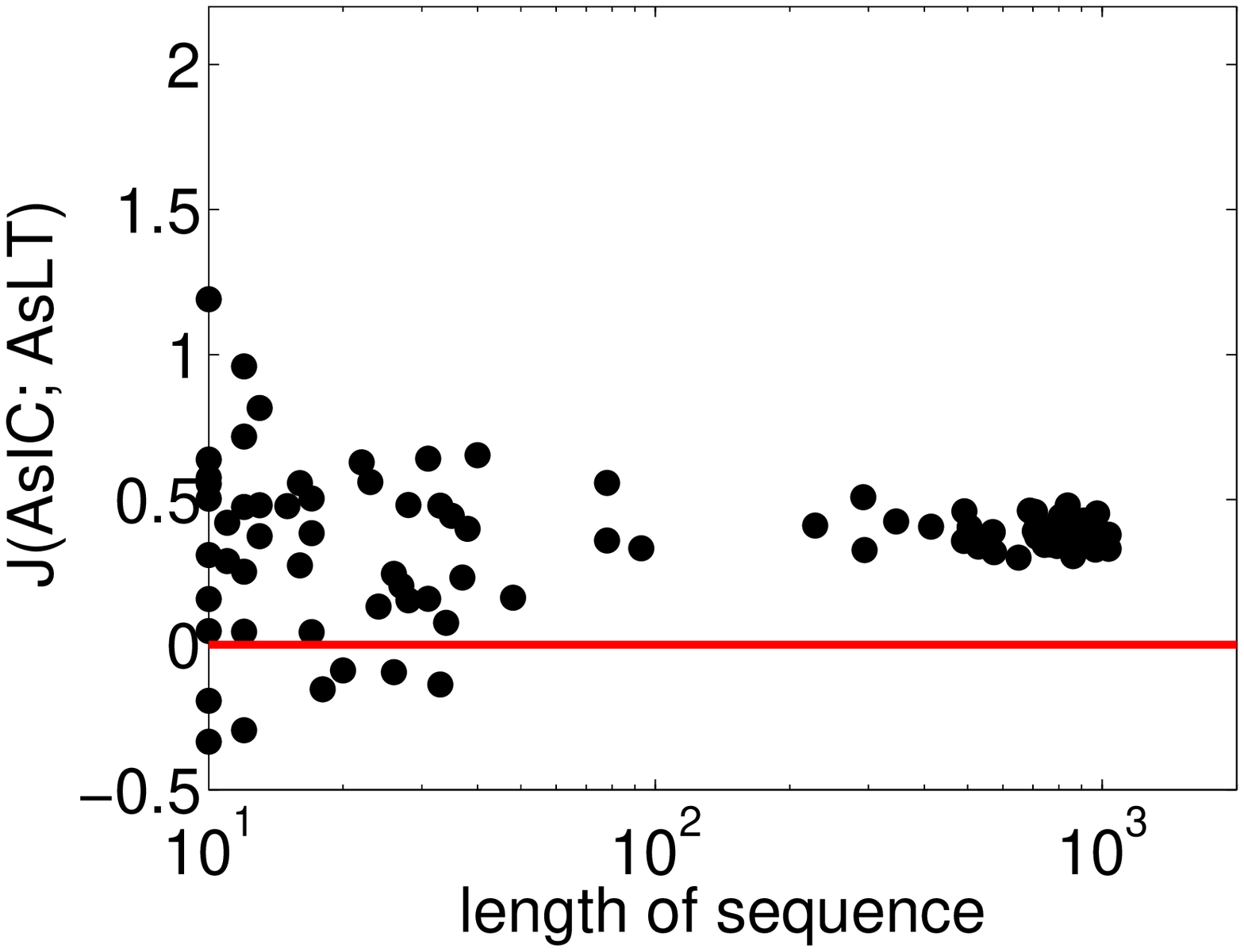}}
\caption{Relation between the length of sequence and 
the accuracy of model selection for a single diffusion sequence
generated from the AsIC model (There are 100 points.)}
\label{icm_dist}
\end{figure*}

\begin{figure*}[!t]
\centering
\subfloat[Blog network \label{ltmBlog}]
{\includegraphics[width=4.8cm]{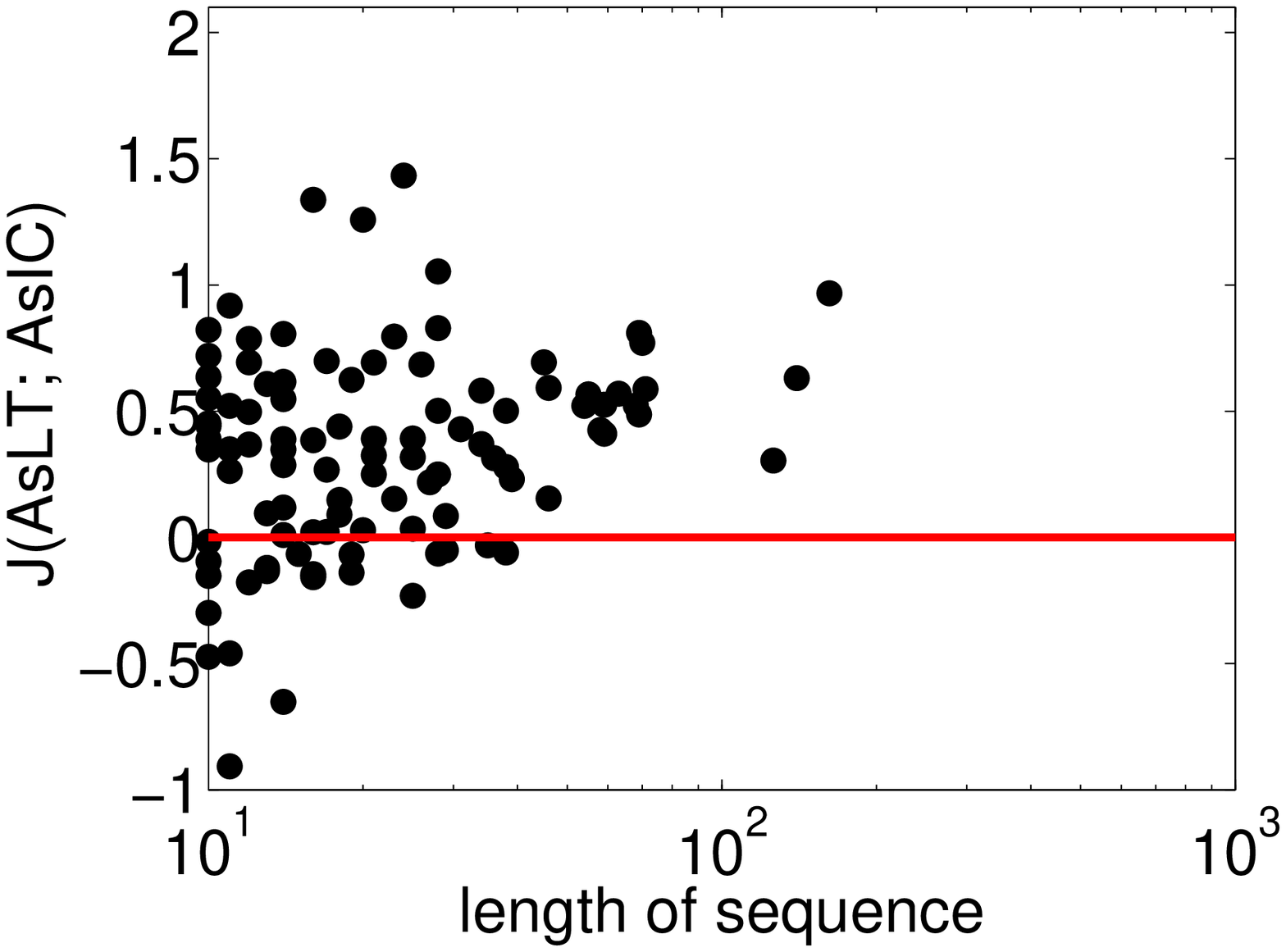}}
\hspace{10mm}
\subfloat[Wikipedia network \label{ltmWiki}]
{\includegraphics[width=4.8cm]{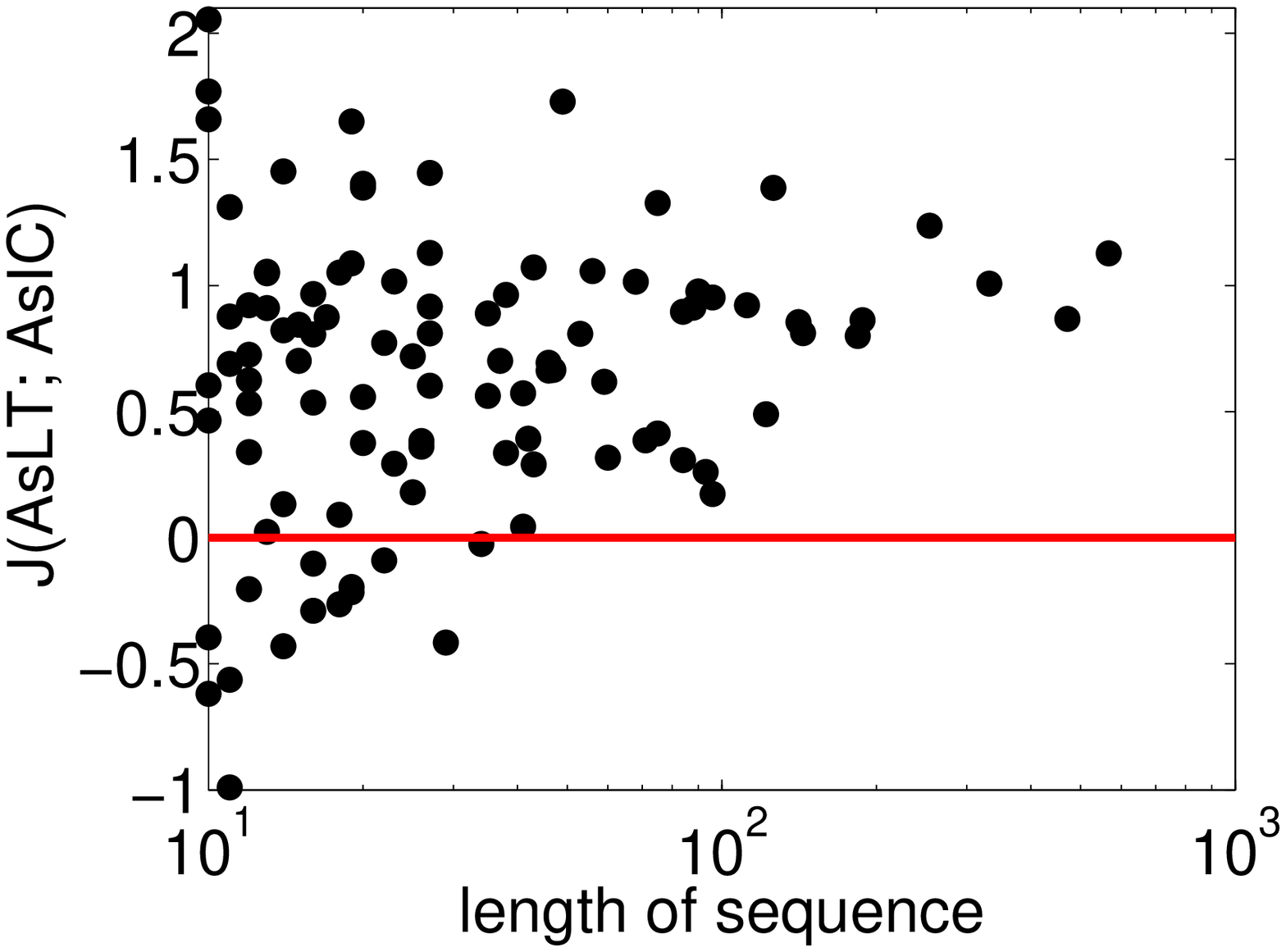}}
\hspace{10mm}
\subfloat[Enron network \label{ltmEnron}]
{\includegraphics[width=4.8cm]{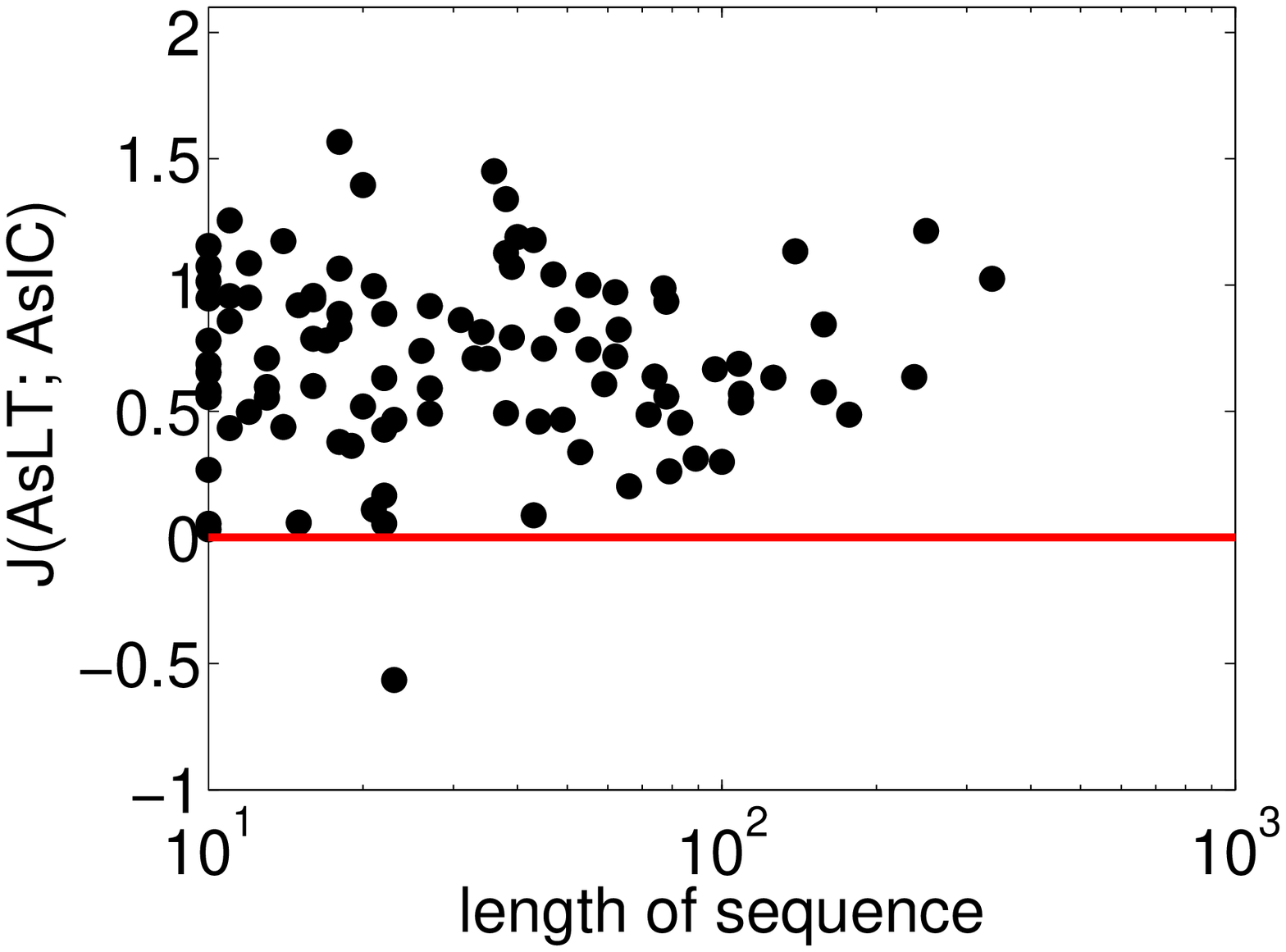}}
\hspace{10mm}
\subfloat[Coauthorship network \label{ltmAuthor}]
{\includegraphics[width=4.8cm]{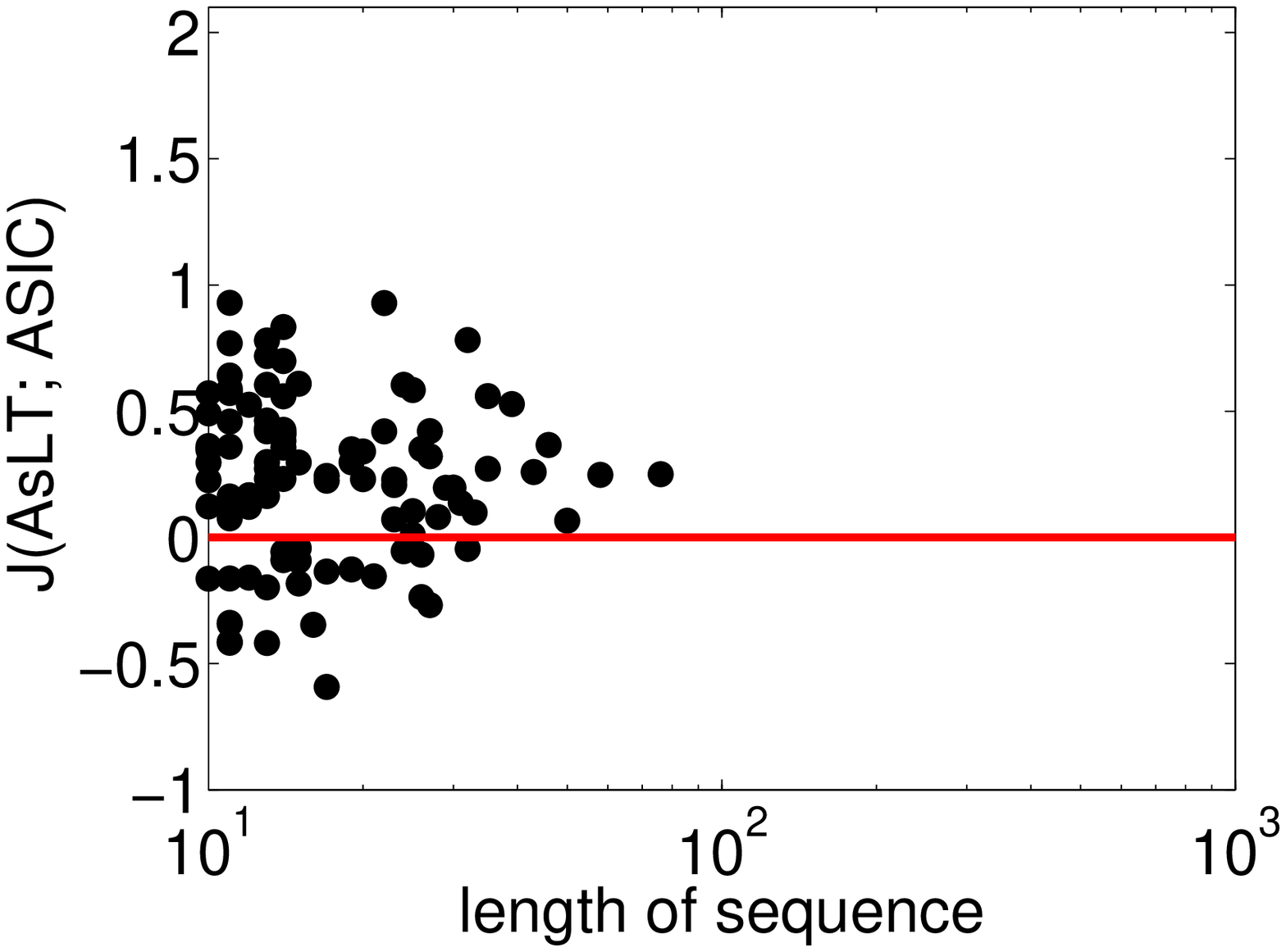}}
\caption{Relation between the length of sequence and
the accuracy of model selection for a single diffusion sequence
generated from the AsLT model (There are 100 points.)}
\label{ltm_dist}
\end{figure*}

Table~\ref{table:ms_result} summarizes the number of times that the
model selection method correctly identified the true model.  The number
within the parentheses is the average length of the diffusion sequences
in the training set.  From these results, we can say that the proposed
method achieved a good accuracy, 90.6\% on average. Especially, for the
Enron network, its estimation was almost perfect. To analyze the
performance of the proposed method more deeply, we investigated the
relation between the length of sequence and the model selection result.
Figure~\ref{icm_dist} shows the results for the case that $D_l$ is
generated by the AsIC model.  Here, the horizontal axis denotes the
length of sequence in each dataset and the vertical axis is the
difference of the average KL divergence defined by $J(AsIC; AsLT) =
{\cal E} (AsLT; D_l) - {\cal E} (AsIC; D_l)$.  Thus,
$J(AsIC; AsLT) > 0$ means that the proposed method correctly estimated
the true model AsIC
because it means

${\cal E} (AsIC; D_l)$ is smaller than ${\cal E} (AsLT; D_l)$.  From the
figure, we can see that there is a correlation between the length of
sequence and the estimation accuracy, and that the misselection occurs
when the length of the sequence is short.  In particular, Wikipedia and
Blog networks have no misselection.  Figure~\ref{ltm_dist} shows the
results for the case that $D_l$ is generated by the AsLT model.  Here,
$J(AsLT; AsIC) = {\cal E} (AsIC; D_l) - {\cal E} (AsLT; D_l)$.  We
notice that the overall accuracy becomes 95.5\% when considering only
the sequences that contain no less than 20 nodes.  This means that the
proposed model selection method is highly reliable for a long sequence
and its accuracy could asymptotically approach to 100\% as the sequence
gets longer.  We can also see from Figures~\ref{icm_dist} and
\ref{ltm_dist} that the results for the AsIC model are better than those
for the AsLT model.  We note that the plots for the diffusion sequences
generated from the AsIC model are shifted to the right in all networks,
meaning that the diffusion sequences are longer for AsIC than for AsLT.
The better accuracy is attributed to this.

\subsection{Evaluation by Real World Blog Data}
\label{b_analysis}

We analyzed the behavior of topics in a real world blog data. Here,
again, we assumed the true model behind the data to be either the AsIC
model or the AsLT model.  Using each pair of the estimated parameters,
$(r_l, p_l)$ for AsIC and $(r_l, q_l)$ for AsLT, we first analyzed the
behavior of people with respect to the information topics by simply
plotting them as a point in $2$-dimensional space.  We next estimated
the true model for each topic by applying the model selection method
described in Section \ref{model-selection}.

\subsubsection{Data Sets and Parameter Setting}

We employed the real blogroll network used by \citeA{saito:acml09},
which was generated from the database of a blog-hosting service in Japan called
{\it Doblog}.
\footnote{ Doblog({\tt http://www.doblog.com/}), provided by NTT Data
  Corp. and Hotto Link, Inc.}  In the network, bloggers are connected to
each other and we assume that topics propagate from blogger $x$ to
another blogger $y$ when there is a blogroll link from $y$ to $x$.  In
addition, according to the work of \citeA{adar}, it is assumed that a
topic is represented as a URL which can be tracked down from blog to
blog. We used the propagation sequences of 172 URLs for this analysis,
each of which has at least 10 time steps. In these 172 URLs some of
them are the same, meaning that there are multiple sequences for the
same topic, i.e., $M_l > 1$. However, as in the analysis of Section
\ref{ms_eval}, we treated them as if $M_l=1$ and used
each sequence independently. The main reason for this is that we want to
investigate whether the same topic propagates in the same way when there
are multiple sequences as well as to test whether the model selection is
feasible from a single sequence data in case of the real data.

\begin{figure*}[!t]
\centering
\subfloat[The AsIC model based method \label{doblog_AsIC}]
{\includegraphics[width=7.5cm]{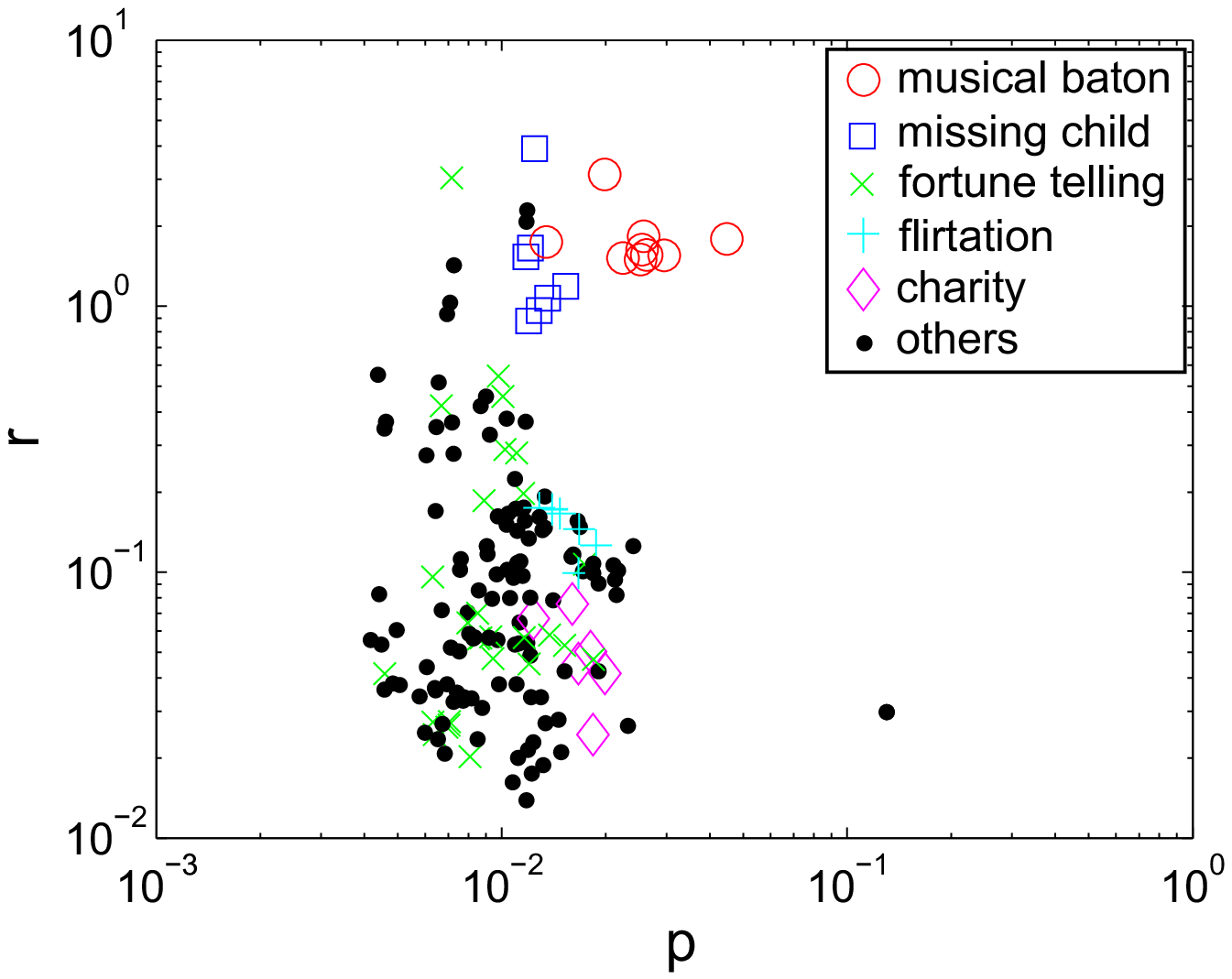}}
\hspace{1mm}
\subfloat[The AsLT model based method \label{doblog_AsLT}]
{\includegraphics[width=7.5cm]{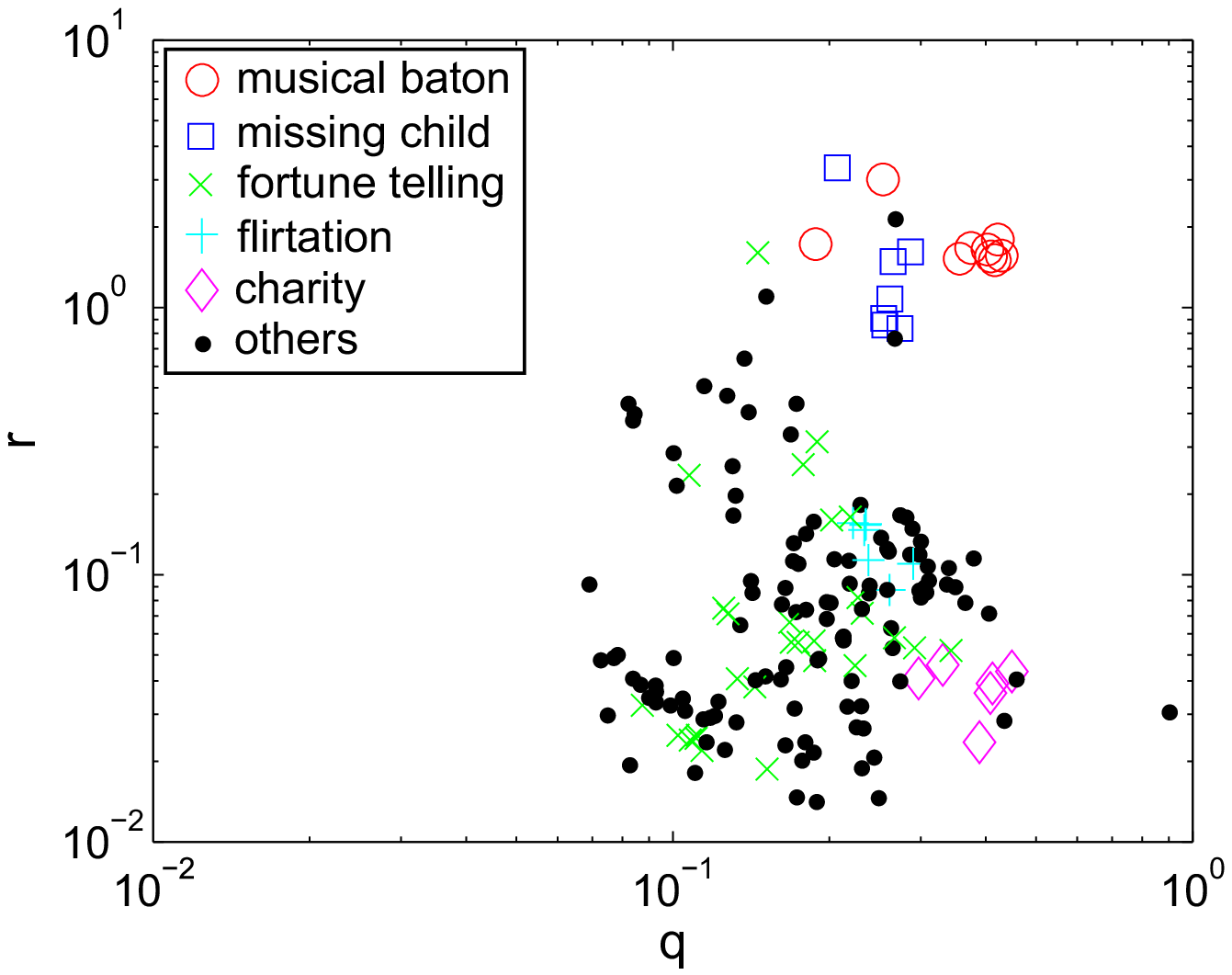}}
\caption{Results for the Doblog database}
\end{figure*}

\subsubsection{Parameter Estimation}
\label{parameter-estimation}

We ran the experiments for each identified URL and obtained the
parameters $p$ and $r$ for the AsIC model based method and $q$ and $r$
for the AsLT model based method. Figures \ref{doblog_AsIC} and
\ref{doblog_AsLT} are the plots of the results for the major URLs
(topics) by the AsIC and AsLT methods, respectively. The horizontal axis
is the diffusion parameter $p$ for the AsIC method and $q$ for the AsLT
method, while the vertical axis is the delay parameter $r$ for both. The
latter axis is normalized such that $r=1$ corresponds to a delay of one
day, meaning $r=0.1$ corresponds to a delay of 10 days.  In these
figures, we used five kinds of markers other than dots, to represent
five different typical URLs: the circle ($\circ$) stands for a URL that
corresponds to the musical baton which is a kind of telephone game on
the Internet (the musical baton),\footnote{It has the following
rules. First, a blogger is requested to respond to five questions
about music by some other blogger (receive the baton) and the
requested blogger replies to the questions and designates the next
five bloggers with the same questions (pass the baton).} the square
($\Box$) for a URL that corresponds to articles about a missing child
(the missing child), the cross ($\times$) for a URL that corresponds to
articles about fortune telling (the fortune telling), the diamond
($\Diamond$) for a URL of a certain charity site (the charity), and the
plus ($+$) for a URL of a site for flirtatious tendency test (the
flirtation). All the other topics are denoted by dots ($\cdot$), which
means they are a mixture of many topics.

The results indicate that in general both the AsIC and AsLT models
capture reasonably well the characteristic properties of topics in a
similar way. We note that the same topic behaves similarly for different
sequences except for the fortune telling. This supports the assumption
we made in Section~\ref{model-selection}. Careful look at the URLs used
to identify the topic of fortune telling indicates that there are
multiple URLs involved and mixing them as a single topic may have been a
too crude assumption. Other interpretation is that people's perception
on this topic is not uniform and varies considerably from person to
person and should be viewed as an exception of the assumption. Behavior
of the other topics is interpretable. For example, the results capture
the urgency of the missing child, which propagates quickly with a
meaningful probability (one out of 80 persons responds).  Musical baton
which actually became the latest craze on the Internet also propagates
quickly (less than one day on the average) with a good chance (one out
of 25 to 100 persons responds).  In contrast non-emergency topics such
as the flirtation and the charity propagate very slowly.  We further
note that the dependency of topics on the parameter $r$ is almost the
same for both AsIC and AsLT, but that on the parameters $p$ and $q$ is
slightly different, e.g., relative difference of musical baton, missing
child and charity. Although $p$ and $q$ are different parameters but
both are the measures that represent how easily the diffusion takes
place. As is shown in Section \ref{ranking}, the influential nodes are
very sensitive to the model used and this can be attributed to the
differences of these parameter values.

\subsubsection{Results of Model Selection}

\begin{table}[!t]
\centering
\caption{Results of model selection for the Doblog dataset}
\vspace{0.2cm}
\label{model_selection_doblog}
\begin{tabular}{c|r|r|r}
\hline
Topic           & Total & AsLT & AsIC\\
\hline
Musical baton   &     9  &  5&    4\\
Missing child   &     7  &  0&    7\\
Fortune telling &    28  &  4&   24\\
Charity         &     6  &  5&    1\\
Flirtation      &     7  &  7&    0\\
Others          &   115  & 11&  104\\
\hline
\end{tabular}
\end{table}

\begin{figure}[t]
\centering
\includegraphics[width=8.4cm]{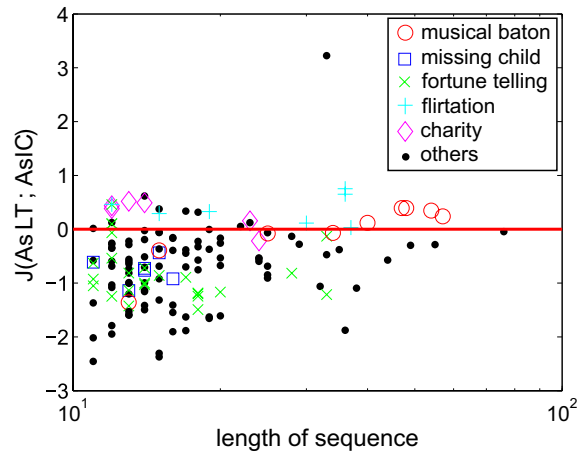}
\caption{The relation between the KL difference and sequence 
length for the Doblog database}
\label{doblog_ltm}
\end{figure}

In the analysis of previous subsection, we assumed that all topics
follow the same diffusion model.  However, in reality this is not true
and each topic should propagate following more closely to either one of
the AsLT and AsIC models. We attempt to estimate the underlying behavior
model of each topic by applying the model selection method described in
Section \ref{model-selection}.  As explained, we treat each sequence
independently and learn the parameters from each sequence, calculate its
KL divergences by Equation~(\ref{criterion}) for both the models, and
compare the goodness.  Table~\ref{model_selection_doblog} and
Figure~\ref{doblog_ltm} summarize the results.  From these results, we
can see that most of the diffusion behaviors on this blog network follow
the AsIC model. It is interesting to note that the model estimated for
the musical baton is not identical to that for the missing child
although their diffusion patterns are very similar (see Section
\ref{parameter-estimation}).  The missing child strictly follows the
AsIC model. This is attributed to its greater urgency. People would post
what they know if they think it is useful without influenced by the
behaviors of their neighbors. For musical baton
Table~\ref{model_selection_doblog} indicates that the numbers are almost
tie (4 vs. 5), but we saw in Section \ref{ms_eval} that the longer
sequence gives a better accuracy, and the models selected in longer
sequences are all AsLT in Figure~\ref{doblog_ltm} for musical baton.
Thus, we estimate that musical baton follows more closely to AsLT. This
can be interpreted that people follow their friends in this
game. Likewise, it is easy to imagine that people would behave similarly
to their neighbors when requested to give a donation. This explains that
charity follows AsLT. The flirtation clearly follows AsLT. People are
attempted to do bad things when their neighbor do so.  Note that there
exists one dot at near the top center in Figure~\ref{doblog_ltm},
showing the greatest tendency to follow AsLT. This dot represents a
typical circle site that distributes one's original news article on
personal events.

\section{Discussion}
\label{discussion}

\citeA{myers:nips10} have recently proposed a method in which the
liklihood is described in somewhat generic way with respect to a given
diffusion dataset for a wide class of IC type information diffusion
models. Their purpose is to infer the latent network structure. On the
other hand, our interest is to explore the salient characteristics of
two contrasting information diffusion models assuming that the
structure is known. Although their purpose is substantially different
from ours, we share with them the common idea of estimating parameters
in information diffusion models.  However, there exist some
mathematically notable differences. The main difference comes from the
derivation of the probability density $h_{m,v}$ that one or more active
parent nodes of a node $v$ succeed(s) in activating $v$ at time $t_{m,v}$
for the $m$-diffusion sequence (see Equation~(\ref{probability1})).  In
order to clarify this point, we denote the corresponding formula used in
\citeA{myers:nips10} by $\tilde{h}_{m,v}$, then $\tilde{h}_{m,v}$ is
expressed as follows:
\begin{eqnarray}
\tilde{h}_{m, v} & = & 1 - \prod_{u \in C_m(t_{m,v})} (1-w(t_{m,v}-t_{m,u})A_{i,j}).  
\label{myers}
\end{eqnarray}
where, according to their terminology, $w(t)$ and $A_{i,j}$ stand for
the transmission time model and the conditional probability of infection
transmission, respectively. Here note that the product term
$w(t_{m,v}-t_{m,u})A_{i,j}$ is equivalent to our formula ${\cal
X}_{m,u,v}$, where ${\cal X}_{m,u,v}$ is defined as the probability
density that a node $u$ activates the node $v$ at time $t_{m,v}$.  (see
Equation~(\ref{calA})).

For an active parent node $u$, the term $(1-w(t_{m,v}-t_{m,u})A_{i,j})$
appearing in Equation~(\ref{myers}) conceptually corresponds to our
formula ${\cal Y}_{m,u,v}$, where ${\cal Y}_{m,u,v}$ is defined as the
probability that the node $v$ is not activated by the node $u$ within
the time-period $[t_{m,u}, t_{m,v})$ (see Equation~(\ref{calB})).  Here
note that from the observed $m$-th diffusion sequence, we know for sure
that the node $u$ could not succeed in activating $v$ during the time
interval $t \in [t_{m,u}, t_{m,v})$.  Namely, our formulation reflects
this observation explicitly in probability estimation, rather than just
subtracting the probability from $1$, as in the expression
$(1-w(t_{m,v}-t_{m,u})A_{i,j})$.  Furthermore, we can transform
Equation~(\ref{calB}) as follows:
\begin{eqnarray}
{\cal Y}_{m,u,v} & = & (1-p_{u,v}) + \int_{t_{m,v}}^{\infty} p_{u,v}
r_{u, v} \exp(- r_{u, v}(t-t_{m,u}))~dt.
\label{calB2}
\end{eqnarray}
Here we can naturally interpret this formula as follows: the first term
of right-hand-side is the probability that the node $u$ fails to
activate $v$, and the second term corresponds to the probability that
the node $u$ succeeds in activating $v$ after the $t_{m,v}$, {\em i.e.},
the fact that the node $v$ is not activated by the node $u$ within the
time-period $[t_{m,u}, t_{m,v})$ means that it has either failed to
activate $v$ at all or succeeded to activate $v$ but the activation time
is outside of the observed time-period.  The basic interpretation of
$\tilde{h}_{m,v}$ is that at least one active parent node activates $v$
at time $t_{m,v}$. Namely, the formulation allows that $v$ is activated
simultaneously by its multiple parent nodes exactly at time $t_{m,v}$,
while our formulation does not consider this possibility. When the
diffusion process unfolds in continuous-time $t$, the probability
measure of such simultaneous activation is zero.
Thus, we employ our $h_{m,v}$ formulation as described in
Equation~(\ref{probability1})). Of course, in case of the discrete-time
modeling, the situation of simultaneous activation by multiple active
parents must be considered adequately.  The objective function for this
case under the discrete-time IC model has been derived
in~\citeA{kimura:ida}.  The major advantage of their method is that it
guarantees a unique optimal solution, whereas ours only guarantees that
it converges to a stationary solution which is not necessarily a global
maximum.  However, it is not clear that a similar approach can be
applied to Linear Threshold type diffusion models.  In addition, as
discussed above and also shown in Section~\ref{alternative-time-delay},
we need to elaborate on the formula for $h_{m,v}$ in order to model the
information diffusion process more accurately reflecting subtle notion
of different time delay models and as much information of observed data
as possible. It is also not clear that the above advantage of their
formulation still holds when the formula for $\tilde{h}_{m,v}$ is
modified accordingly. Our view is that their formulation can be a useful
technique for inferring latent network structure, but it has limitation
if we use it to explore the salient characteristics of different
diffusion models. In this sense, we believe that our approach based on
the EM-like learning algorithm remains vital and useful for a wide class
of information diffusion models.

We started with general description for the parameter values but had to
introduce drastic simplification in experimental evaluations both for
synthetic datasets and real world datasets. The results in Section
\ref{parameter-estimation} implies that the assumption of topics being a
decisive factor for diffusion parameter values seems to be plausible,
which in turn justifies the use of the same parameter values for
multiple sequence observation data if they are talking on the same
topic. However, as one counter example is observed (fortune telling),
this is definitely not true in general. Finding a small number of
factors, {\em e.g.}, important node attributes, from which the parameter
values can be estimated in good accuracy is a crucial problem. Learning
such dependency is easy as exemplified in \citeA{saito:ismis11} once
such factors are identified and the real world data for such factors ara
available as part of observed information diffusion data.

As we explained in Section \ref{ranking}, the ranking results that
involve detailed probabilistic simulation are very sensitive to the
underlying model which is assumed to generate the observed data. In
other words, it is very important to select an appropriate model for the
analysis of information diffusion from which the data has been generated
if the node characteristics are the main objective of analysis,
e.g., such problems as the influence maximization problem
~\cite{kempe:kdd,kimura:dmkd}, a problem at a more detailed
level. However, it is also true that the parameters for the topics that
actually propagated quickly/slowly in observation converged to the
values that enable them to propagate quickly/slowly on the model,
regardless of the model chosen.  Namely, we can say that the difference
of models does not have much influence on the relative difference of
topic propagation which indeed strongly depends on topic itself. Both
models are well defined and can explain this property at this level of
abstraction. Nevertheless, the model selection is very important if we
want to characterize how each topic propagates through the network.


One of the objectives of this paper is to understand the behavioral
difference between the AsIC model and the AsLT model. The analysis in
Section \ref{behavior-difference-results} is based on the network
structures taken from real world data. We feel more mathematical-oriented
treatment is needed to qualitatively understand the behavior
difference of these two models for a wide class of graphs from various
perspectives, {\em e.g.}, types of graphs: regular vs random, graphs with
different characteristics: power-law, small-worldness, community structure, etc.

There are other studies that deal with topic dependent information
diffusion.  Recent study by \citeA{romero:www11} discusses differences
in the diffusion mechanism across different topics. They experimentally
obtain from the observation data the probability $p(k)$ that a node gets
activated after its active parents failed to activate it $k-1$ times in
succession, and model the diffusion process using $p(k)$ under the SIR
(Susceptible/Infectious/Recover) setting. Their finding is that the
shape of $p(k)$ differs considerably from one topic to another, which
is characterized by two factors, stickness (maximum value of $p(k)$) and
persistency (rate of $p(k)$'s decay after the peak), and that the
repeated exposures to a topic are particularly crucial when it is in some
way controversial or contentious. Another recent study on Twitter by
\citeA{bakshy:wsdm11} attempts to quantify a node's influence degree
(the number of nodes that a seed node (initial node) can activate by
learning a regression tree using various node's attributes such as
no. of followers, no. of friends, no. of tweets, past influence degree
and content related features. To their surprise none of the content
related attributes are selected in the learned regression tree. They
attribute this to the fact that most explanations of success tend to
focus only on observed success, which invariably represent a small and
biased sample of the total population. They conclude that individual
level predictions of influence is unreliable, and it is important to
rely on average performance. Both studies approach the similar problem
from different angles. There are many factors that need be considered
and much more work is needed to understand this problem.

\section{Conclusion}

We deal with the problem of analyzing information diffusion process in a
social network using probabilistic information diffusion models. There
are two contrasting fundamental models that have been widely used by
many people: Independent Cascade model and Linear Threshold model. These
are modeled based on two different ends of the spectrum. The IC model is
sender-centered ({\em information push style model}) where the
information sender tries to push information to its neighbors, whereas
the LT model is receiver-centered ({\em information pull style model}
where the information receiver tries to pull information. We extended
these two contrasting models (called AsIC and AsLT) by incorporating
asynchronous time delay to make them realistic enabling effective use of
observed information diffusion data. Using these as the basic tools, we
challenged the following three problems: 1) to clarify how these two
contrasting models differ from or similar to each other in terms of
information diffusion, 2) to devise effective algorithms to learn the
model itself from the observed information diffusion data, and 3) to
identify which model is more appropriate to explain for a particular
topic (information) to diffuse/propagate.

We first showed that there can be variations to each of these two models
depending on how we treat time delay. We identified there are two kinds
of time delay: link delay and node delay, and the latter is further
divided into two categories: override and non-override. We derived the
liklihood function, the probability density to generate the observed
data for each model.  Choosing one particular time delay model, we
showed that the model parameters are learnable from a limited amount of
observation by deriving the parameter update algorithm for both AsIC and
AsLT that maximizes the likelihood function which is guaranteed to
converge and performs stably. We also proposed a method to select a
model that better explains the observation based on its predictive
accuracy. To this end, we devised a variant of hold-out training
algorithm applicable to a set of sequential data and a method to select
a better model by comparing the predictive accuracy using the KL
divergence.

Extensive evaluations were performed using both synthetic data and real
data. We first showed using synthetic data with the network structures
taken from four real networks that there are considerable behavioral
difference between the AsIC and the AsLT models, and gave a qualitative
account of why such difference is brought. We then experimentally
confirmed that the proposed parameter update algorithm converges to the
correct values very stably and efficiently, it can learn the parameter
values even from a single observation sequence if its length is
reasonably long, it can estimate the influential nodes quite accurately
whereas the frequently used centrality heuristics performs very poorly,
the influential nodes are very sensitive to the model used, and the
proposed model selection method can correctly identify the diffusion
models by which the observed data is generated. We further applied the
methods to the real blog data and analyzed the behavior of topic
propagation. The relative propagation speed of topics, {\em i.e.}, how
far/near and how fast/slow each topic propagates, that are derived from
the learned parameter values is rather insensitive to the model
selected, but the model selection algorithm clearly identifies the
difference of model goodness for each topic. We found that many of the
topics follow the AsIC model in general, but some specific topics have
clear interpretations for them being better modeled by either one of the
two, and these interpretations are consistent with the model selection
results. There are numerous factors that affect the information
diffusion process, and there can be a number of different
models. Understanding the behavioral difference of each model, learning
these models efficiently from the available data and selecting the
correct model are a big challenge in social network analysis and this
work is the first step towards this goal.

\acks{This work was partly supported by Asian Office of Aerospace Research and
Development, Air Force Office of Scientific Research under Grant
No. AOARD-11-4111, and JSPS Grant-in-Aid for Scientific Research (C)
(No. 23500194).
}

\appendix
\section*{Appendix A. Learning Algorithm for AsIC model}
\label{asic-learning}

Maximizing ${\cal L}(\bbr, \bbp;{\cal D}_M)$ is equivalent to maximizing
its logarithm.
Let $\bar{\bbr} = ({\bar r}_{u,v})$ and $\bar{\bbp} = ({\bar
p}_{u, v})$ be the current estimates of $\bbr$ and $\bbp$, respectively.
Taking $\log$ of $ h_{m,v}$ involves $\log$ of sum of ${\cal X}_{m,u,v}
({\cal Y}_{m,u,v})^{-1}$, which is problematic. To get around this
problem, we
define $\alpha_{m,u,v}$ for each $(v, t_{m,v}) \in D_m$ and $u$ $\in$
${\cal B}_{m,v}$, by
$$
\alpha_{m,u,v} \ = \
{\cal X}_{m,u,v} ({\cal Y}_{m,u,v})^{-1} \, 
\left/  \,
\sum_{z \in {\cal B}_{m,v}} {\cal X}_{m,z,v} ({\cal Y}_{m,z,v})^{-1}.
\right.
$$
Let
$\bar{\cal X}_{m,u,v}$, $\bar{\cal Y}_{m,u,v}$,
$\bar{h}_{m,v}$, and $\bar{\alpha}_{m,u,v}$
denote the values of
${\cal X}_{m,u,v}$, ${\cal Y}_{m,u,v}$,
$h_{m,v}$, and $\alpha_{m,u,v}$
calculated by using $\bar{\bbr}$ and $\bar{\bbp}$,
respectively.

From Equations~(\ref{probability1}), (\ref{probability5}) and (\ref{objective}),
we can transform our objective function
${\cal L} (\bbr, \bbp; {\cal D}_M)$ as follows:
\begin{equation}
\log {\cal L}(\bbr, \bbp;{\cal D}_M)
=  Q(\bbr, \bbp; \bar{\bbr}, \bar{\bbp})
- H(\bbr, \bbp; \bar{\bbr}, \bar{\bbp}),
\label{AsIC:transformation0}
\end{equation}
where $Q(\bbr, \bbp; \bar{\bbr}, \bar{\bbp})$ is defined by
\begin{eqnarray*}
Q(\bbr, \bbp; \bar{\bbr}, \bar{\bbp}) = \sum_{m = 1}^{M} \left(
\sum_{v \in C_m} Q_{m,v}
+\sum_{v \in C_m} \sum_{w \in F(v) \setminus C_m} \log(1-p_{v,w})
\right),
\nonumber \\
Q_{m,v} = \sum_{u \in {\cal B}_{m,v}} \log \left( {\cal Y}_{m,u,v} \right)
+ \sum_{u \in {\cal B}_{m,v}} {\bar \alpha}_{m,u,v}
\log \left( {\cal X}_{m,u,v} ({\cal Y}_{m,u,v})^{-1} \right)
\end{eqnarray*}
and
$H(\bbr, \bbp; \bar{\bbr}, \bar{\bbp})$ is defined by
\begin{equation}
H(\bbr, \bbp; \bar{\bbr}, \bar{\bbp}) = \sum_{m = 1}^{M}
\sum_{v \in C_m} \sum_{u \in {\cal B}_{m,v}}
{\bar \alpha}_{m,u,v} \log \alpha_{m,u,v}.
\label{transformation}
\end{equation}
Since $H(\bbr, \bbp; \bar{\bbr}, \bar{\bbp})$ is maximized
at $\bbr = \bar{\bbr}$ and $\bbp = \bar{\bbp}$
from Equation~(\ref{transformation}),\footnote{This can be easily
  verified using the Lagrange multipliers method with the constraint $\sum_{u
  \in {\cal B}_{m,v}}\alpha_{m,u,v}=1$.}
we can increase
the value of ${\cal L}(\bbr, \bbp; {\cal D}_M)$
by maximizing $Q(\bbr, \bbp; \bar{\bbr}, \bar{\bbp})$
(see Equation~(\ref{AsIC:transformation0})).
Note here that $Q$ is a convex function with respect to $\bbr$ and
$\bbp$, and thus the convergence is guaranteed.  
Here again we have a problem of $\log$ of sum for $\log {\cal Y}_{m,u,v}$.
In order to cope with this problem, we
transform $\log {\cal Y}_{m,u,v}$ in the same way as we introduced
$\alpha_{m,u,v}$, and define $\beta_{m,u,v}$ by
$$
\beta_{m,u,v} = p_{u,v} \exp(- r_{u, v}(t_{m,v}-t_{m,u}))  \, / \, {\cal Y}_{m,u,v}.
$$ 
Finally, we obtain the following update formulas of our estimation
method as the solution which maximizes $Q(\bbr, \bbp; \bar{\bbr},
\bar{\bbp})$:
\begin{eqnarray*}
r_{u, v} & = &
\frac{
\sum_{m \in {\cal M}^+_{u,v}} \bar{\alpha}_{m,u,v}
}
{
\sum_{m \in {\cal M}^+_{u,v}} (\bar{\alpha}_{m,u,v} + (1-\bar{\alpha}_{m,u,v}) \bar{\beta}_{m,u,v}) (t_{m,v} - t_{m,u})
},
\\
p_{u, v} & = &
\frac{1}{|{\cal M}^+_{u,v}| + |{\cal M}^-_{u,v}|}
\sum_{m \in {\cal M}^+_{u,v}} (\bar{\alpha}_{m,u,v} + (1-\bar{\alpha}_{m,u,v}) \bar{\beta}_{m,u,v}),
\end{eqnarray*}
where ${\cal M}^+_{u, v}$ and ${\cal M}^-_{u, v}$ are defined by
\begin{eqnarray*}
{\cal M}^+_{u, v} & = & \{ m \in \{1, \cdots, M\}; \
v \in C_m, \ u \in {\cal B}_{m,v} \}, 
\\
{\cal M}^-_{u, v} & = & \{ m \in \{1, \cdots, M\}; \
u \in C_m, \ v \in \partial C_m\}.
\end{eqnarray*}
Note that we can regard our estimation method as a variant of the EM
algorithm.  We want to emphasize here that each time iteration proceeds
the value of the likelihood function never decreases and the iterative
algorithm is guaranteed to converge due to the convexity of $Q$.

\section*{Appendix B. Learning Algorithm for AsLT model}
\label{aslt-learning}

An iterative parameter update algorithm similar to the AsIC model can be
derived for the AsLT model, too. We first define 
$\phi_{m,u,v}$ for each $v \in C_m$ and $u \in {\cal B}_{m,v}$, 
$\varphi_{m,u,v}$ for each $v \in \partial C_m$ 
and $u$ $\in$ $\{v\}$ $\cup$ $B(v)$ $\setminus$ ${\cal B}_{m,v}$,
and $\psi_{m,u,v}$ for each $v \in \partial C_m$ 
and $u \in {\cal B}_{m,v}$, respectively by the following formulas. 
\begin{eqnarray*}
\phi_{m,u,v} & = & q_{u,v} r_{u,v} \exp(-r_{u,v}(t_{m,v}-t_{m,u})) \, / \, h_{m,v},\\
\varphi_{m,u,v} & = & q_{u,v} \, / \, g_{m,v}, \\
\psi_{m,u,v} & = & q_{u,v} \exp(-r_{u,v}(T_m-t_{m,u})) \, / \, g_{m,v}. 
\label{posterior1}
\end{eqnarray*}
Let $\bar{\bbr} = ({\bar r}_{v})$ and $\bar{\bbq} = ({\bar q}_{u, v})$ 
be the current estimates of $\bbr$ and $\bbq$, respectively.
Similarly, let
$\bar{\phi}_{m,u,v}$, $\bar{\varphi}_{m,u,v}$, and $\bar{\psi}_{m,u,v}$
denote the values of
${\phi}_{m,u,v}$, ${\varphi}_{m,u,v}$, and ${\psi}_{m,u,v}$
calculated by using $\bar{\bbr}$ and $\bar{\bbq}$,
respectively.

From Equations~(\ref{AsLT:probability1}), (\ref{AsLT:probability5}) and
(\ref{AsLT:objective}),
we can transform
${\cal L} (\bbr, \bbq; {\cal D}_M)$ as follows:
\begin{equation}
\log {\cal L}(\bbr, \bbq;{\cal D}_M)
=  Q(\bbr, \bbq; \bar{\bbr}, \bar{\bbq})
- H(\bbr, \bbq; \bar{\bbr}, \bar{\bbq}),
\label{transformation0}
\end{equation}
where $Q(\bbr, \bbq; \bar{\bbr}, \bar{\bbq})$ is defined by 
\begin{eqnarray}
Q(\bbr, \bbq; \bar{\bbr}, \bar{\bbq}) = \sum_{m = 1}^{M} \left(
\sum_{v \in C_m} Q_{m,v}^{(1)}
+\sum_{v \in \partial C_m} Q_{m,v}^{(2)}
\right),
\label{qfunction}
\end{eqnarray}
\begin{eqnarray*}
Q_{m,v}^{(1)} & = & \sum_{u \in {\cal B}_{m,v}} \bar{\phi}_{m,u,v} 
\log ( q_{u, v} r_v \exp(- r_v(t_{m,v}-t_{m,u}))) \nonumber \\
Q_{m,v}^{(2)} & = & 
\sum_{u \in \{v\} \cup B(v) \setminus {\cal B}_{m,v}} \bar{\varphi}_{m,u,v} 
\log ( q_{u, v})
+ \sum_{u \in {\cal B}_{m,v}} \bar{\psi}_{m,u,v} 
\log ( q_{u, v} \exp(- r_v(T_{m}-t_{m,u}))).
\end{eqnarray*}
It is easy to see that $Q(\bbr, \bbq; \bar{\bbr}, \bar{\bbq})$
is convex with respect to $\bbr$ and $\bbq$, 
and
$H(\bbr, \bbp; \bar{\bbr}, \bar{\bbq})$ is defined by
\begin{eqnarray}
H(\bbr, \bbq; \bar{\bbr}, \bar{\bbq}) = \sum_{m = 1}^{M} \left(
\sum_{v \in C_m} H_{m,v}^{(1)}
+\sum_{v \in \partial C_m} H_{m,v}^{(2)}
\right),
\label{hfunction}
\end{eqnarray}
\begin{eqnarray*}
H_{m,v}^{(1)} & = &
\sum_{u \in {\cal B}_{m,v}} \bar{\phi}_{m,u,v} 
\log ( \phi_{m,u,v} ),\\
H_{m,v}^{(2)} & =  &
\sum_{u \in \{v\} \cup B(v) \setminus C_m} \bar{\varphi}_{m,u,v} 
\log ( \varphi_{m,u,v} )
+ \sum_{u \in {\cal B}_{m,v}} \bar{\psi}_{m,u,v} 
\log ( \psi_{m,u,v}).
\end{eqnarray*}
Since $H(\bbr, \bbq; \bar{\bbr}, \bar{\bbq})$ is maximized
at $\bbr = \bar{\bbr}$ and $\bbq = \bar{\bbq}$
from Equation~(\ref{hfunction}),
we can increase
the value of ${\cal L}(\bbr, \bbq; {\cal D}_M)$
by maximizing $Q(\bbr, \bbq; \bar{\bbr}, \bar{\bbq})$
(see Equation~(\ref{transformation0})).

Thus, we obtain the following update formulas of our estimation method
as the solution which maximizes $Q(\bbr, \bbq; \bar{\bbr},
\bar{\bbq})$ with respect to $\bbr$ :
\begin{eqnarray*}
r_{u,v}  & = & \left (
\sum_{m \in {\cal M}^{(1)}_{v}} 
\sum_{u \in {\cal B}_{m,v}} \bar{\phi}_{m,u,v}
\right ) \\
& & 
\times \ \left (
\sum_{m \in {\cal M}^{(1)}_{v}} 
\sum_{u \in {\cal B}_{m,v}} \bar{\phi}_{m,u,v} (t_{m,v}-t_{m,u})
+ \sum_{m \in {\cal M}^{(2)}_{v}} 
\sum_{u \in {\cal B}_{m,v}} \bar{\psi}_{m,u,v} (T_{m}-t_{m,u}) \right )^{-1}
\end{eqnarray*}
where ${\cal M}^{(1)}_{v}$ and ${\cal M}^{(2)}_{v}$ are defined by
\begin{eqnarray*}
{\cal M}^{(1)}_{v} & = & \{ m \in \{1, \cdots, M\}; \
v \in C_m \},\\
{\cal M}^{(2)}_{v} & = & \{ m \in \{1, \cdots, M\}; \
v \in \partial C_m \}.
\end{eqnarray*}
As for $\bbq$, we have to take the constraints $q_{v, v} +
\sum_{u \in B(v)} q_{u, v} = 1$ into account for each $v$, which
can easily be made using the Lagrange multipliers method, and we obtain
the following update formulas of our estimation method:
\begin{eqnarray*}
q_{u, v} & \propto &
\sum_{m \in {\cal M}_{u,v}^{(1)}} \bar{\phi}_{m,u,v}
+ \sum_{m \in {\cal M}_{u,v}^{(2)}} \bar{\varphi}_{m,u,v}
+ \sum_{m \in {\cal M}_{u,v}^{(3)}} \bar{\psi}_{m,u,v},\\
q_{v, v} & \propto &
\sum_{m \in {\cal M}_{v}^{(2)}} \bar{\varphi}_{m,v,v}
\end{eqnarray*}
where ${\cal M}^{(1)}_{u, v}$, ${\cal M}^{(2)}_{u, v}$ 
and ${\cal M}^{(3)}_{u, v}$ are defined by
\begin{eqnarray*}
{\cal M}^{(1)}_{u, v} & = & \{ m \in \{1, \cdots, M\}; \
v \in C_m, \ u \in {\cal B}_{m,v} \},
\\
{\cal M}^{(2)}_{u, v} & = & \{ m \in \{1, \cdots, M\}; \
v \in \partial C_m, \ u \in B(v) \setminus {\cal B}_{m,v} \},
\\
{\cal M}^{(3)}_{u, v} & = & \{ m \in \{1, \cdots, M\}; \
v \in \partial C_m, \ u \in {\cal B}_{m,v}\}.
\end{eqnarray*}
The actual values are obtained after normalization. Here again, the
convergence is guaranteed.

\end{document}